\documentclass[fleqn, usenatbib]{mnras}

\usepackage[T1]{fontenc}
\usepackage{ae, aecompl}

\usepackage{graphicx}        
\usepackage{amsmath}         
\usepackage{amssymb}         
\usepackage{booktabs}        
\usepackage{scalerel}        
\usepackage{xcolor}          
\usepackage{upquote}         
\usepackage{subcaption}      
\usepackage{multirow}        
\usepackage{pdflscape}       
\usepackage{microtype}       
\usepackage{orcidlink}       

\usepackage{listings} 
\definecolor{key1}{RGB}{122, 5, 137}
\definecolor{key2}{RGB}{34, 19, 163}

\newcommand*{\dpi}{600} 

\newcommand*{\orcid}[1]{\;\!\orcidlink{#1}\;\!}

\newcommand*{\units}[1]{\scalebox{0.8}{(#1)}}

\newcommand*{\Sun}{\protect\scalebox{0.7}{$\odot$}}

\newcommand*{\Range}[2]{[#1\,\text{,}\,#2]}
\newcommand*{\pd}[1]{\times\!10^{#1}}

\newcommand{\norm}[1]{\left\lVert #1 \right\rVert}

\newcommand*{\SM}[1]{{\scaleto{\rm #1}{3.5pt}}}

\newcommand*{\var}[1]{\mbox{\footnotesize$(#1)$}}

\newcommand*{\Gband}{\mbox{$G\:\!\text{-band}$} }

\newcommand*{\BV}{{B\textnormal{-}\:\:\!\!\!V}}

\newcommand*{\BPRP}{{G_{\rm BP}\!-\!G_{\rm RP}}}
\newcommand*{\FeH}{{\rm [Fe\!\!\;/\!\!\;H]}}
\newcommand*{\MH}{{\rm [M\!\!\;/\!\!\;H]}}

\newcommand*{\DKL}{ {\rm D_{\protect\scalebox{0.65}{KL}}} }

\newcommand*{\nbody}{\hbox{\textit{N}\!\!\:-body}}
\newcommand*{\stube}{\hbox{S\hspace{0.075em}-\hspace{-0.025em}Tube}}

\newcommand*{\GtrSim}{\smallrel\gtrsim}
\newcommand*{\LessSim}{\smallrel\lesssim}

\newcommand*{\Approx}{\smallrel\sim}

\makeatletter
\newcommand*{\smallrel}[2][.8]{%
  \mathrel{\mathpalette{\smallrel@{#1}}{#2}}%
}
\newcommand*{\smallrel@}[3]{%
  \sbox0{$#2\vcenter{}$}%
  \dimen@=\ht0 %
  \raise\dimen@\hbox{%
    \scalebox{#1}{%
      \raise-\dimen@\hbox{$#2#3\m@th$}%
    }%
  }%
}

\DeclareMathOperator{\erf}{erf}
\DeclareMathOperator{\mean}{mean}
\DeclareMathOperator{\std}{std}
\DeclareMathOperator{\median}{median}

\DeclareMathOperator{\IO}{I_{\:\!0}}

\definecolor{mycolour}{HTML}{D62728}

\newcommand*{\Ht}{\protect\scalebox{0.65}{${\rm H}$}}
\newcommand*{\RL}{\protect\scalebox{0.65}{${\rm RL}$}}

\newcommand*{\St}{\protect\scalebox{0.55}{{\rm S}}}

\title[Inverse Time Integration]{Constraining the potential of the Milky Way using stellar streams and the Inverse Time Integration method}
\author[C. G. Palau, W. Wang, J. Han]{
Carles G. Palau\orcid{0000-0002-7583-534X}$^{1,2}$\thanks{E-mail: cgpalau@sjtu.edu.cn},
Wenting Wang\orcid{0000-0002-5762-7571}$^{1,2}$, 
Jiaxin Han\orcid{0000-0002-8010-6715}$^{1,2}$\thanks{E-mail: jiaxin.han@sjtu.edu.cn}
\\
$^{1}$Department of Astronomy, Shanghai Jiao Tong University, 800 Dongchuan Road, Shanghai 200240, China\\
$^{2}$Shanghai Key Laboratory for Particle Physics and Cosmology, 800 Dongchuan Road, Shanghai 200240, China\\
}

\date{Accepted XXX. Received YYY; in original form ZZZ}

\pubyear{2024}

\begin{document}
\label{firstpage}
\pagerange{\pageref{firstpage}--\pageref{lastpage}}
\maketitle

\begin{abstract}

We develop a method for constraining the potential of the Milky Way using stellar streams with a known progenitor. The method expresses the stream in angle-action coordinates and integrates the orbits of the stars backwards in time to obtain the stripping point positions of the stream stars relative to the cluster. In the potential that generated the stream, the stars return approximately to the cluster centre. In a different potential, they are redirected to different locations. The free parameters of the model are estimated by maximising the degree of clustering of the stripping point distribution. We test this method with the stellar stream of the globular cluster M68 (NGC~4590). We use an \nbody\ code to simulate the stream and generate a realistic star sample using a model of the \textit{Gaia} selection function. We also simulate the expected observational uncertainties, and estimate the heliocentric distances and radial velocities of the stream stars from the cluster orbit. Using this sample of stars, we recover the values of four free parameters characterising the potential of the disc and the dark halo to an accuracy of about 8 per cent of the correct values. We show that this accuracy is improved up to about 2 per cent using the expected end-of-mission \textit{Gaia} data. In addition, we obtain a strong correlation between the mass of the disc and the dark halo axis ratio, which is explained by the fact that the stream flows close and parallel to the disc plane.

\end{abstract}


\defcitealias{2008gady.book.....B}{BT08}
\defcitealias{2019MNRAS.488.1535P}{PM19}



\section{Introduction}

In recent years, many stellar streams have been discovered around the Milky Way \citep{2023MNRAS.520.5225M}. These streams are formed by stars stripped by tidal forces from a progenitor cluster or dwarf galaxy orbiting the Milky Way as a satellite. The most prominent of these is the stream from the Sagittarius dwarf spheroidal galaxy \citep[e.g.][]{1994Natur.370..194I}. Some other streams are the result of the total dissolution of their progenitor. This is the case of the GD-1 stellar stream, which probably originated from a dissolved globular cluster \citep{2006ApJ...643L..17G}. Similar structures of smaller mass and size are expected to be ubiquitous in the inner region of the Galaxy, but they are generally too faint to be observed. Nevertheless, a few dozen of these streams have recently been discovered near the Sun \citep[e.g.][]{2021ApJ...914..123I}, using astrometric and photometric data provided by the \textit{Gaia} mission \citep{2016AandA...595A...1G}.

As the stream stars escape, they roughly follow the orbit of the progenitor cluster \citep[e.g.][]{2010MNRAS.401..105K, 2012MNRAS.420.2700K, 2013MNRAS.433.1813S}. In some cases, the streams show clear signs of external perturbation. For example, the Orphan-Chenab stream is deviated from its original trajectory by the gravitational pull of the Large Magellanic Cloud (LMC) \citep[e.g.][]{2019MNRAS.487.2685E, 2023MNRAS.521.4936K}. But if there is no significant perturbation, a stream is an excellent tracer of the orbit of the progenitor, and this orbit can be used to constrain the potential of the Galaxy. The most useful streams for this purpose are long, thin, and dynamically cold, and have a known progenitor. This is the case of the streams generated by globular clusters such as Palomar 5 \citep{2001ApJ...548L.165O} or M68 (NGC~4590) \citep[][hereafter \citetalias{2019MNRAS.488.1535P}]{2019MNRAS.488.1535P}.

In general, three different approaches have been used to constrain the potential of the host galaxy using stellar streams. They differ in the sense of time integration. In the first method, an initial configuration of the system consisting of progenitor and host galaxy is assumed. The progenitor is then integrated forward in time to the present within the potential of the host galaxy, and the resulting stream is compared with the observational data. By changing the potential, the final state can be modified to give the best fit to the observations. This method allows us to constrain the potential with streams with missing data, for example streams where only sky coordinates are available. However, it requires the assumption of an initial configuration of the system.

Direct \nbody\ simulations are the most accurate method to obtain the final configuration \citep[e.g.][]{2010ApJ...714..229L}, but they are computationally expensive. This limits the ability to fully explore a multidimensional parameter space. Several methods have been developed to speed up the simulation by introducing different assumptions. For example, avoiding \nbody\ interactions by releasing particles from the Lagrange points \citep[e.g.][]{2012MNRAS.420.2700K, 2014MNRAS.445.3788G} or exploiting the symmetries of the stream in angle-action coordinates \citep{2014MNRAS.443..423S, 2016ApJ...833...31B}. In addition, several simplifications of the stream morphology have been used to reduce the computational time under certain conditions. For example, the assumption of a constant internal dispersion and offset of the stream with respect to the orbit of the cluster \citep{2023MNRAS.524.2124P}, the reduction of the stream to a one-dimensional model delineating an orbit \citep[e.g.][]{2010ApJ...712..260K, 2019MNRAS.486.2995M}, or as a collection of orbits \citep{2022ApJ...940...22N}.

A second method can be defined by an inverse time integration of the system, from its present state to an initial configuration \citep{1999ApJ...512L.109J, 2014ApJ...794....4P}. The main advantage is that the initial configuration corresponds to all stars within the progenitor and does not require further assumptions. This configuration can be recovered by integrating in the correct potential. A simulated model of the stellar stream is therefore not required. However, this method requires full phase-space information of the progenitor and the stream stars, which is often not available.

Finally, a third approach avoids time integration by working with integrals of motion such as energy or angular momentum \citep[e.g.][]{2012ApJ...760....2P, 2015ApJ...801...98S, 2021MNRAS.502.4170R, 2024MNRAS.532.2657B}. In this case, it is assumed that the degree of clustering of the integrals of motion of the stream stars is maximised for the correct potential of the host galaxy. This method does not require the existence of a progenitor, but assumes the existence of the integrals of motion, for example, by assuming an axisymmetric time-independent potential. It also requires the full phase-space information of the stars. Similarly, the integration can be avoided by calculating the angle-action coordinates of the observed stars \citep{2013MNRAS.433.1826S}. In the correct potential, the stream is linearly symmetric, indicating that all the stars are coming from the same source.

In this paper, we develop a variant of the second method, which we call Inverse Time Integration (\texttt{invi}). The main difference from \texttt{Rewinder} \citep{2014ApJ...794....4P} is that we integrate backwards in time in angle-action coordinates instead of usual coordinates. In our case, the integration is straightforward because the stars follow linear uniform motions. In addition, knowing the current phase-space position of the progenitor, we can determine the time integration from the relative separation of the stream stars from the cluster. A disadvantage of this approach is that the methods used to compute angles and frequencies require a static potential. This limits the inclusion of perturbations that have been shown to be important in modelling some stellar streams such as the Galactic bar \citep[e.g.][]{2016ApJ...824..104P, 2016MNRAS.460..497H, 2017MNRAS.470...60E, 2017NatAs...1..633P} or the LMC \citep[e.g.][]{2023Galax..11...59V, 2025ApJ...978...79B}.

We explore the capabilities of the \texttt{invi} method by applying it to the case of the globular cluster M68. This cluster is located in the North Galactic Hemisphere at about 36 deg from the plane of the disc. Its stream was first identified by \citet{2019ApJ...872..152I} using the \texttt{STREAMFINDER} method \citep{2018MNRAS.477.4063M} and named it Fj\"orm. It was then identified by \citetalias{2019MNRAS.488.1535P} as the stream generated by M68. This stream is one of the longest known and flows almost parallel to the Galactic disc, at about 5 kpc from the plane of the disc. Due to observational limitations such as distance, foreground contamination, etc., we can only observe a section of the leading arm. However, this section is long and thin, spanning about 190 deg across the North Galactic Hemisphere, and it is close to the Sun, at about 5 kpc. In addition, its projection on the sky is such that it is sensitive to variations in the potential of the Galaxy, especially the shape of the dark halo and the mass of the disc \citep{2023MNRAS.524.2124P}.

The characteristics of the M68 stream, together with the precise knowledge of the six phase-space coordinates of the cluster, its mass and density profile, make this system the optimal candidate for testing our method. With this aim in mind, we structure this paper as follows: In Section~\ref{sec2} we present the properties of M68 and its stellar population. We simulate its evolution within a potential model of the Milky Way using an \nbody\ code, and describe the orbit of the cluster and the stream produced. In Section~\ref{stream_aaf} we introduce the angle-action framework and describe the stream in this system of coordinates. In Section~\ref{invi} we introduce the \texttt{invi} method and the methodology for constraining the Galactic potential. We apply this method to the simulated stream and describe the results. In Section~\ref{obs_sample} we generate a realistic observational sample of stars by applying a model of the \textit{Gaia} selection function and simulating the observational uncertainties. We also develop a method to estimate the heliocentric distance and the radial velocity of the stars from the orbit of the cluster. In Section~\ref{opt_obs} we test the \texttt{invi} method with a realistic observational sample of stream stars and estimate the precision with which we can recover the parameters of the potential used in the simulation. Finally, we present the conclusion in Section~\ref{conclusion}.

\section{\texorpdfstring{$\boldsymbol{N}$}{N}-Body simulation of the M68 stellar stream}\label{sec2}

In this section we describe the simulation of the M68 stellar stream that we use to determine which properties of the Galactic potential can be accurately recovered with the \texttt{invi} method. First, we generate a realistic stellar population of the globular cluster M68, including mass, \textit{Gaia} $\Gband$ magnitude, $\BPRP$ colour index and other stellar parameters such as metallicity (Section~\ref{sim_stell_pop}). These parameters are needed to evaluate the \textit{Gaia} selection function and to estimate the astrometric and spectroscopic observational uncertainties of the \textit{Gaia} catalogue (Section~\ref{obs_sample}). We also simulate a realistic position and velocity distribution of the stars within the cluster using a self-consistent phase-space model (Section~\ref{phs_dist}). We compute the orbit of the cluster and the tidal stripping process using a \nbody\ code (Section~\ref{nbody_code}), within an axisymmetric three-component potential model of the Milky Way (Section~\ref{mw_pot}). Finally, we describe the orbit of the cluster and the morphology of the simulated stellar stream (Section~\ref{sim_orb} and~\ref{sim_str}).

\subsection{Synthetic stellar population}\label{sim_stell_pop}

We generate a realistic stellar population by using the CMD~3.7 web interface, which implements the theoretical isochrones from PARSEC 1.2S \citep{2012MNRAS.427..127B,2014MNRAS.444.2525C,2015MNRAS.452.1068C,2014MNRAS.445.4287T} and COLIBRI S\_37 \citep{2017ApJ...835...77M,2019IAUS..343..269P,2019MNRAS.485.5666P} in the EGDR3 photometric system \citep{2021AandA...649A...3R}. We use all the default options, but with a mass-loss on the red giant branch according to the Reimers formula with $\eta_{\rm\, Reimers} = 0.25$. This value is optimised to improve the fit to the observed stars at the extreme of the red giant branch. We assume a single population with an age of $12$ Gyr and an initial metallicity of $\MH=-2.2$ ($Z=9.812\pd{-5}$), which is the lower limit allowed by PARSEC. These values are compatible with $12.00\pm0.25$ Gyr from \citet{2013ApJ...775..134V} and $\FeH=-2.32\pm0.12$ from \citet{2023MNRAS.519..192W}, assuming the approximation M=Fe. The values of the parameters characterising this model and their observational estimates, when available, are given in Table~\ref{M68_properties_table}.

\begin{table}
\caption[]{M68 (NGC 4590) globular cluster model properties and observational estimates: age, metallicity, current and initial dynamical mass, number of stars, and lowered isothermal density profile parameters.}
\begin{center}
\begin{tabular}{llllc}
\toprule
\multicolumn{4}{l}{\textbf{Age and Metallicity}}&\\
\midrule
Age&\units{Gyr}&$12$&$12.00\pm0.25$&[1]\\
$\FeH$&-&$-2.2$&$-2.32\pm0.12$&[2]\\
\midrule
\multicolumn{4}{l}{\textbf{Total Mass and Number of Stars}}&\\
\midrule
$M_{\rm gc}$&\units{$10^5\,{\rm M}_{\odot}$}&&$1.23\pm0.12$&[3]\\
$M_{\rm ini}$&\units{$10^5\,{\rm M}_{\odot}$}&$1.25$&&\\
$N$&-&$432\,384$&\\
\midrule
\multicolumn{4}{l}{\textbf{Lowered Isothermal Model} (\texttt{King})}&\\
\midrule
$W_{\rm 0}$&-&5.17&$5.17\pm0.08$&[4]\\
$g$&-&2.46&$2.46\pm0.04$&[4]\\
$r_{\rm half}$&\units{pc}&$5.74$&$5.74\pm0.05$&[4]\\
$r_{\rm core}$&\units{pc}&$2.56$&&\\
$r_{\rm trun}$&\units{pc}&$281.67$&&\\
$\sigma_{\rm v}$&\units{km s$^{-1}$}&$3.43$&&\\
\bottomrule
\end{tabular}
\end{center}

\begin{tabular}{l}
\textbf{References:}\\
\text{[1]}: \citet{2013ApJ...775..134V}\\
\text{[2]}: \citet{2023MNRAS.519..192W}\\
\text{[3]}: \citet{2018MNRAS.478.1520B}\\
\text{[4]}: \citet{2019MNRAS.485.4906D}\\
\end{tabular}
\label{M68_properties_table}
\end{table}

The masses of the initial stellar sample follow a \citet{2001MNRAS.322..231K,2002Sci...295...82K} Initial Mass Function corrected for unresolved binaries. The final simulated population has a total of $N = 432\,384$ stars with masses ranging from 0.1 to 0.792 ${\rm M}_{\odot}$, and a total mass of $M_{\rm ini} = 1.25\pd{5}$ ${\rm M}_{\odot}$. We choose a higher value than the mean of the current dynamical mass of $M_{\rm gc} = 1.23\pm0.12\pd{5}$ ${\rm M}_{\odot}$ from \citet{2018MNRAS.478.1520B}\footnotemark to account for the cluster mass-loss due to tidal stripping.

\footnotetext{\url{https://people.smp.uq.edu.au/HolgerBaumgardt/globular/}}

\begin{figure}
\includegraphics[width=1.0\columnwidth]{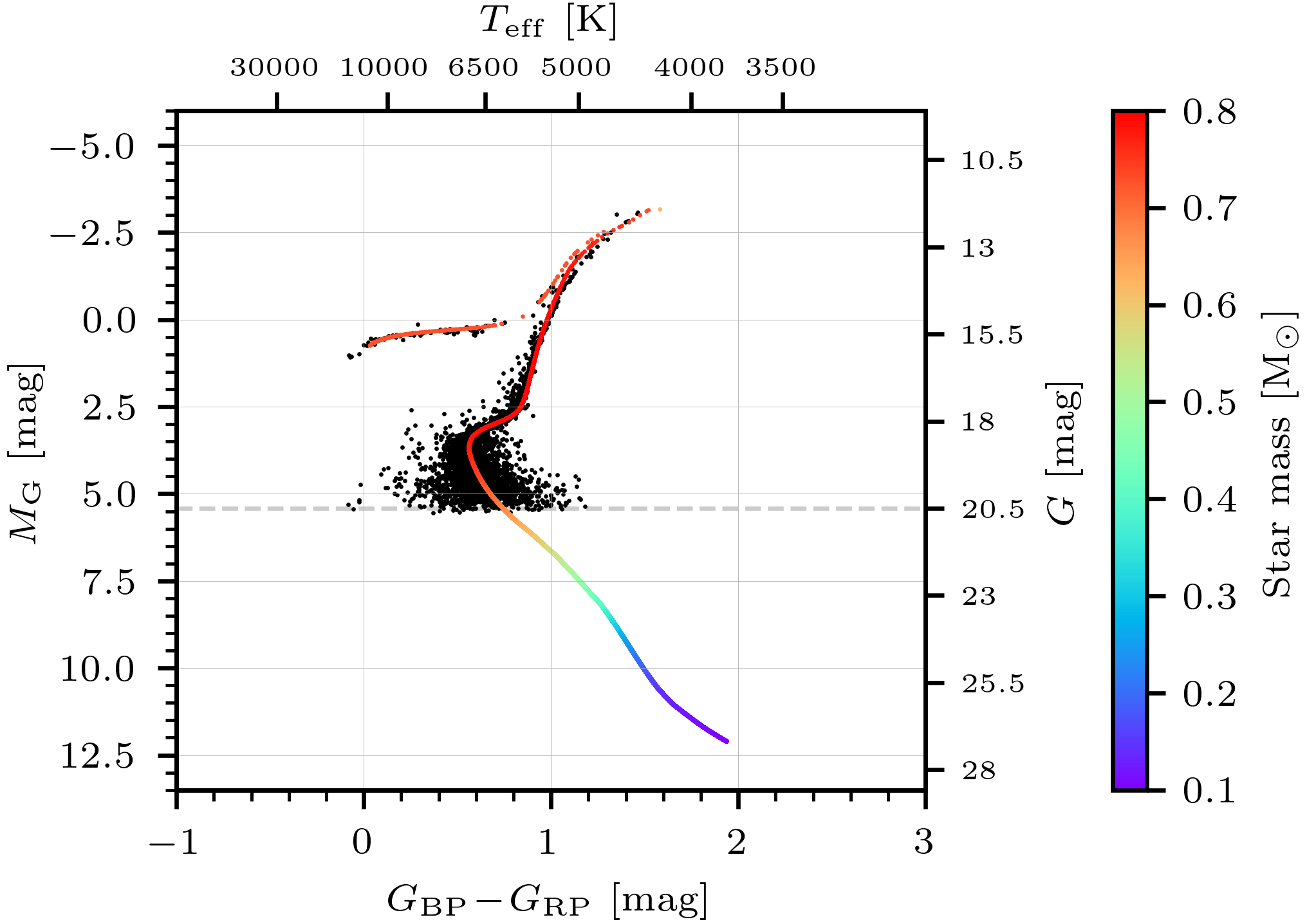}
\caption{Hertzsprung-Russell diagram of the synthetic population of the globular cluster M68, generated by PARSEC/COLIBRI. Each coloured dot represents a star in the synthetic population, with colour proportional to its mass. The black dots are GDR3 stars selected from the cluster. We indicate the \textit{Gaia} colour index $\BPRP$, the absolute magnitude $M_{\rm G}$, the apparent magnitude $G$ assuming a distance of 10.404 kpc for all the stars, and the effective temperature $T_{\rm eff}$. The dashed horizontal grey line marks the \textit{Gaia} observational limit of $G=20.5$ mag.}
\label{HR_syntetic_population}
\end{figure}

Figure~\ref{HR_syntetic_population} shows the colour index $\BPRP$ and absolute magnitude $M_{\rm G}$ of the synthetic population generated by CMD 3.7. We compare this population with the observed Gaia Data Release 3 (GDR3) stars at the location of M68 passing through the cuts described in Appendix~\ref{App1}. We assume that these stars are located at $10.404$ kpc from the Sun (Section~\ref{sim_orb}) and apply the dust reddening correction described in Appendix~\ref{App2}. We plot these stars in black. We obtain a synthetic population in good agreement with the observed stars, but with a systematic deviation of $\Approx0.14$ mag at the extreme of the red and asymptotic giant branches, probably due to an inaccurate bolometric correction.

\subsection{Phase-space stellar distribution}\label{phs_dist}

We model the phase-space distribution of the stars in the globular cluster M68 using a generalisation of a spherical lowered isothermal model (\S4.3.3.c, \citet{2008gady.book.....B}, hereafter \citetalias{2008gady.book.....B}). Here we use the variant specified in Equation 1 of \citet{2015MNRAS.454..576G}, assuming that the distribution function is isotropic. We neglect the rotation of the cluster because the observational measurements are consistent with zero \citep[e.g.][]{2023MNRAS.519..192W}. This model is specified by the dimensionless central potential $W_{\rm 0}$, which determines the central concentration, and a radial scale factor such as the half-mass radius $r_{\rm half}$ or the truncation radius $r_{\rm trun}$. It also depends on the sharpness of the truncation in energy $g$, which determines the extension of the cluster envelope. For $g=1$ and $g=2$, it reduces to the King and Wilson models respectively.

For the initial conditions of the \nbody\ simulation, we take the mean value of the estimates from \citet{2019MNRAS.485.4906D} and list them in Table~\ref{M68_properties_table}. The cluster is characterised by a $W_{\rm 0}=5.17$, a half-mass radius $r_{\rm half}=5.74$ pc, and a truncation $g=2.46$. The value of the core radius $r_{\rm core}$, the truncation radius $r_{\rm trun}$, and the velocity dispersion $\sigma_{\rm v}$, which we list in Table~\ref{M68_properties_table}, are calculated numerically using the Python package \texttt{astro-limepy} \citep{2015MNRAS.454..576G}. We use the current values of these parameters as initial conditions, since they will remain almost constant throughout the entire evolution of the cluster. Finally, we generate the sample of phase-space initial conditions for the $N$ simulated stars using the random sampling method provided in the Python package \texttt{agama} \citep{2019MNRAS.482.1525V}. In this package, the phase-space distribution we use is referred to as \texttt{King}.

\subsection{\texorpdfstring{$\boldsymbol{N}$}{N}-Body integration code} \label{nbody_code}

We treat each star as an \nbody\ particle and compute its orbit within the cluster and the potential of the Milky Way using the code \textsc{PeTar} \citep{2020MNRAS.497..536W}. This code is designed to simulate gravitationally bound collisional stellar systems, where many multiple systems (binaries, triples, etc.) and close encounters are important for the dynamical evolution. It achieves high performance by combining three integration methods: The long-range interactions between stars are computed by the Barnes-Hut tree method, and the orbits are integrated with a second-order symplectic leap-frog integrator. The short-range interactions between stars and the centres of mass of multiple systems are computed using a fourth-order Hermite integrator, and the hyperbolic encounters, binaries and hierarchical few-body systems are computed using the Slow-Down Algorithmic Regularization method \citep[SDAR,][]{2020MNRAS.493.3398W}.

The accuracy of the long-range and short-range interactions is determined by two parameters. The tree time step \texttt{s} represents the accuracy of the long-range interactions, and the outer boundary of changeover radii \texttt{r} determines the accuracy of the short-range interactions. A smaller value of \texttt{r} indicates that fewer particles are treated by the Hermite or SDAR methods. We set $\texttt{s}=1/1024$ and $\texttt{r}=0.03$. The final position of the stars depends significantly on these values. In general, the orbit of a star within the cluster changes completely if it is part of a multiple group, such as a binary system, or not. This determines its possible escape time and position along the stream. However, general properties such as the phase-space distribution of the cluster, the morphology of the stream or the number of escapees, converge for multiple simulations if the interactions are computed with sufficient accuracy \citep{2021arXiv210410843W}. We have verified that convergence is achieved for the chosen values of the configuration parameters.

The code \textsc{PeTar} is implemented using the Framework for Developing Particle Simulator \citep[FDPS,][]{2016PASJ...68...54I,2018PASJ...70...70N}, which provides parallelization methods for particle-tree construction and long-range interaction calculations. We run the algorithm on 72 MPI processors, each with 1 CPU thread. In addition, it also includes stellar evolution models and external potentials from the Python package \texttt{galpy} \citep{2015ApJS..216...29B}. We neglect the stellar evolution and include a three-component potential model of the Galaxy.

\subsection{Milky Way potential model}\label{mw_pot}

We define the Milky Way potential model in terms of \texttt{galpy} densities and potentials because we use this library throughout this paper to compute the angle-action coordinates of the stars. The model of the Galaxy consists of three axisymmetric components: A spherical bulge modelled with a \texttt{PowerSphericalPotentialwCutoff} which includes an exponential cut-off, a disc with one \texttt{MiyamotoNagaiPotential} \citep{1975PASJ...27..533M} component, and a spherical dark halo following a Navarro, Frenk \& White \citep[NFW,][]{1996ApJ...462..563N} density profile modelled with a \texttt{TriaxialNFWPotential}. The density and potential models and their parameters are listed in Table~\ref{MW_table}. This model is similar to \texttt{BovyMWPotential2014} \citep{2015ApJS..216...29B}, but with a dark halo that is about 20 per cent more massive. Such a mass increase improves the fit of the simulated M68 stellar stream to the observed data (\citetalias{2019MNRAS.488.1535P}). The combination of the three components gives an overall oblate potential, but converges to spherical with increasing distance from the disc.

\begin{table}
\caption[]{Components of the Milky Way model. The density and potential models are defined as in \texttt{galpy}, and unspecified parameters are taken as defaults.}
\begin{center}
\begin{tabular}{p{1.7em}lllc}
\toprule
\multicolumn{4}{l}{\textbf{Bulge:} \texttt{PowerSphericalPotentialwCutoff}}\\
\midrule
\texttt{amp}&&\units{$10^{6}\,{\rm M}_{\Sun}\,{\rm kpc}^{-3}$}&$5.3$&\\
\texttt{alpha}&&-&$1.8$&\\
\texttt{r1}&&\units{kpc}&$8$&\\
\texttt{rc}&&\units{kpc}&$1.9$&\\
\midrule
\multicolumn{4}{l}{\textbf{Disc}: \texttt{MiyamotoNagaiPotential}}\\
\midrule
\texttt{amp}&$M_{\rm d}$&\units{$10^{10}\,{\rm M}_{\odot}$}&$6.8$&\\
\texttt{a}&$a_{\rm d}$&\units{kpc}&$3$&\\
\texttt{b}&$b_{\rm d}$&\units{kpc}&$0.28$&\\
\midrule
\multicolumn{4}{l}{\textbf{Dark Halo:} \texttt{TriaxialNFWPotential}}\\
\midrule
\texttt{amp}&$\rho_{\rm h}$&\units{$4\pi\:\!$\texttt{a}$^3\:\!10^{7}\,{\rm M}_{\Sun}$}&$1.05$&\\
\texttt{a}&$a_{\rm h}$&\units{kpc}&$16$&\\
\bottomrule
\end{tabular}
\end{center}
\label{MW_table}
\end{table}

\subsection{Simulated orbit of the cluster}\label{sim_orb}

We numerically integrate the orbit of the cluster within the potential model of the Milky Way to determine its class and degree of chaoticity. The initial condition of the integration is the mean value of the observational estimates of the current phase-space position of the cluster. The value of each component and its uncertainties are listed in Table~\ref{M68_orbital_parameters_table}. We take the heliocentric distance $r_{\rm h}=10.404$ kpc from \citet{2021MNRAS.505.5957B}, which has an estimated uncertainty of one per cent. We take the sky coordinates and the proper motions measured with EGDR3 data from \citet {2021MNRAS.505.5978V}, and the radial velocity measured with the 2dF/AAOmega instrument from \citet{2023MNRAS.519..192W}. To transform this position from the Heliocetric (ICRS) to the Galactocentric coordinate system, we use the mean of the estimates of the position of the Sun, given in Appendix~\ref{App3}.

\begin{table}
\caption[]{Orbital parameters of the globular cluster M68: Current phase-space position in heliocentric coordinates (ICRS), where $\mu_{\alpha *}\equiv\mu_\alpha\cos\var{\delta}$, and spherical galactocentric radius and time with respect to the present of the pericentre passages. The pericentres are estimated assuming the potential model described in Section~\ref{mw_pot}.}

\begin{center}
\begin{tabular}{lllc}
\toprule
\multicolumn{3}{l}{\textbf{ICRS}}&\textbf{Ref.}\\
\midrule
$r_{\rm h}$&\units{kpc}&$10.404^{+0.100}_{-0.099}$&[1]\\[0.1em]
$\delta$&\units{deg}&$-26.744$&[2]\\
$\alpha$&\units{deg}&$189.867$&[2]\\
$v_r$&\units{km s$^{-1}$}&$-92.07^{+1.60}_{-1.42}$&[3]\\[0.1em]
$\mu_\delta$&\units{mas yr$^{-1}$}&$1.779\pm0.024$&[2]\\[0.1em]
$\mu_{\alpha *}$&\units{mas yr$^{-1}$}&$-2.739\pm0.024$&[2]\\[0.1em]
\midrule
\multicolumn{2}{l}{\textbf{Pericentre}}&\multicolumn{1}{l}{$\boldsymbol{r_{\rm peri}}\quad$\units{kpc}}&\multicolumn{1}{l}{$\boldsymbol{t_{\rm peri}}\quad$\units{Myr}}\\
\midrule
\multicolumn{2}{l}{1}&\multicolumn{1}{l}{$9.20$}&\multicolumn{1}{l}{$-1353.59$}\\
\multicolumn{2}{l}{2}&\multicolumn{1}{l}{$9.25$}&\multicolumn{1}{l}{$-896.12$}\\
\multicolumn{2}{l}{3}&\multicolumn{1}{l}{$9.18$}&\multicolumn{1}{l}{$-439.64$}\\
\bottomrule
\end{tabular}
\end{center}

\begin{tabular}{l}
\textbf{References:}\\
\text{[1]}: \citet{2021MNRAS.505.5957B}\\
\text{[2]}: \citet{2021MNRAS.505.5978V}\\
\text{[3]}: \citet{2023MNRAS.519..192W}\\
\end{tabular}
\label{M68_orbital_parameters_table}
\end{table}

We assume that the cluster is a test particle and integrate its orbit within the Milky Way potential defined in Section~\ref{mw_pot} using \texttt{galpy}, which solves the system of differential equations numerically using a 6th-order symplectic integrator method (\texttt{symplec6\_c}). The cluster follows a periodic orbit with eccentricity $e\simeq0.639$, oscillating between the pericentre $r_{\rm peri} \simeq 9.2$ kpc and the apocentre $r_{\rm apo} \simeq 31.9$ kpc, where $r$ is the spherical galactocentric radius. The approximate periods are: $T_{r}\approx T_{R}\simeq456.9$, $T_{\phi}\simeq651.3$, and $T_{z}\simeq622.9$ Myr, expressed in a cylindrical galactocentric coordinate system $(R,\phi,z)$.

We measure the stochasticity of the orbit using a finite-time estimate of the Lyapunov exponent $\lambda_{{\rm L}}$ (\S3.7.3.c, \citetalias{2008gady.book.....B}). It measures the growth of a deviation vector between the orbit and another infinitely close orbit. For regular orbits, the magnitude of the deviation vector grows linearly in time, and for chaotic orbits, it grows exponentially. We use the implementation in \texttt{agama} obtaining $\lambda_{{\rm L}}=0$. This indicates that no exponential growth of the deviation vector has been detected and therefore, the orbit is regular.

\begin{figure*}
\includegraphics[width=1.0\textwidth]{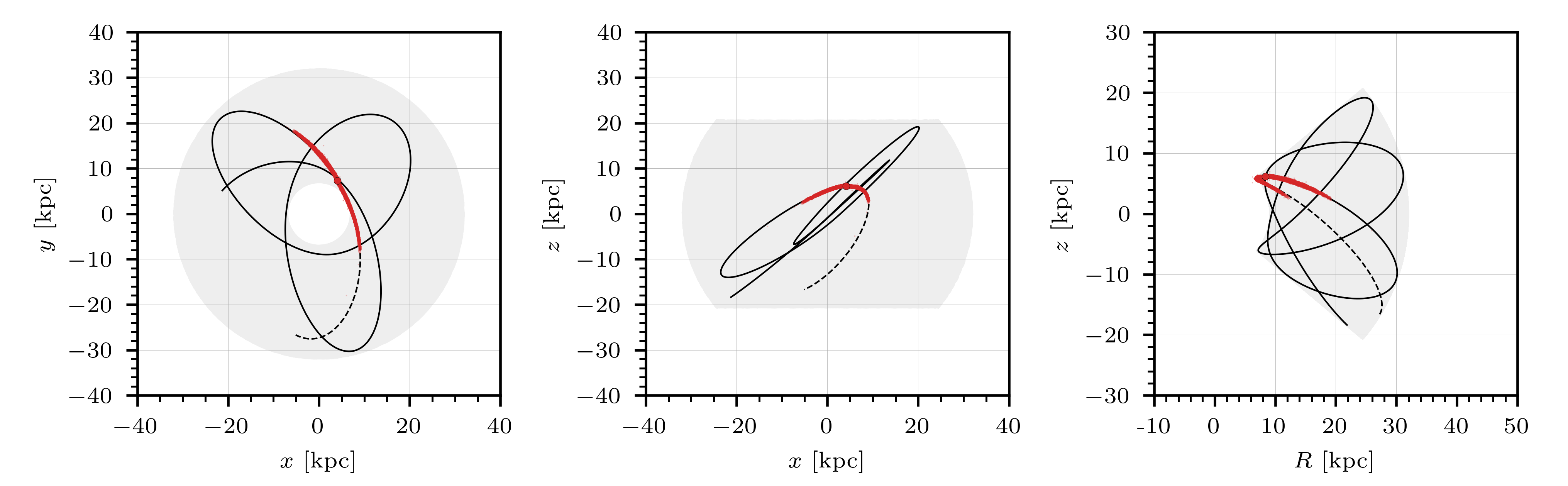}
\caption{Orbit of the globular cluster M68. The large red dot marks the current position of the cluster, and the solid black line its orbit during the last 1.5 Gyr. The dashed black line marks its forward orbit during 250 Myr. The projection of the volume covered by the entire orbit is shown as a grey shaded area. The simulated stellar stream stars are shown as small red dots. \textit{Left and Middle:} Galactic plane and perpendicular plane in cartesian coordinates. \textit{Right:} Perpendicular plane in cylindrical coordinates.}
\label{orbit}
\end{figure*}

To determine the orbit class, we calculate in cartesian coordinates the frequencies with the largest amplitude in the Fourier spectrum using the NAFF algorithm \citep{1992PhyD...56..253L} implemented in the Python package \texttt{naif} \citep{2023ApJ...955...38B}. We obtain $F_x=F_y\simeq9.647$ and $F_z\simeq10.088$ rad Gyr$^{-1}$, with the ratio $F_y/F_x = 1$ and an incommensurable ratio $F_z/F_x \simeq 1.046$. This indicates that the orbit is not resonant. The orbit does not cross the symmetry axis $z$ and the potential model of the Galaxy is oblate, so the orbit is a short-axis tube or \stube. Figure~\ref{orbit} shows the projection of the volume covered by the orbit of the cluster with a grey surface.

\subsection{Simulated stellar stream}\label{sim_str}

To generate the initial distribution of the stars in the \nbody\ simulation, we integrate the orbit of the cluster within the potential of the Milky Way (Section~\ref{mw_pot}) over a period $T=1.5$ Gyr back in time from its current position. We choose this time interval because it approximates the length of the observed stream \citepalias{2019MNRAS.488.1535P}. Relative to the endpoint of the orbit, we distribute a sample of $N$ stars with the realistic mass distribution and following the phase-space model of the cluster introduced in Sections~\ref{sim_stell_pop} and~\ref{phs_dist}. Finally, we integrate the \nbody\ system forward in time during 1.5 Gyr using the code \textsc{PeTar} to obtain a simulation of the stellar stream at the present time. In this paper we locate the origin of the time frame ($t=0$) at the present time, so the simulation stars at $t=-T$.

In Figure~\ref{orbit} we show the orbit of the cluster during the last 1.5 Gyr as a solid black line, and the orbit during 0.25 Gyr forward in time as a dashed black line. The left and middle panels show the Galactic plane ($x\:\!\text{-}\:\!y$) and the plane perpendicular to the disc ($x\:\!\text{-}\:\!z$) in cartesian coordinates. The right panel shows the $R\:\!\text{-}\:\!z$ plane in cylindrical coordinates. Along the simulated orbit, the cluster crosses the Galactic disc five times, passing through three pericentres and three apocentres. We list the cylindrical radius and the time of each pericentre in Table~\ref{M68_orbital_parameters_table}.

During this evolution, the stars are stripped from the cluster by the tidal forces caused by the gravitational field of the Milky Way. In Figure~\ref{orbit} we show the current position of the cluster as a large red dot, and the escaped stars as small red dots. The resulting stream is a long and thin structure, and consists of two arms, one leading and one trailing, which are symmetrical with respect to the cluster. Each arm is about 20 kpc long and runs almost parallel to the Galactic plane, at a distance of about 5 kpc, in the North Galactic hemisphere. The leading arm passes through the pericentre and its visible section, which is defined in Section~\ref{obs_section}, is about 5 kpc from the Sun.

In Table~\ref{number_stars} we list the number of stars in each component of the stream, classified according to the criteria defined in Appendix~\ref{App4}. The cluster loses about 1.31 per cent of its initial stars, which corresponds to a rate of about $800$ and $1\,000$ stars per pericentre passage for each arm. In total, there are $5\,471$ stars in the stream, half of which correspond to each arm. We also identify $210$ low-mass stars that are lost by evaporation and ejection. In the following section we describe the stream in angle-action coordinates.

\begin{table}
\caption[]{Number of stars for each component of the simulated stream. Mean and standard deviation of a random sample of stars passing through the \textit{Gaia} selection function, and stars that additionally have magnitude $G<20$ and $G<18$ mag. The standard deviation is omitted from the last two columns if it is zero. The three internal streams are labelled with a number, and the observable section of the stream is defined by the stars with declination $\delta > -8$ deg.}

\begin{center}
\begin{tabular}{lrrrr}
\toprule
&\textbf{Simu.}&\textbf{Sel. Func.}&$\boldsymbol{G\!\smallrel<\!20}$ &$\!\!\!\boldsymbol{G\!\smallrel<\!18}$\\
\midrule
\textbf{Total}                     &$432\,384$ &$21\,286\!\pm\!45$ &$12\,052\!\pm\!13$ &$1\,300$\\
\midrule
$\!\!\!\quad$\textbf{Cluster}      &$426\,703$ &$20\,873\!\pm\!45$ &$11\,865\!\pm\!13$ &$1\,265$\\
\midrule
$\!\!\!\quad$\textbf{Stream}       &$5\,471$   &$411 \pm 5$      &$187$            &$35$\\
\midrule
$\!\!\!\quad\quad$\textbf{Leading} &$2\,720$   &$322 \pm 4$      &$150$            &$24$\\
$\!\!\!\quad\quad\quad$\textbf{1}  &$825$      &$109 \pm 2$      &$56$             &$8$\\
$\!\!\!\quad\quad\quad$\textbf{2}  &$1\,008$   &$128 \pm 2$      &$59$             &$11$\\
$\!\!\!\quad\quad\quad$\textbf{3}  &$800$      &$81 \pm 2$       &$34$             &$5$\\
$\!\!\!\quad\quad\quad$\textbf{-}  &$87$       &$3 \pm 1$        &$1$              &$0$\\
\midrule
$\!\!\!\quad\quad$\textbf{Trailing}&$2\,751$   &$89 \pm 3$       &$37$             &$11$\\
$\!\!\!\quad\quad\quad$\textbf{1}  &$846$      &$14 \pm 1$       &$6$              &$1$\\
$\!\!\!\quad\quad\quad$\textbf{2}  &$1\,028$   &$37 \pm 2$       &$14$             &$2$\\
$\!\!\!\quad\quad\quad$\textbf{3}  &$794$      &$34 \pm 1$       &$15$             &$6$\\
$\!\!\!\quad\quad\quad$\textbf{-}  &$83$       &$3 \pm 1$        &$2$              &$2$\\
\midrule
$\!\!\!\quad$\textbf{Scapees}      &$210$      &$2 \pm 1$        &$0$              &$0$\\
\midrule
\textbf{Obs. Sec.}                 &$1\,752$   &$237 \pm 3$      &$116$            &$19$\\
\bottomrule
\end{tabular}
\end{center}

\label{number_stars}
\end{table}

\section{The simulated stream in Angle-Action coordinates}\label{stream_aaf}

The dynamics of a stellar stream and its structure is greatly simplified when it is described in angle-action coordinates \citep{2011MNRAS.413.1852E,2014ApJ...795...95B, 2015MNRAS.452..301F}. In this system of coordinates, the phase-space position is denoted by $\{\theta,J\}$, where $\theta_i$ are the angles, $J_i$ are the actions and the sub-index $i$ indicates each one of the three dimensions. By definition, the angles and actions form a canonical coordinate system such that the Hamiltonian is independent of the angles, i.e. $H\var{J}$ (\S3.5, \citetalias{2008gady.book.....B}). Consequently, by applying Hamilton's equations, it follows that the actions are integrals of motion and the angles can be integrated analytically, giving:
\begin{equation}\label{int_orb}
\theta_i\var{t} = \varOmega_i \;\! t + \theta_i\var{0},
\end{equation}
where $\varOmega_i \equiv \partial H / \partial J_i$ are called frequencies. In this way, the orbit of the particles is characterised by three constants $J_i$ and a uniform linear motion in $\theta_i$ in a three-dimensional subspace. If we restrict to periodic orbits, the angles can be scaled within $\theta_i\in[0,2\pi)$, with the orbital period:
\begin{equation}
T_i = \frac{2\pi}{\varOmega_i}.
\end{equation}

The angle-action coordinates are only defined for regular orbits, and can only be calculated analytically for highly symmetric potentials (\S3.5.2 and \S3.5.3, \citetalias{2008gady.book.....B}). Several methods have been developed to numerically calculate actions, angles and frequencies in general potentials \citep{2016MNRAS.457.2107S}. In this paper, we will use a variant of the torus modelling technique \citep{1990MNRAS.244..634M} introduced by \citet{2014arXiv1407.1688F} and named the Averaged Generating Function (AvGF). This method requires an integrated orbit and allows us to calculate actions for a general potential. The AvGF method numerically constructs a generating function that maps the analytic angle-actions of an auxiliary potential to the angle-actions of the real potential. The auxiliary potential should closely resemble the real one. For loop orbits, such as the orbit of M68, an Isochrone potential is used. The same auxiliary potential is used for all stream stars, as they follow a similar orbit. The main limitation of this method is that the errors cannot be reduced arbitrarily for resonant trapped orbits. In Appendix~\ref{App5} we discuss the most relevant resonances that the cluster can undergo for different halo models and how this affects the application of the \texttt{invi} method.

We use the implementation of this method included in the Python package \texttt{galpy}, which contains an extension for the calculation of frequencies in addition to angles and actions \citep{2014ApJ...795...95B}. We set the following free parameters of the \texttt{actionAngleIsochroneApprox} method: scale length of the Isochrone potential $\texttt{b}=4.976$ kpc, $\texttt{maxn}=4$, and integrate the orbit during $\texttt{tintJ}=5$ Gyr with $\texttt{integrate\_method}=\texttt{symplec4\_c}$. We compute the actions $J_r$ and $J_z$, which quantify the oscillations in the spherical radius and vertical dimension, and $J_\phi$ which is equal to the angular momentum $L_z$. Thus, we denote $i \equiv\{r,\, \phi,\, z\}$.

For the globular cluster M68 we obtain the following angles, actions and frequencies at the present time:
\begin{equation}
\begin{tabular}{lll}
$\theta_{0}$ &\hspace{-0.7em}$\simeq$\hspace{0.3em} $(\phantom{0}6.043,\, \phantom{-}0.505,\, \phantom{0}1.580)$ & \hspace{-0.5em}rad \\
$J_{0}$ &\hspace{-0.7em}$\simeq$\hspace{0.3em} $(\phantom{0}0.935,\, -2.441,\, \phantom{0}0.814)$ & \hspace{-0.5em}kpc$^{2}$ Myr$^{-1}$\\
$\varOmega_{0}$ &\hspace{-0.7em}$\simeq$\hspace{0.3em} $(13.751,\, -9.647,\, 10.088)$ & \hspace{-0.5em}rad Gyr$^{-1}$.\\
\end{tabular}
\end{equation}
To quantify the inaccuracies of the AvGF method configured with the previous parameters, we evaluate the actions and frequencies along the orbit of the cluster during about $40$ radial periods. We calculate the coefficient of variation ${\rm c}_{\rm v}$ or the ratio between the standard deviation and the mean of each variable. We get ${\rm c}_{\rm v}\var{J}\simeq(0.07,\, 0,\, 0.12)$ per cent and ${\rm c}_{\rm v}\var{\varOmega}\simeq(8.29,\, 4.06,\, 3.50)\pd{-5}$ per cent. These inaccuracies are very small compared to the errors in angles and actions introduced by observational uncertainties (Section~\ref{gaia_obs_un}). We therefore consider them to be negligible.

If the separation between the stripped stars and the cluster is large enough, the mutual gravitational interaction can be neglected, and the motion of the stars depends only on the potential of the host galaxy. The relative motion of the stars with respect to the cluster can then be approximated by a first-order Taylor expansion of the Hamiltonian at the centre of the cluster (\S8.3.3, \citetalias{2008gady.book.....B}). This approximation is valid for small separations in the action space. In this case, we have for each stream star:
\begin{equation}\label{Hessian}
 \Delta\varOmega_i \approx \frac{\partial^2 H}{\partial J_i \, \partial J_j} \bigg\rvert_{J_{0}} \Delta J_j \equiv D_{ij} \:\! \Delta J_j,
\end{equation}
where $\Delta$ indicates the relative position of the star with respect to the cluster centre for each variable. Defining $t_{\rm s}$ as the stripping time of a stream star (Eq.~\ref{ts}), the angular separation of the star from the cluster increases approximately linearly with time:
\begin{equation}\label{lin_grow}
 \Delta\theta_i \approx  D_{ij} \:\! \Delta J_j \;\! (t-t_{\rm s}) + \Delta\theta_i\var{t_{\rm s}} \hfill t\geqslant t_{\rm s}.
\end{equation}

The Hessian matrix $D$, defined in Eq.~\ref{Hessian}, determines the direction along which the stream spreads in angle space. We can calculate the principal directions (eigenvectors) by diagonalising this matrix. We use these directions to define a rotated reference frame aligned with the stream, taking the eigenvectors in the order defined by their eigenvalues $\lambda_j$ such that: $|\lambda_1| > |\lambda_2| > |\lambda_3|$. We denote by $\{\bar{\theta}, \bar{J}, \bar{\varOmega}\}$ the phase-space coordinates and frequencies in the reference frame defined by the eigenvectors. Since $D$ as a function of the actions is unknown, we compute its diagonal form numerically from the simulation of the stream as described in Appendix~\ref{App6}. The value of the numerically estimated eigenvalues is:
\begin{equation}\label{eigen}
\begin{tabular}{lll}
$\lambda$ &\hspace{-0.7em}$\simeq$\hspace{0.3em} $(-10.08,\, -0.30,\,\, 0.24)$ & \hspace{-0.5em}mrad kpc$^{-2}$. \\
\end{tabular}
\end{equation}

\begin{figure*}
\begin{subfigure}[b]{1.0\textwidth}
\includegraphics[width=1.0\textwidth]{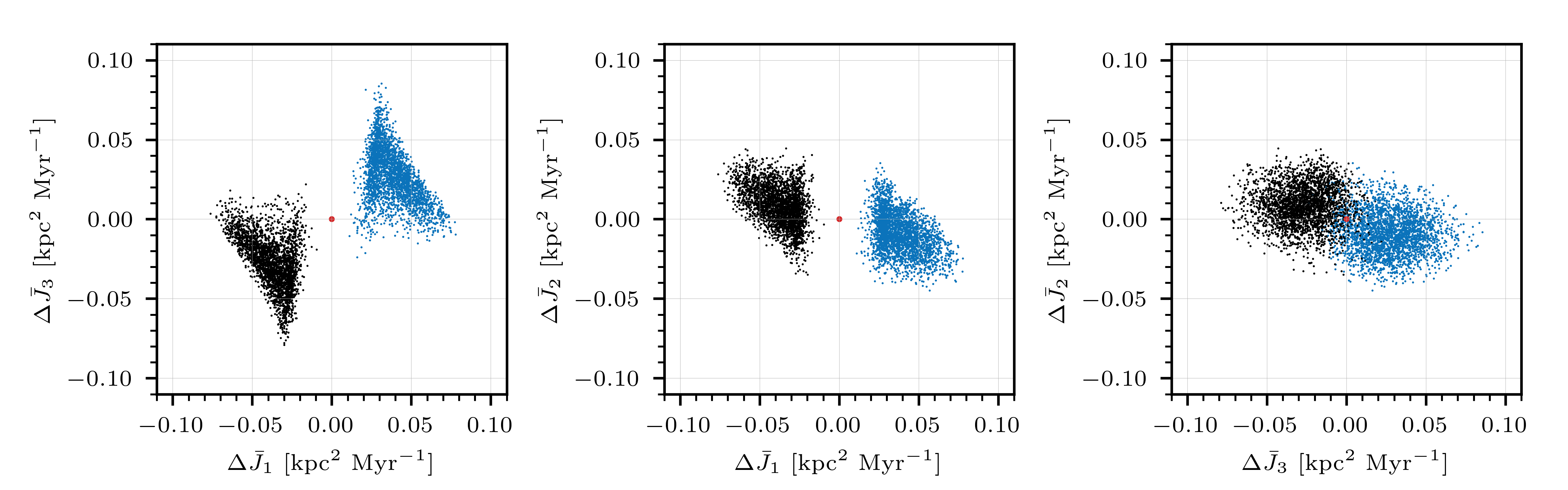}
\caption{}
\label{actions}
\end{subfigure}
\vfill
\begin{subfigure}[b]{1.0\textwidth}
\includegraphics[width=1.0\textwidth]{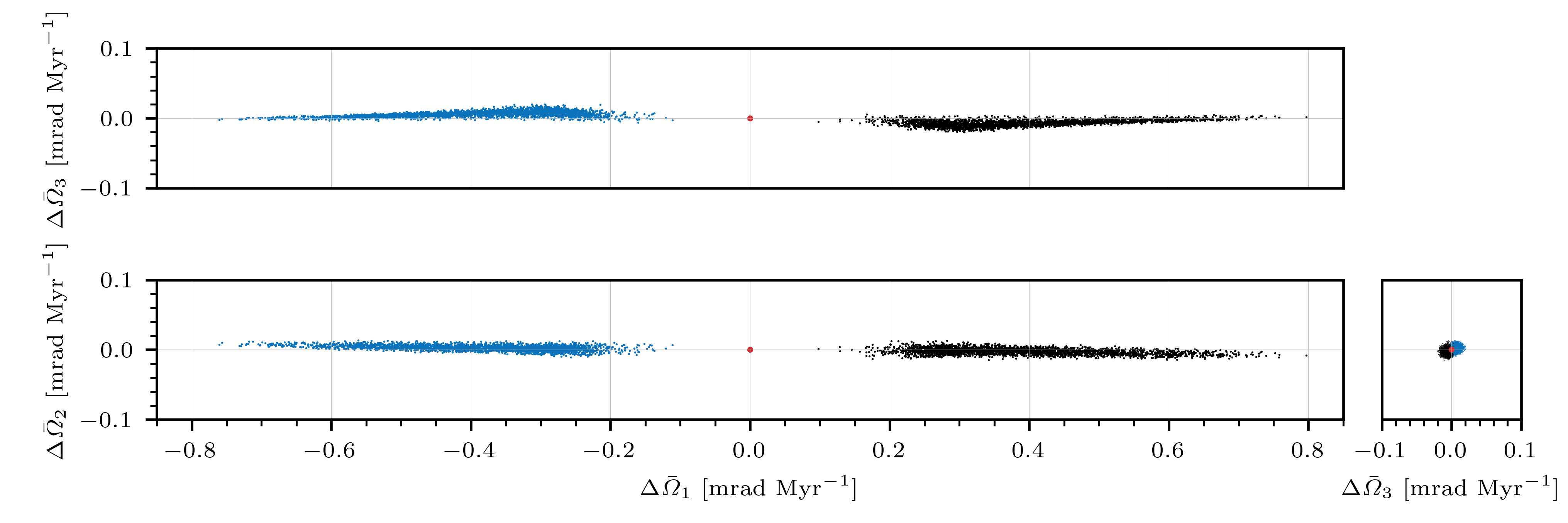}
\caption{}
\label{freqs}
\end{subfigure}
\vfill
\begin{subfigure}[b]{1.0\textwidth}
\includegraphics[width=1.0\textwidth]{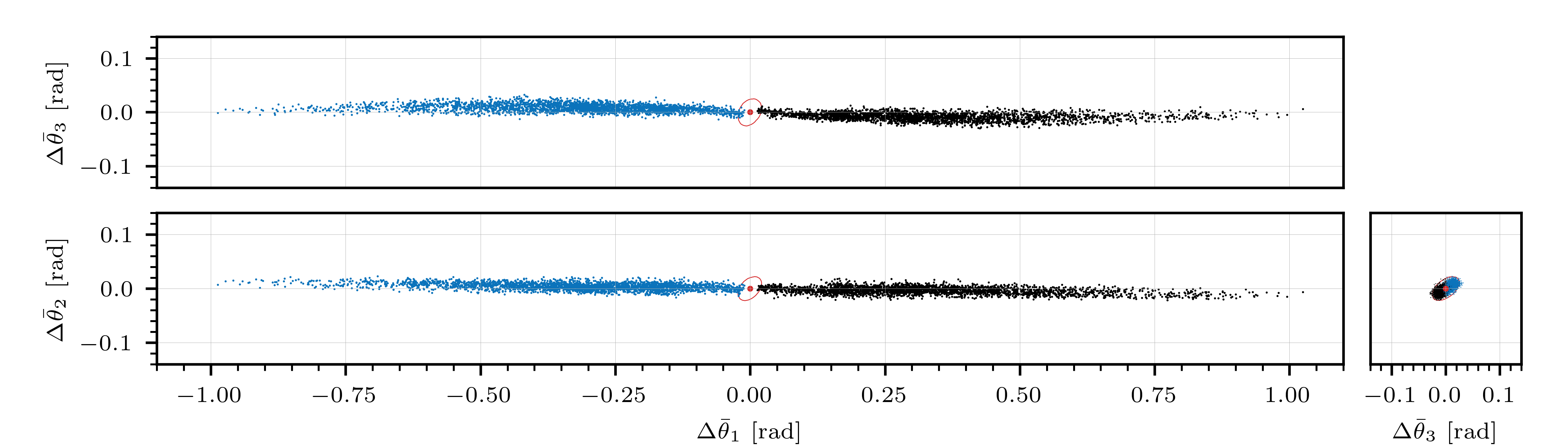}
\caption{}
\label{angles}
\end{subfigure}
\caption{Actions (a), frequencies (b), and angles (c) of the M68 simulated stream stars in the reference frame defined by the eigenvectors that diagonalize the Hessian matrix (Eq.~\ref{Hessian}). The black dots indicate the leading arm, and the blue dots the trailing arm. The centre of the cluster is marked by a large red dot. In the (c) panels, the area containing 68 per cent of the cluster's stars is marked with a red contour line.}
\label{fig:stream_aaf}
\end{figure*}

We show the stream in the rotated reference frame in Figure~\ref{fig:stream_aaf}. We plot the stream stars of the leading arm as black dots, and the stars of the trailing arm as blue dots. We also mark the position of the cluster with a large red dot. In the upper panels (Fig.~\ref{actions}) we observe that the dispersion of the stream stars in action space is similar in magnitude along the three principal axes. Thus, from the ratio of the numerically calculated eigenvalues in Eq.~\ref{eigen} and Eq.~\ref{Hessian}, we estimate that the stars have a frequency along the principal axis that is between 34 and 42 times greater than along the perpendicular axes. According to Eq.~\ref{lin_grow}, this implies that the stream grows mainly in one direction, forming a long and thin structure. We show the distribution of the frequencies in the middle panels (Fig.~\ref{freqs}), and the distribution of angles in the lower panels (Fig.~\ref{angles}). The principal axis of the stream is misaligned with respect to the orbit of the cluster, at an angle of $\varphi\simeq1.72$ deg. This angle depends on the potential of the host galaxy and, for a globular cluster like M68, is independent of the mass of the progenitor \citep{2013MNRAS.433.1813S}.

\section{Inverse time integration method}\label{invi}

At the beginning of its evolution, the system formed by all the bound stars in the cluster is in a state of low entropy. The entropy of the system increases as the stars are heated up and stripped from the cluster by tidal shocks. Once the potential of the progenitor and the interactions between the stars are negligible, the stripped stars follow a reversible evolution within the potential of the host galaxy. Given the current position of the stream stars, the only way to recover the original state of low entropy is to integrate their orbits backwards in time in the same potential that drove their reversible evolution. If the potential is not the same, the stars do not return to the progenitor, and a configuration with higher entropy than the original is obtained.

The stream stars are ejected around the Lagrange points of the progenitor cluster (\S8.3.3, \citetalias{2008gady.book.....B}). The position of the Lagrange points relative to the cluster varies along the orbit, with a minimum at the pericentre and a maximum at the apocentre. For M68 they oscillate in the interval $\Approx \Range{70}{165}$ pc, or equivalently at $\Approx \Range{0}{1.5}$ mrad in angle space from the cluster centre. As shown in Figure~\ref{angles}, this distance is three orders of magnitude smaller than the angular separation between the cluster and the stream stars along the principal axis of the stream. We can therefore neglect the separation between the cluster and the Lagrange points and assume that the stars are ejected from the centre of the cluster. Also neglecting the effect of the potential of the progenitor during the stripping process, we estimate the time since the particle was stripped $\delta t$ as:
\begin{equation}\label{delta_t}
 \delta t \approx \frac{\norm{\Delta\theta}}{\norm{\Delta\varOmega}},
\end{equation}
where we assume a Euclidean metric to compute the norm. Since we have defined the past time as negative, the stripping time is approximately equal to:
\begin{equation}\label{ts}
t_{\rm s} \approx -\delta t.
\end{equation}

\begin{figure}
\includegraphics[width=1.0\columnwidth]{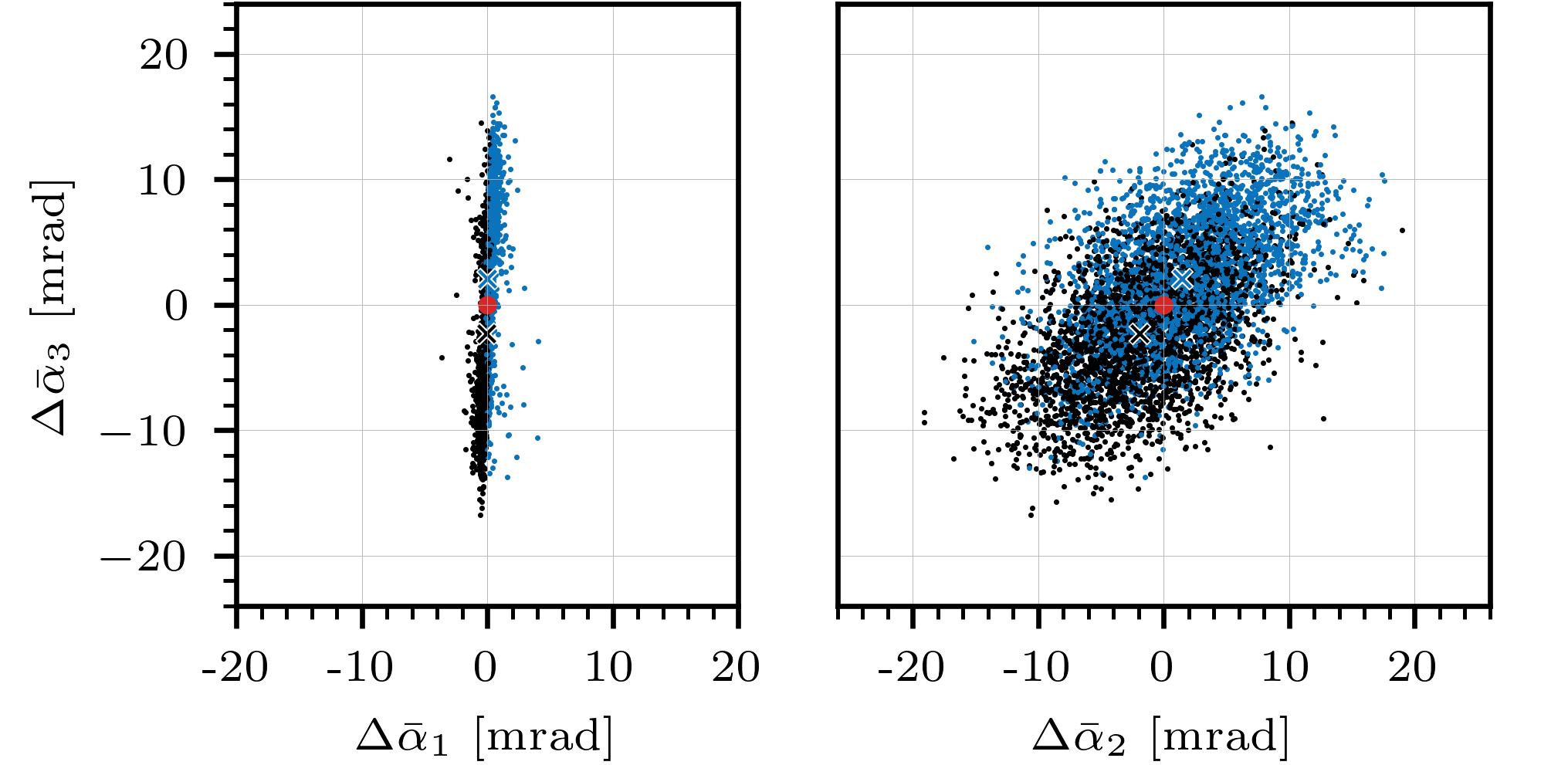}\caption{Positions of the stripping points of the stream stars with respect to the cluster centre in the reference frame defined by the principal axes of the stream. The stripping points corresponding to the stars of the leading arm are shown as black dots, and the stars of the trailing arm as blue dots. The medians of these distributions are shown as crosses, and the position of the cluster is shown as a large red dot. \textit{Left:} Principal axis of the stream $\Delta\bar{\varOmega}_1$ and the perpendicular direction $\Delta\bar{\varOmega}_3$. \textit{Right:} Plane perpendicular to the principal axis.}
\label{invi_alpha}
\end{figure}

Integrating the orbit of a stream star a $\delta t$ time interval backwards from its current position using Eq.~\ref{int_orb}, we obtain its stripping position with respect to the cluster centre:
\begin{equation}\label{alpha}
 \Delta\alpha_i \equiv \Delta\theta_i - \Delta\varOmega_i \, \delta t.
\end{equation}
In Figure~\ref{invi_alpha} we plot the stripping positions of the leading arm stars as black dots and the trailing arm stars as blue dots. We show the distributions in the rotated coordinate frame defined by the principal axes of the stream introduced in Section~\ref{stream_aaf}. In this reference frame, the angle and frequency of the stars in the subspace $\{\Delta\bar{\theta}_1, \Delta\bar{\varOmega}_1\}$ $\gg$ $\{\Delta\bar{\theta}_i, \Delta\bar{\varOmega}_i\}$ for $i=\{2,3\}$. Thus, $\delta t \approx \Delta\theta_1 / \Delta\varOmega_1 $, and by Eq.~\ref{alpha}:
\begin{equation}\label{alpha_approx}
\Delta\bar{\alpha}_1 \approx 0.
\end{equation}
This implies that the particles are ejected roughly from a plane perpendicular to the stream, passing through the centre of the cluster. We show this plane in the upper left panel of Figure~\ref{invi_alpha}, and mark the centre of the cluster with a large red dot. This plane has a small tilt angle. This is explained because Eq.~\ref{alpha_approx} is only valid for stars with small perpendicular separations from the principal axis of the stream. The position of the stripping points on this plane is shown in the right panel. The distribution is stretched along the directions defined by the frequencies $\Delta\bar{\varOmega}_2$ and $\Delta\bar{\varOmega}_3$. This means that the stars in the leading arm tend to be skewed towards negative values and the stars in the trailing arm tend to be skewed towards positive values. The centre of each distribution is also skewed in the same direction. In Figure~\ref{invi_alpha}, we mark with crosses the median position of the stripping points of each arm.

\begin{figure}
\includegraphics[width=1.0\columnwidth]{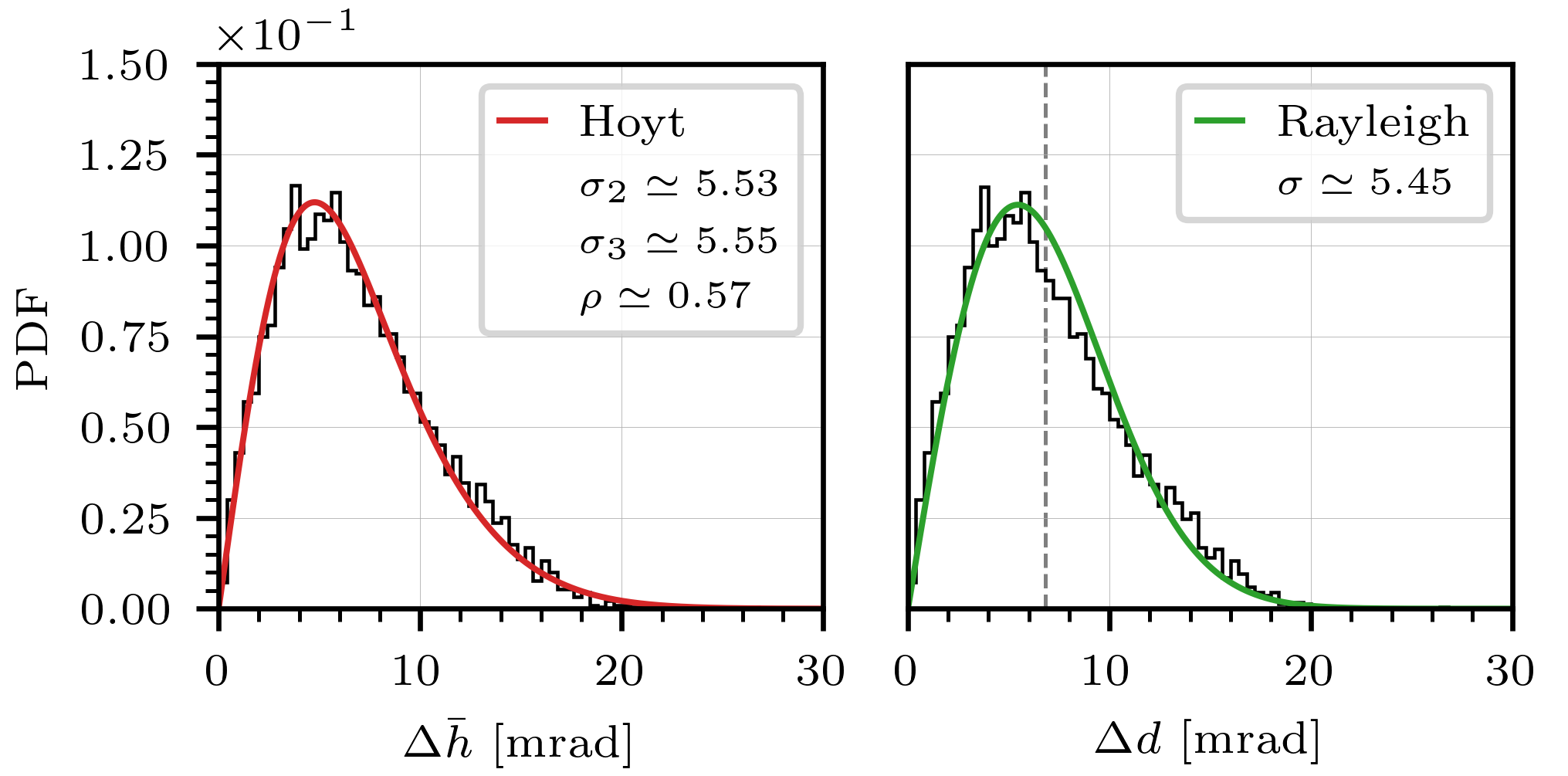}
\caption{\textit{Left:} Histogram of the distances from the cluster centre to the stripping points of the stream stars in the plane perpendicular to the principal axis of the stream. The red line shows the Hoyt distribution, characterised by the standard deviations $\sigma_i$ expressed in mrad and the correlation coefficient $\rho$ given in the legend. \textit{Right:} Histogram of the distances from the cluster centre to the stripping points of the stream stars. The green line shows the best-fitting Rayleigh distribution. Its parameter $\sigma$ is given in mrad in the legend. The vertical grey dashed line marks the mean of the distribution at $\mu_0 \simeq 6.83$ mrad.}
\label{hoyt_ray}
\end{figure}

We define the distance from the cluster centre to the stripping points on the plane perpendicular to the principal axis of the stream as:
\begin{equation}
 \Delta\bar{h}\equiv\sqrt{(\Delta\bar{\alpha}_2)^2+(\Delta\bar{\alpha}_3)^2}.
\end{equation}
In the left panel of Figure~\ref{hoyt_ray} we show a histogram of the sample of distances obtained from the simulated stream. We use the Kolmogorov-Smirnov test to check that the sample of distances, taking both arms together, is compatible with being generated by a Hoyt distribution. In Appendix~\ref{App9} we define this distribution as a function of the covariance matrix of a correlated bivariate Gaussian distribution centred at the origin. We assume that the sample of stripping points on the plane follows this Gaussian to calculate the parameters of the Hoyt distribution. We obtain a $\text{p\:\!-value}\simeq0.12$, indicating that this hypothesis is not rejected at the 95 per cent confidence level ($\alpha=0.05$). This implies that the stripping points are consistent with being generated by a correlated bivariate Gaussian distribution. We plot the Hoyt distribution in red and list its parameters in the legend of the left panel of Figure~\ref{hoyt_ray}.

The mean of the distribution of distances on the plane $\mu_{\rm h}$ can be approximated following Eq. 8.109 from \citetalias{2008gady.book.....B} as:
\begin{equation}\label{alpha_mass}
\mu_{\rm h} \equiv\mean\var{\Delta\bar{h}} \Approx \frac{5}{8}\!\left(\frac{M_{\rm gc}}{M} \right)^{1/3},
\end{equation}
where $M_{\rm gc}$ is the mass of the globular cluster (Table~\ref{M68_properties_table}) and $M$ is the mass of the Milky Way enclosed within a galactocentric radius $r<r_{\rm peri}$. For M68 $\mu_{\rm h} \simeq 6.8$ mrad. Since $M_{\rm gc}$ decreases by $0.34$ per cent at each pericentre passage, the distribution of stripping points remains approximately constant throughout the simulated evolution of the cluster. We therefore consider its variation over time to be negligible.

\begin{figure}
\includegraphics[width=1.0\columnwidth]{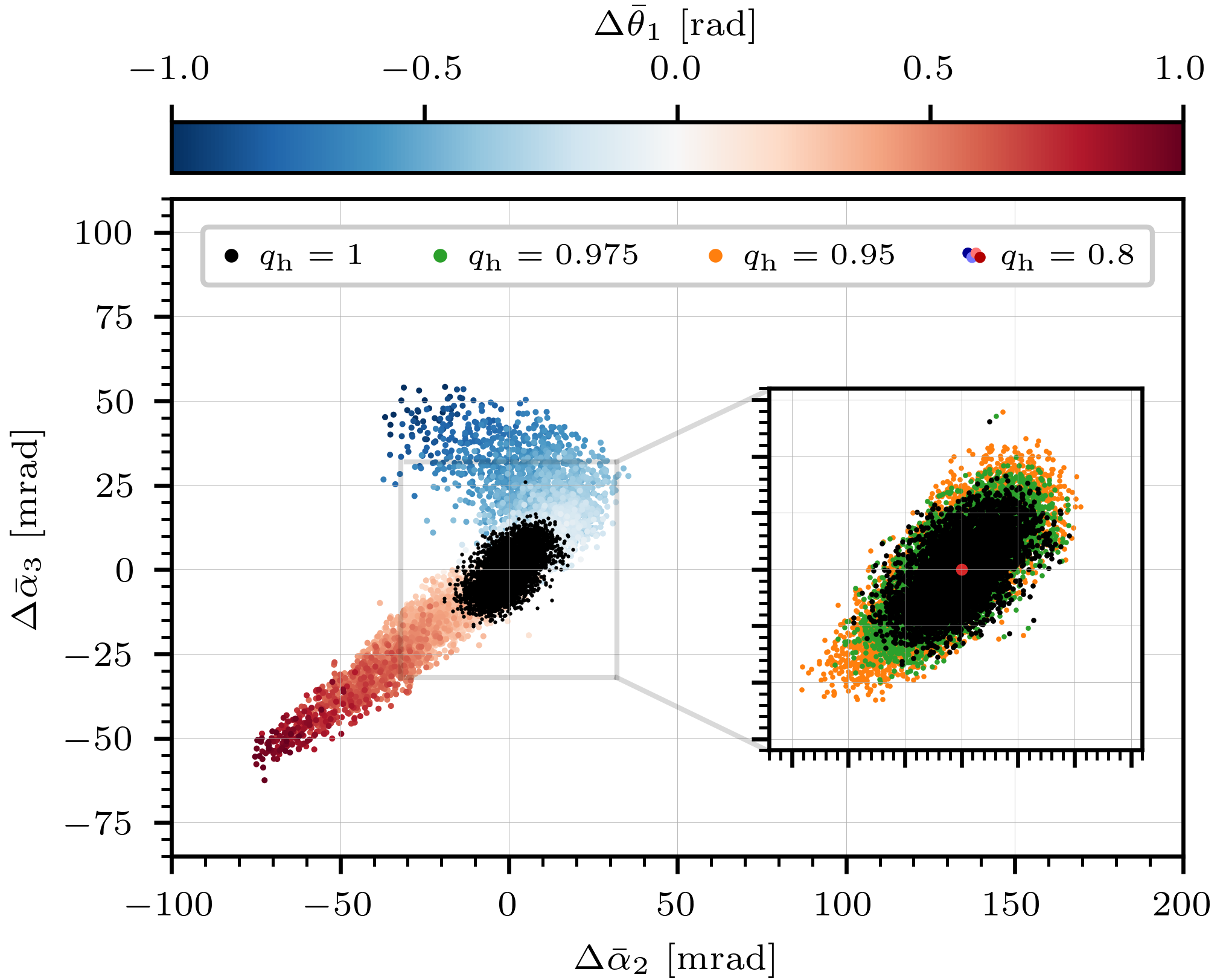}
\caption{Positions of the stripping points in the plane perpendicular to the principal axis of the stream for different dark matter halo density axis ratios $q_{\rm h}$. The reference configuration $q_{\rm h}=1$ is shown as black dots. For $q_{\rm h}=0.8$ the stars in the leading arm are shown as red dots and the stars in the trailing arm are shown as blue dots. The colour brightness is proportional to the position along the principal axis $\Delta\bar{\varOmega}_1$. \textit{Floating panel:} Green and orange dots correspond to the different degrees of oblateness given in the legend. The large red dot marks the position of the globular cluster.}
\label{invi_alpha_q}
\end{figure}

The distribution of stripping points is sensitive to variations in the potential used to compute the angle-action coordinates of the stream stars. As an example, in Figure~\ref{invi_alpha_q}, we show the distribution of $\Delta\bar{\alpha}_2$ and $\Delta\bar{\alpha}_3$ for different dark halo configurations. We define the dark halo density axis ratio as:
\begin{equation}\label{axis_ratio}
q_{\rm h} \equiv \frac{a_3}{a_1},
\end{equation}
where we assume that the scale length in the plane of the Galactic disc $a_1$ is equal to $a_2$ (axisymmetric), and define $a_3$ as the scale length along the perpendicular axis. We reparameterise the \texttt{TriaxialNFWPotential} model from \texttt{galpy} so that the new amplitude is $\texttt{amp}/q_{\rm h}$ and $\texttt{c}=q_{\rm h}$. In this way, the parameter $q_{\rm h}$ determines only the shape of the halo, not its mass.

We compare the spherical configuration used in the simulation (black dots: $q_{\rm h}=1$) to different degrees of oblateness ($q_{\rm h}<1$). For small deviations from the spherical shape (green dots: $q_{\rm h}=0.975$), the distribution of stripping points expands slightly. For larger deviations (orange dots: $q_{\rm h}=0.95$) it expands significantly more. The outliers in the lower left corner of the floating panel of Figure~\ref{invi_alpha_q} allow us to easily distinguish the new distribution from the original. For larger deviations of $q_{\rm h}$, the distribution of stripping points is unstable. In the main panel of the same figure we show the distribution obtained for an oblate halo of $q_{\rm h}=0.8$. The colour gradient indicates the position of the stars along the stream. The stars at the tip of the leading arm (red dots) and the trailing arm (blue dots) deviate significantly, reaching distances of about $70$ mrad from the centre of the cluster. In the following section, we define a method to quantify the difference between the original distribution and those obtained for arbitrary potentials, and use it to constrain the potential of the Milky Way.

\subsection{Methodology for constraining the Galactic potential}\label{const_pot}

In general, we can quantify the difference between two probability distributions $p\var{x}$ and $q\var{x}$ using the Kullback-Leibler divergence \citep[e.g.][]{book:91893105}, which is defined as:
\begin{equation}\label{dkl_eq}
\DKL \equiv \int_{-\infty}^{\infty} p\var{x} \!\: \log\!\left( \frac{p\var{x}}{q\var{x}} \right) dx.
\end{equation}
The $\DKL=0$ if $p\var{x}=q\var{x}$ and $\DKL>0$ if $p\var{x}\neq q\var{x}$. In our case, we can compare the distribution of stripping points of a sample of observed stream stars computed with an arbitrary potential to a model. This model corresponds to the expected distribution of stripping points in the correct potential of the Milky Way, given the properties of the progenitor cluster. By minimising the $\DKL$, we can obtain the potential that best fits this model.

In principle, under certain assumptions, the expected distribution of stripping points can be modelled in the reference frame defined by the principal axes of the stream. For example, assuming that the stripping points follow a Gaussian distribution with scale proportional to $\mu_{\rm h}$ according to Eq.~\ref{alpha_mass}. In practice, the exact distribution for the real stream is not known, and the orientation of the reference frame is difficult to determine. It can be determined by diagonalising the Hessian matrix (Eq.~\ref{Hessian}), but this requires the Hamiltonian expressed as a function of the actions, which is not available for a general potential. Alternatively, it can be determined from a numerical simulation of the stream as in Appendix~\ref{App6}, but this is usually computationally expensive.

These problems can be avoided by maximising the degree of clustering of the stripping points, rather than modelling their expected distribution \citep[e.g.][]{2024arXiv240707947B, 2025MNRAS.538.1442L}. In this case, the $\DKL$ can be used to maximise the divergence between the distribution of stripping points $p\var{x}$ determined from a sample of observed stars $S$ and a uniform distribution $q\var{x}=U$. Substituting into Eq.~\ref{dkl_eq}, we obtain:
\begin{equation}\label{dkl_uni}
\DKL\var{S} = -h\var{S}-\log\!\var{U},
\end{equation}
where $h\var{S}$ is the differential entropy of $p\var{x|S}$ and $U$ is an arbitrary constant. If $h$ is also a function of a set of parameters $\kappa$ that characterise the potential of the Milky Way, the value that maximises the degree of clustering $\hat{\kappa}$ can be obtained by maximising the $\DKL$:
\begin{equation}\label{max_dkl}
\hat{\kappa} \equiv \underset{\kappa}{\mathrm{argmax}}\:\! \big( \DKL\var{\kappa|S} \big).
\end{equation}

In general, it is more convenient to maximise the exponential of the $\DKL$ to avoid negative numbers, since the differential entropy can be either positive or negative. The same maximum is guaranteed because the exponential is a monotonically increasing function. In addition, we can set up the arbitrary constant in Eq.~\ref{dkl_uni} to zero. Under these conditions, Eq.~\ref{max_dkl} becomes a function of the entropy:
\begin{equation}\label{max_dkl_2}
\hat{\kappa} = \underset{\kappa}{\mathrm{argmax}}\:\! \Big( e^{-h\scalebox{0.75}{\var{\kappa|S}}} \Big).
\end{equation}

Estimating the entropy of a random sample requires the evaluation of the probability density function $p\var{x|S}$ that generated the sample. In our case, the number of stars available to estimate the distribution of stripping points is small due to the magnitude constraints imposed by the selection function (Section~\ref{gaia_sf}) and the presence of foreground star contamination that obscures large parts of the stream \citepalias{2019MNRAS.488.1535P}. In addition, the observational errors (Section~\ref{gaia_obs_un}) tend to increase the entropy of the sample and the number of outliers (Section~\ref{obs_strpp}). Limiting the sample to the brightest stars to select only those with smaller observational uncertainties further reduces the number of stars available. For the observable section of the M68 stellar stream we expect to observe $\Approx116$ stars with \textit{Gaia} magnitude $G<20$, and $\Approx19$ with $G<18$ (Table~\ref{number_stars}). These quantities are insufficient to accurately determine $p\var{x|S}$, which is a three-dimensional distribution.

We can increase the density of stripping points by integrating two dimensions, or equivalently, by calculating the Euclidean distance from the stripping points to the cluster centre, which we define as:
\begin{equation}\label{delta_d}
\Delta d \equiv \norm{\Delta\alpha}.
\end{equation}
Several methods have been developed to numerically estimate the entropy of a one-dimensional probability density function from a random sample \citep{AlizadehNoughabi2015}. In general, these methods can be slow and numerically unstable for small samples. Alternatively, we can assume a distribution for $\Delta d$ that converges to the expected distribution of stripping points for the correct potential. In this case, if the entropy can be evaluated analytically, the evaluation is faster and numerically stable for small star samples and small variations of the potential.

Considering that the stripping points are approximately distributed following a two-dimensional Gaussian distribution, as shown in Figure~\ref{hoyt_ray}, we can approximate $\Delta d$ by a Hoyt distribution. Neglecting the correlation and assuming equal variances, the Hoyt distribution reduces to a Rayleigh distribution (Appendix~\ref{App9}). This simplification is convenient because we can calculate the entropy of a Rayleigh distribution analytically. In the right panel of Figure~\ref{hoyt_ray} we plot the histogram of the distances $\Delta d$ of the simulated stream stars in black and the best-fitting Rayleigh distribution in green. The fitted scale $\sigma$ is proportional to the mean according to Eq.~\ref{mean_ray} and its numerical value is given in the legend. We observe that the Rayleigh distribution correctly approximates the distribution of the distances of the stripping points.

The differential entropy is calculated by replacing Eq.~\ref{mean_ray} into Eq.~\ref{entropy_ray}, obtaining:
\begin{equation}\label{entropy}
 h\var{S} = 1 + \frac{\gamma}{2} + \log\:\!\!\left(\frac{\mu\var{S}}{\sqrt{\pi}}\right),
\end{equation}
where the mean of the distribution of stripping points distances is defined as:
\begin{equation}\label{mu}
\mu\equiv\mean\!\var{\Delta d}.
\end{equation}
Finally, substituting Eq.~\ref{entropy} into Eq.~\ref{max_dkl_2} and simplifying the constants, we obtain:
\begin{equation}\label{max_dkl_mean}
\hat{\kappa} = \underset{\kappa}{\mathrm{argmin}} \:\!\! \left(\:\!\mu\var{\kappa|S} \:\!\right).
\end{equation}
In this way, the mean distance of the stripping points from the cluster centre is defined as the loss function. The parameters that better characterise the potential of the Milky Way are those that minimise this function.

The observational uncertainties on the phase-space coordinates of the stream stars (Section~\ref{gaia_obs_un}) significantly affect the estimated angles and frequencies used to compute the distribution of stripping points (Section~\ref{obs_strpp}). In general, the errors tend to increase the scatter of $\Delta d$. The presence of outliers can bias the result of the optimisation. In such a case, it is convenient to estimate the mean of $\Delta d$ with the median, as this statistic is robust in the presence of outliers. We therefore define $\nu\equiv\median\!\var{\Delta d}$, which by Eq.~\ref{mean_ray} and~\ref{median_ray} is proportional to the mean as $\mu=\nu\sqrt{\pi/(4\log\!\!\:\var{2})}$, and substitute it into Eq.~\ref{entropy}. By simplifying the constants, we obtain an analogue to Eq.~\ref{max_dkl_mean} to determine the best-fitting parameters:
\begin{equation}\label{max_dkl_median}
\hat{\kappa} = \underset{\kappa}{\mathrm{argmin}} \:\!\! \left(\:\!\nu\var{\kappa|S} \:\!\right).
\end{equation}
The accuracy of the determination of the best-fitting parameters, their correlations and possible systematic biases obtained with the \texttt{invi} method are examined in the following section.

\subsection{Optimisation of the free parameters}\label{opt_par}

The computational cost of determining the best-fitting configuration is proportional to the number of free parameters to optimise and the time required to evaluate the loss function. The most computationally intensive part of the \texttt{invi} method is the calculation of the angles and frequencies of the stream stars. The parameters of the \texttt{actionAngleIsochroneApprox} method given in Section~\ref{stream_aaf} have been optimised to achieve a compromise between speed and accuracy. However, it is not possible with our computational resources to fully explore all the parameters of the model. We therefore fix the parameters characterising the Sun and the globular cluster, and concentrate on the Galactic potential.

The stellar stream of M68 is located in the inner region of the Milky Way, which we define as the volume within a galactocentric radius $r<15$ kpc. The stream covers a significant part of this volume as shown in Figure~\ref{orbit}. According to the model defined in Section~\ref{mw_pot}, the entire Galactic bulge is contained within this volume, but its gravitational influence on the stream is weak compared to the disc, since its mass is about $\Approx 13$ times smaller. We therefore consider the bulge to be negligible compared to the disc, and fix all its parameters. In contrast, we take the mass $M_{\rm d}$ and the scale length $a_{\rm d}$ of the Miyamoto-Nagai disc as free parameters in the optimisation. We also neglect the vertical scale length of the disc $b_{\rm d}$ because it is about 20 times smaller than the separation between the stream and the plane of the disc, which is $\Approx5$ kpc.

\begin{figure}
\includegraphics[width=1.0\columnwidth]{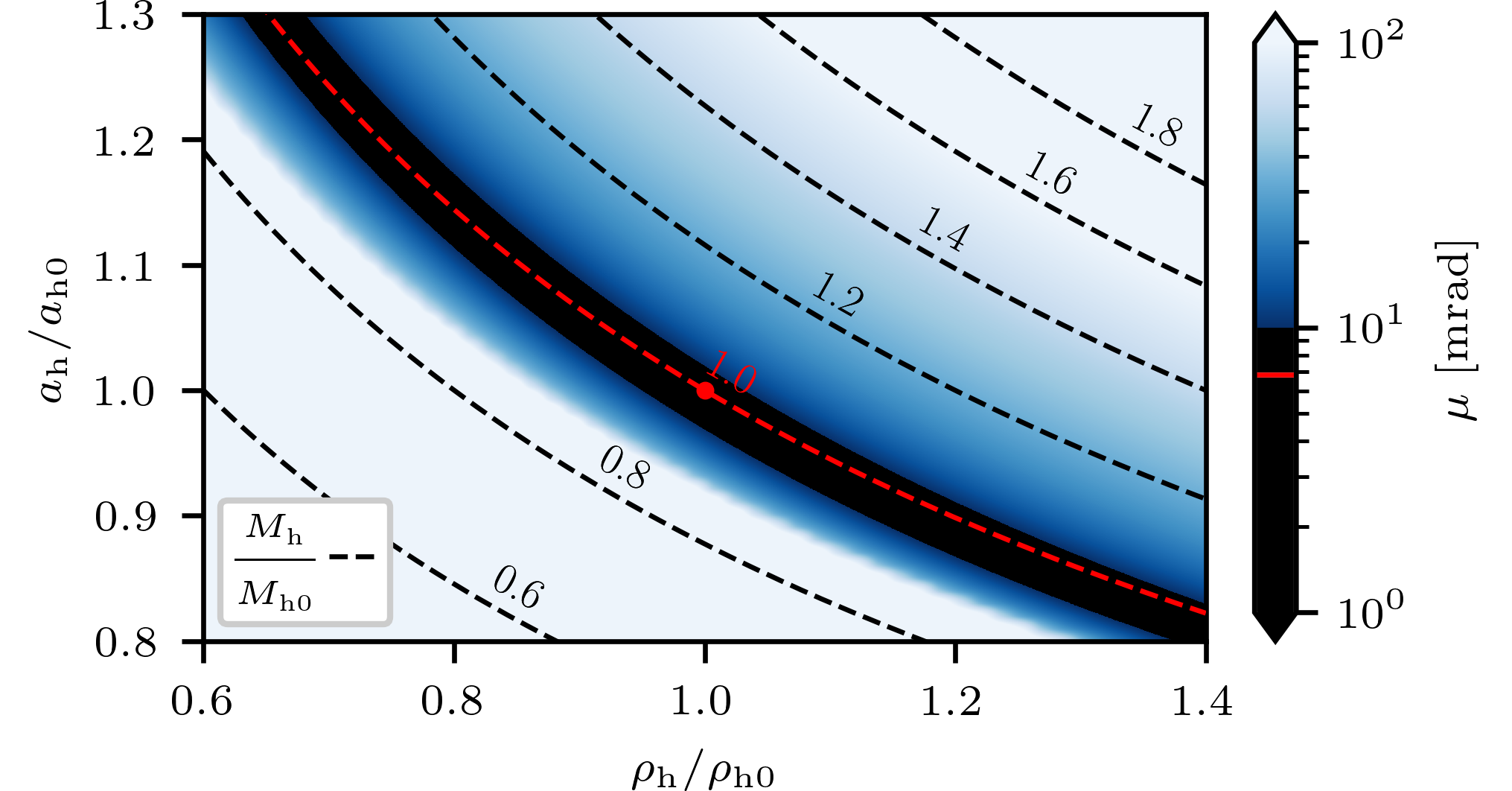}
\caption{Mean of the distribution of the distances of the stripping points of the stream stars $\mu$ as a function of the scale density $\rho_{\rm h}$ and the scale length $a_{\rm h}$ of the dark halo for $q_{\rm h}=1$. The dashed lines mark the configurations of equal mass. The parameters are normalised by the reference values so that the configuration used in the simulation is equal to one. This configuration is marked with a large red dot, and the configurations with the same mass as the reference are marked with a red dashed line.}
\label{halo_2d}
\end{figure}

The dominant component in the inner region of the Galaxy is the dark halo. Its mass within this volume is $M_{\rm h}~\Approx~10^{11} M_{\odot}$, which is about $1.75$ times larger than the mass of the disc within the same volume. In Section~\ref{mw_pot} we assumed that the dark halo follows an NFW density profile. This model depends on several free parameters which are significantly degenerate. To determine whether the stream has sufficient constraining power to break the degeneracies, we study the mean of the stripping point distance distribution $\mu$ defined in Eq.~\ref{mu} for different halo configurations. In Figure~\ref{halo_2d}, we plot $\mu$ as a function of the scale density $\rho_{\rm h}$ and the scale length $a_{\rm h}$ for $q_{\rm h}=1$. The parameters are normalised by the reference values, which are those used in the \nbody\ simulation and given in Table~\ref{MW_table}. We also mark the reference potential with a red dot, and show the equal mass configurations as dashed black lines with the reference highlighted in red.

The region of the parameter space where $\mu\leqslant10$ mrad is shown in black. These configurations are similar to the reference mean $\mu_0 \simeq 6.83$ mrad, and have the same mass (dashed red line). There is no well-defined minimum within this region, its exact value depends on marginal differences between the positions of the stripping points, which can vary significantly as a function of the random sample of stream stars. This means that $\rho_{\rm h}$ and $a_{\rm h}$ cannot be constrained simultaneously. This degeneracy can be broken by fitting multiple streams together or by introducing other observational constraints, such as the Galactic rotation curve. We consider such solutions beyond the scope of this study, and fix $\rho_{\rm h}$ because it does not determine the shape of the halo, but is directly proportional to its mass. We therefore assume that the halo follows an NFW density profile and choose $a_{\rm h}$ and $q_{\rm h}$ as free parameters.

\subsubsection{Optimisation using different star samples}\label{opt_diff_samples}

We use different samples of $S$ simulated stream stars to optimise the parameters $\kappa=\{M_{\rm d},\, a_{\rm d},\, q_{\rm h},\, a_{\rm h}\}$ that characterise the potential of the Milky Way. We optimise using the \texttt{Nelder-Mead} algorithm implemented in the \texttt{scipy} Python package \citep{2020SciPy-NMeth}. The results $\hat{\kappa}$ are presented normalised by the values given in Table~\ref{opt_table}, so that the reference potential used to simulate the stream is $\hat{\kappa}=1$.

The first column of Table~\ref{opt_table} shows the results of optimising with Eq.~\ref{max_dkl_mean} using all the simulated stream stars ($S=5471$). We obtain a systematic bias of up to $\Approx5$ per cent for $q_{\rm h}$. This is caused by the fact that the two arms of the stream completely overlap on the plane of stripping points for some configurations of the potential. In such cases, the centres of the arms, marked with crosses in Figure~\ref{invi_alpha}, coincide with the centre of the cluster, which is marked with a red dot. These configurations have a higher degree of clustering than the reference and therefore bias the result. In Appendix~\ref{App10} we develop a methodology to minimise this systematic bias, assuming that the orientation of the plane of stripping points can be estimated for each potential. We list the corrected values in the second column of Table~\ref{opt_table}. In this case, the systematic deviations are less than 2 per cent for all parameters.

\begin{table}
\caption[]{Results of the optimisation of free parameters $\hat{\kappa}$. The columns below $S=5471$ show the results obtained for the entire sample of simulated stream stars. The columns below $S=100$ show the mean and standard deviation of the results obtained with subsamples of 100 stars. The results obtained by minimising the mean of the stripping point distance distribution are labelled with Eq.~\ref{max_dkl_mean}, and the results including the correction for the systematic bias are labelled App.~\ref{App10}. Estimates of the halo mass $\hat{M}_{\rm h}$ and the mean stripping point distance $\hat{\mu}$ are included. All results are normalised by the values used in the simulation so that the reference potential is $\hat{\kappa}=1$.}
\begin{center}
\begin{tabular}{ccccc}
\toprule
 \boldmath$\hat{\kappa}$ & \multicolumn{2}{c}{\boldmath$S=5471$} & \multicolumn{2}{c}{\boldmath$S=100$} \\[0.25em]
 & Eq.~\ref{max_dkl_mean} & App.~\ref{App10} & Eq.~\ref{max_dkl_mean} & App.~\ref{App10} \\
\midrule
$\hat{M}_{\rm d}$ & $1.043$ & $1.017$ & $1.039\pm0.038$ & $1.027\pm0.038$ \\[0.05em]
$\hat{a}_{\rm d}$ & $1.023$ & $1.014$ & $1.024\pm0.032$ & $1.016\pm0.042$ \\[0.05em]
$\hat{q}_{\rm h}$ & $1.054$ & $1.004$ & $1.052\pm0.032$ & $1.011\pm0.033$ \\[0.05em]
$\hat{a}_{\rm h}$ & $0.994$ & $0.999$ & $0.995\pm0.008$ & $0.998\pm0.009$ \\
\midrule
$\hat{M}_{\rm h}$ & $0.990$ & $0.999$ & $0.992\pm0.013$ & $0.996\pm0.015$ \\[0.05em]
$\hat{\mu}$       & $0.926$ & $-$ & $0.912\pm0.047$ & $-$ \\
\bottomrule
\end{tabular}
\end{center}
\label{opt_table}
\end{table}

\begin{figure}
\includegraphics[width=1.0\columnwidth]{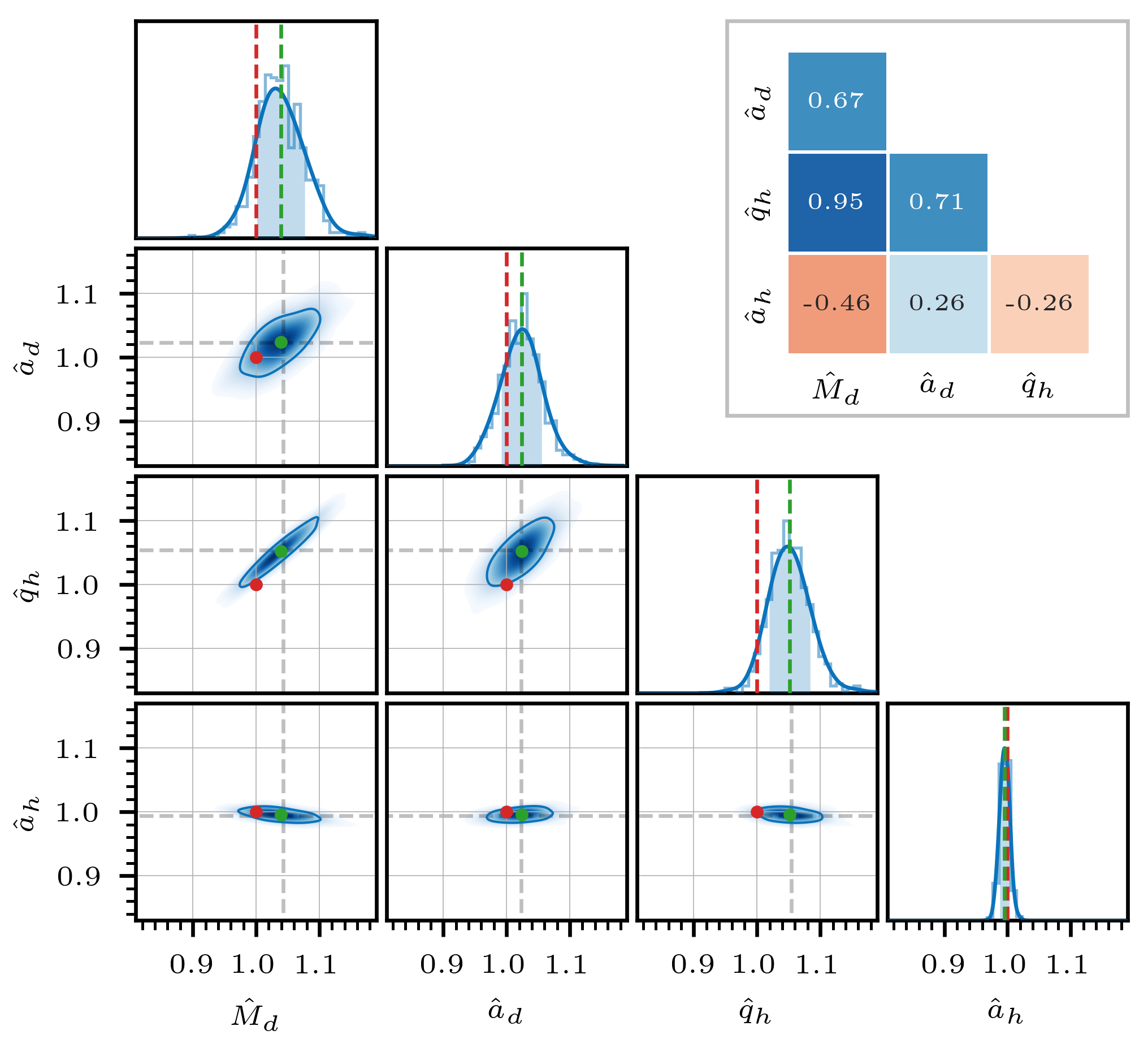}
\caption{Marginal distributions of the results of the optimisation of Eq.~\ref{max_dkl_mean} using subsamples of $S=100$ stars taken from the simulated stream stars. The distributions are estimated using a KDE method. The values of the parameters are normalised so that the reference potential is $\hat{\kappa}=1$. \textit{Diagonal panels:} The solid blue line shows the estimated distribution and the light blue line the histogram of the results. The vertical dashed green line indicates the mean, and the standard deviations are marked by a shaded blue area. The vertical red line indicates the reference potential. \textit{Inferior panels:} The solid blue line marks the limit containing 68 per cent of the distribution. The large green dot marks the mean and the large red dot marks the reference potential. The dashed grey lines mark the values of the optimal configuration obtained with the entire stream ($S=5471$). \textit{Floating panel:} Pearson correlation coefficients.}
\label{corner}
\end{figure}

We also study the results of optimising with Eq.~\ref{max_dkl_mean} for smaller samples of stars. In general, the magnitude of the statistical error depends on the number of stars in the sample and can depend on the distribution of the stars along the stream. Here, as an example, we take random subsamples of $S=100$ stars, which is approximately $1.8$ per cent of the entire simulated stream. We optimise 545 samples, and show in Figure~\ref{corner} the marginal distributions of the optimal configurations, estimated using a Kernel Density Estimation (KDE) method. In the diagonal panels, we include a histogram of the results, and mark the means with vertical green dashed lines and the standard deviations with blue shaded areas. The numerical values are listed in the third column of Table~\ref{opt_table}. In the inferior panels, the grey dashed lines mark the result obtained for the entire stream. We observe that the same systematic bias is present in this case. This bias can be corrected by the method described in Appendix~\ref{App10}. The corrected results are given in the fourth column of Table~\ref{opt_table}. The uncertainties of the free parameters are in the range of $\Approx\Range{1}{4}$ per cent.

In the floating panel of Figure~\ref{corner} we show the Pearson correlation coefficient, which measures the linear correlation between two sets of data. The most significant correlation is between the mass of the disc $M_{\rm d}$ and the dark matter halo axis ratio $q_{\rm h}$, with a coefficient of $0.95$. This is explained by the fact that the stream is close to the disc, almost parallel to the Galactic plane. If the mass of the disc is overestimated, then no dark matter is required near the disc, and prolate configurations of the halo are favoured. If the mass is underestimated, then more dark matter must be introduced to reproduce the reference potential. In this case, oblate halos are favoured. This strong correlation implies that a realistic model of the disc is crucial to obtain a correct estimate of $q_{\rm h}$ from real stellar data.

\section{Simulation of realistic star samples}\label{obs_sample}

In this section we present the methodology for simulating stream star samples as observed by the \textit{Gaia} satellite. These samples are used to test the \texttt{invi} method under realistic conditions. First, we introduce the \textit{Gaia} selection function to determine which simulated stars are likely to be observed. We also define the observable section of the stream by eliminating the stars from regions of the sky where the stream is obscured by foreground star contamination. We then simulate the astrometric, photometric, and spectroscopic uncertainties expected for the GDR3 catalogue. As a consequence of the large parallax uncertainties and the small number of stars with available spectroscopic data, we develop a method to estimate the heliocentric distances and radial velocities of the stream stars from the orbit of the cluster. Finally, the simulated observations are generated by sampling the sky coordinates and proper motions from the estimated error distributions, and taking the distance and radial velocity from the orbit of the cluster.

\subsection{Simulation of the GDR3 selection function}\label{gaia_sf}

We determine which of the simulated stream stars are likely to be observed by the \textit{Gaia} satellite using the model of the \textit{Gaia} selection function included in the Python package \texttt{gaia-unlimited}\footnote{\url{https://gaia-unlimited.org/}}. We apply the model \texttt{DR3SelectionFunctionTCG} \citep{2023AandA...669A..55C}, which provides an estimate of the probability of a source being included in the GDR3 catalogue as a function of its sky coordinates and $G$ magnitude. We take the sky coordinates and distance of each star from the \nbody\ simulation (Section~\ref{sim_str}), and calculate its apparent magnitude from the synthetic stellar population (Section~\ref{sim_stell_pop}), including the effect of dust extinction (Appendix~\ref{App2}).

We generate random samples of stars passing through the selection function and list the mean value and standard deviation in Table~\ref{number_stars}. About $411$ of the stream stars are expected to pass through the selection function cut, which is $\Approx7.5$ per cent of the total stream stars. About $80$ per cent of these stars are located in the leading arm because it is closer to the Sun, at $\Approx5$ kpc, while the trailing arm is beyond $10$ kpc. The uncertainty in passing through this cut is only significant for the faintest stars, which are a small fraction of $\Approx0.2$ per cent of the total. In general, the methods used to separate stream members from foreground stars require low photometric uncertainties, which are only available for the brightest stars \citepalias[e.g.][]{2019MNRAS.488.1535P}. In this way, we also include in Table~\ref{number_stars} the number of stars of magnitude $G<20$ mag that pass the selection function. This provides a more realistic estimate of the number of stars in the GDR3 catalogue that can be identified as stream members.

\subsubsection{Observable section of the stream}\label{obs_section}

We define the observable section of the stream as the part with declination $\delta > -8$ deg \citepalias{2019MNRAS.488.1535P}. This part covers more than a half of the leading arm and can be clearly separated from the foreground because the stream stars have a much larger proper motion. This section is also close to the Sun and is projected onto the halo, in an area of low star density. Below the declination threshold, the stream is obscured by foreground stars. In this region, the proper motions of the stream stars are similar to those of the foreground stars, which makes their separation difficult. In addition, the stream is projected close to the disc, where the density of foreground stars is much higher. In our simulation, the observable part is populated by $1\,752$ stars, of which an average of $237$ pass through the selection function and about $116$ are likely to be identified as stream members.

\subsubsection{GDR3 selection function for radial velocities}\label{rv_sel_func}

We use the \texttt{gaia-unlimited} package to estimate the number of expected stars with measured radial velocity in the \textit{Gaia} catalogue. We use the \texttt{DR3RVSSelectionFunction} model, which gives the probability of observing radial velocities in GDR3 as a function of the sky coordinates, $G$, and $G_{\rm RP}$ magnitudes of the source. For the entire stream we expect $3 \pm 2$ stars passing the selection function with measured radial velocity and $1 \pm 1$ in the observable section. It is also possible to obtain spectroscopic data from other surveys that include stars that are considered to be members of the stream. The Survey of Surveys dataset \citep{2022AandA...659A..95T} includes measurements from SEGUE \citep{2009AJ....137.4377Y} on about 9 candidate stream stars, and \cite{2021ApJ...914..123I} on about 25 stars. In both cases most of the stars are concentrated within the right ascension interval $\alpha\Approx\Range{193}{210}$ deg, which is near the part of the stream closest to the Sun (Section~\ref{rh_rv_est}). For this simulation, this interval corresponds to an angle along the principal axis of the stream of $\Delta\bar{\theta}_1\simeq\Range{0.30}{0.45}$ deg. This part is considerably shorter than the observable section of the stream, covering only about 20 per cent of its length.

\subsection{Simulation of the GDR3 observational uncertainties}\label{gaia_obs_un}

We estimate the astrometric and spectroscopic observational uncertainties of the simulated stream stars as if they were included in the \textit{Gaia} catalogue. We use the performance models provided by the \textit{Gaia} Mission\footnote{\url{https://www.cosmos.esa.int/web/gaia/science-performance\#astrometric\%20performance}} and implemented in the Python toolkit \texttt{pygaia}\footnote{\url{https://pypi.org/project/pygaia/}}. We estimate the uncertainties as expected in the GDR3 catalogue under the assumption that they are uncorrelated. We discuss the mean of the estimates for the 116 stream stars of magnitude $G<20$ mag located in the observable section of the stream.

We obtain a mean uncertainty in the sky coordinates of $\sigma_{\delta},\, \sigma_{\alpha} \Approx 0.25 \pm 0.15$ mas, which we consider negligible for the application of the \texttt{invi} method. The estimated observational uncertainty of the proper motions is $\sigma_{\mu_\delta},\, \sigma_{\mu_{\alpha *}}\Approx0.30 \pm 0.15$ mas~yr$^{-1}$. The error grows exponentially with the $G$ magnitude as $\sigma_{\mu_\delta},\, \sigma_{\mu_{\alpha *}} \!\Approx \exp\:\!\bigl(0.725\,G - 15\bigr)$ mas~yr$^{-1}$. It is therefore possible to significantly reduce the mean uncertainties by restricting the sample to the brightest stars. In Table~\ref{number_stars} we list the number of stars of magnitude $G<18$ mag that pass the selection function for each component of the stream. In this case, the average uncertainty for the 19 stars in the observable section is $\sigma_{\mu_\delta},\, \sigma_{\mu_{\alpha *}} \!\Approx 0.08 \pm 0.03$ mas~yr$^{-1}$. In Appendix~\ref{App7} we study the effect of these uncertainties on the determination of the angle-action coordinates and the stripping time of the stream stars.

For the parallax, we get $\sigma_\varpi \simeq 0.32 \pm 0.17$ mas. This error is too large to provide accurate heliocentric distance estimates. The average distance error derived for the observable stars is $\sigma_{r_{\rm h}} \Approx 4.7 \pm 4.6$ kpc. On the other hand, the estimated mean uncertainty of the radial velocity in GDR3 is small, of $\Approx2.1$ km~s$^{-1}$, but the number of stars is insufficient to apply the \texttt{invi} method. The inclusion of the measurements from the surveys discussed in Section~\ref{rv_sel_func} does not provide a significant improvement, as they cover only a short part of the observable section of the stream, and the radial velocities provided have large observational uncertainties of average $\Approx 8$ km~s$^{-1}$. We therefore develop a method for estimating the distance and radial velocity of the stars from the fact that they belong to a stream with a known progenitor.

\subsection{Distance and radial velocity estimation}\label{rh_rv_est}

The heliocentric distance $r_{\rm h}$ and radial velocity $v_r$ of the M68 stream stars can be estimated from the orbit of the progenitor cluster, calculated assuming a Galactic potential. This gives a much better estimate than the measurements from the \textit{Gaia} satellite, since the phase-space position of the globular cluster is known with high precision (Table~\ref{M68_orbital_parameters_table}) and the stream follows the orbit closely (Figure~\ref{orbit}). Note that if the position of the cluster and the potential of the Galaxy are not known accurately, the estimates obtained by this method may be significantly biased.

To estimate the distance and radial velocity, we determine the closest point on the cluster's orbit to each star. In this case, we compute the orbit of the cluster forward in time, since the observed section of the stream corresponds to the leading arm. The closest point is calculated by minimising the angular distance:
\begin{equation}
 \beta = 2\arcsin\!\left(\frac{c}{2}\right),
\end{equation}
where $c$ is the chord distance between the star and the points of the orbit. We then assign to each star the $r_{\rm h}$ and $v_r$ of the closest point on the orbit, and correct for the systematic offset between the stream and the orbit. The correction is made by adding the mean offset of the stream stars obtained for each section of the stream. The means and standard deviations of the offsets are listed in Table~\ref{str_orb_offset}.

\begin{table}
\caption[]{Mean and standard deviation of the difference between the heliocentric distance $r_{\rm h}$ and radial velocity $v_r$ of the stream stars estimated from the orbit of M68 and the simulated value. The values are given for the leading and trailing arms and for the observable section of the stream.}
\begin{center}
\begin{tabular}{lrr}
\toprule
& \multicolumn{1}{c}{\textbf{Offset} \boldmath$r_{\rm h}$} & \multicolumn{1}{c}{\textbf{Offset} \boldmath$v_r$} \\
& \multicolumn{1}{c}{\units{pc}} & \multicolumn{1}{c}{\units{km s$^{-1}$}} \\
\midrule
\textbf{Leading}  & $-165.22\pm65.33\phantom{0}$ & $-4.22\pm1.91$ \\[0.25em]
\textbf{Trailing} & $308.78\pm135.68$ & $4.26\pm2.50$ \\[0.25em]
\textbf{Obs. Sec.}& $-159.76\pm60.67\phantom{0}$ & $-4.45\pm1.80$ \\
\bottomrule
\end{tabular}
\end{center}
\label{str_orb_offset}
\end{table}

\begin{figure}
\includegraphics[width=1.0\columnwidth]{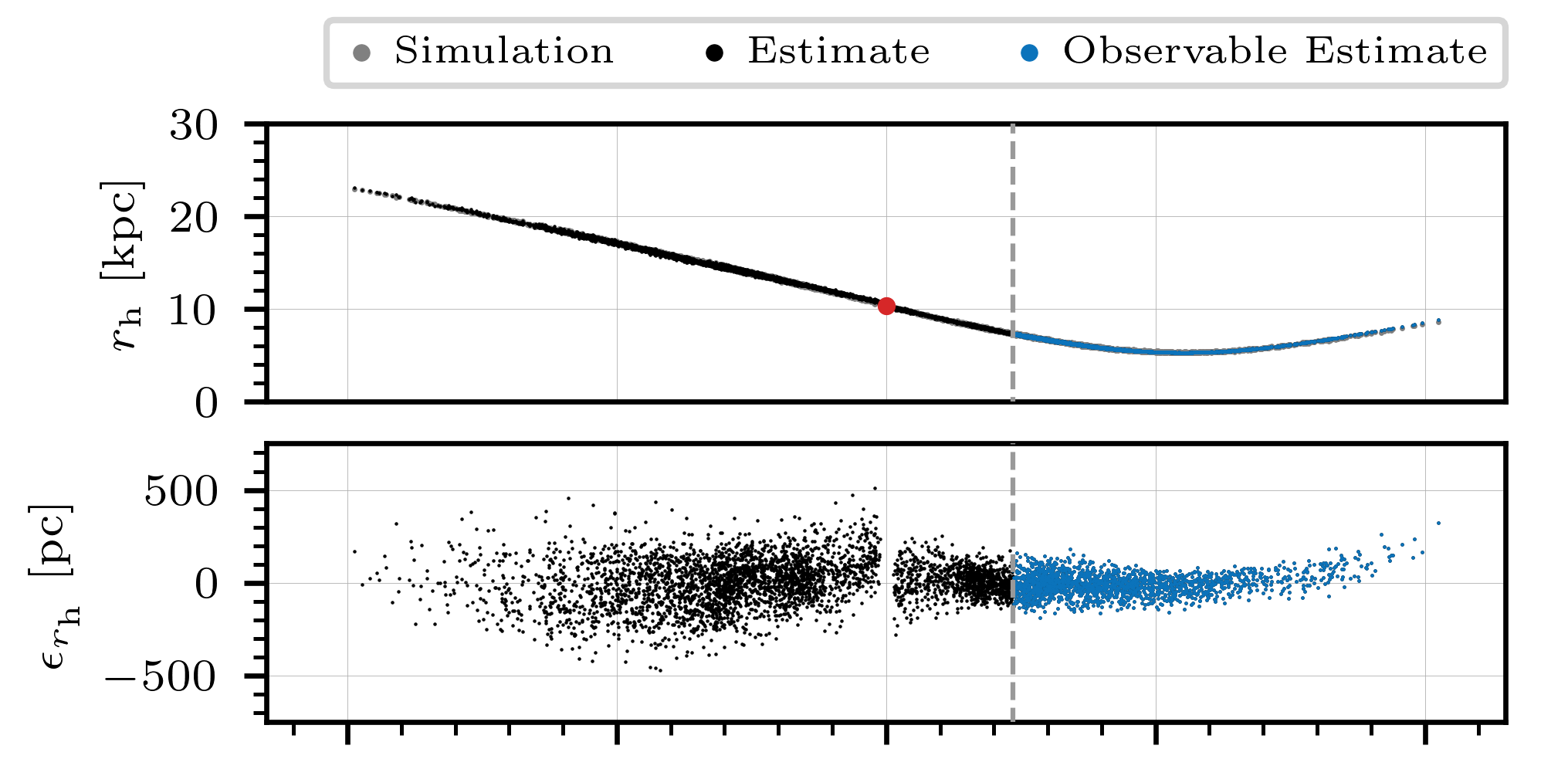}
\includegraphics[width=1.0\columnwidth]{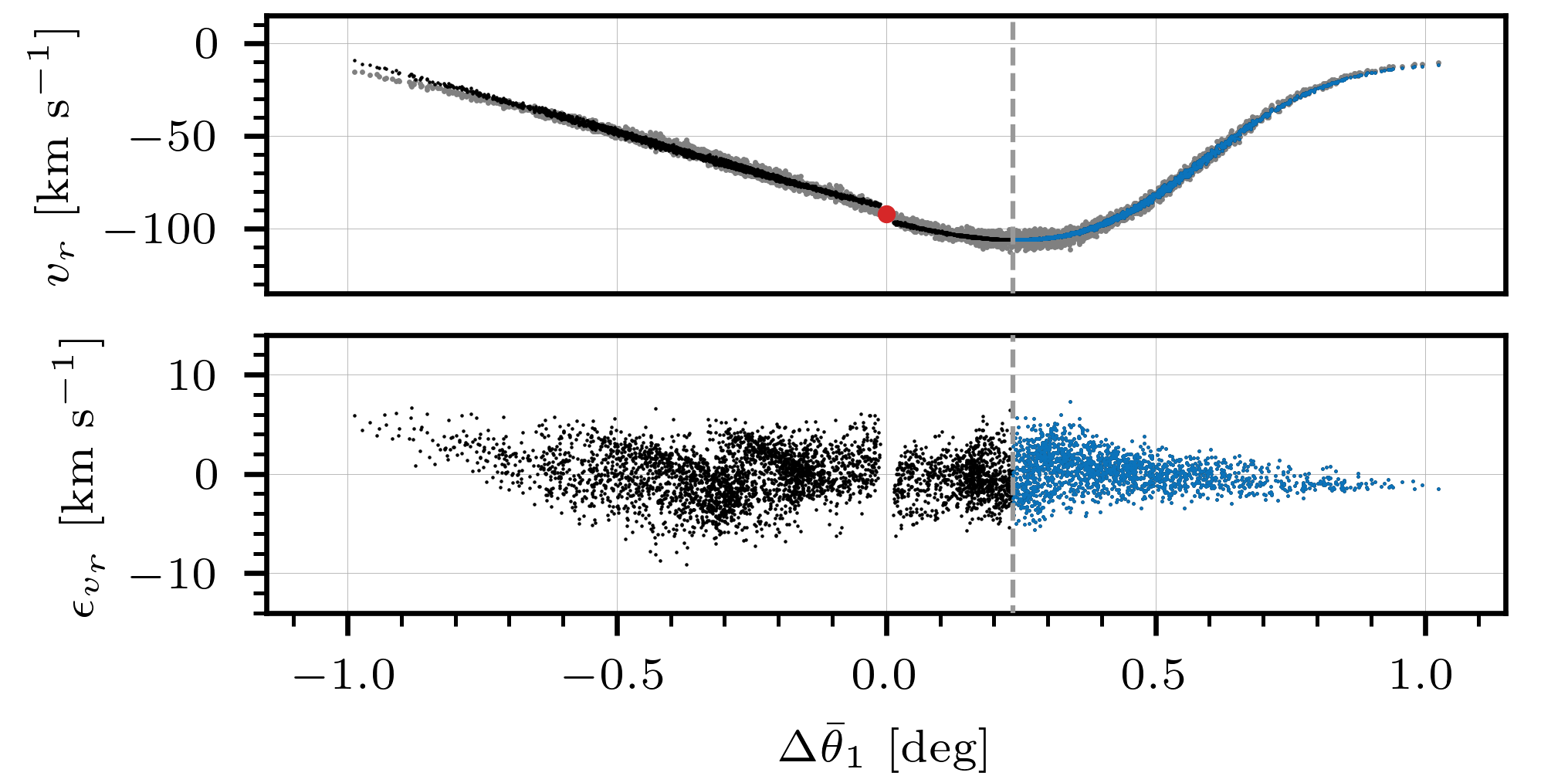}
\caption{Heliocentric distance $r_{\rm h}$ (\textit{First panel}) and radial velocity $v_r$ (\textit{Third panel}) of the stream stars as a function of the angle along the principal axis of the stream $\Delta\bar{\theta}_1$. The small grey dots indicate the exact values from the simulation, and the coloured dots the estimated values from the orbit of the cluster. The observable section of the stream is shown as blue dots, and the rest as black dots. The two parts of the stream are separated by a vertical dashed line. The large red dot indicates the position of the cluster. \textit{Second and Third panels:} Difference between the estimates and the simulated values $\epsilon$.}
\label{A1_r_rv}
\end{figure}

In Figure~\ref{A1_r_rv} we plot the heliocentric distance (first panel) and the radial velocity (third panel) as a function of the angle along the principal axis of the stream $\Delta\bar{\theta}_1$. The large red dot indicates the position of the cluster, the grey dots the simulated stream stars, and the coloured dots the estimated values using this method. The observable section of the stream is shown in blue and the rest of the stream in black. The second and third panels show the difference between the estimates and the simulated values. We define this magnitude as the error $\epsilon$. We observe that the estimates are in good agreement with the simulated values along the entire observable section of the stream. We obtain a systematic error in $r_{\rm h}$ only for the stars with $\Delta\bar{\theta}_1\GtrSim0.8$ rad.

For the 116 stars in the observable section of the stream, this method gives estimates with standard deviation errors of $\sigma_{r_{\rm h}} \simeq 53.9$ pc and $\sigma_{v_{r}} \simeq 1.7$ km s$^{-1}$. These values are a lower bound on the uncertainty, as they are obtained using the Galactic potential used in the simulation and the exact position of the M68 cluster. In Appendix~\ref{App7} we study the effect of these errors on the determination of the angle-action coordinates and the stripping time of the stream stars.

\subsection{Generation of realistic star samples}\label{real_sample}

The simulated observations used to test the \texttt{invi} method are based on the 116 ($G<20$ mag), and the 19 ($G<18$ mag) stars that pass the \textit{Gaia} selection function and are located in the observable section of the stream. We generate the simulated observations by sampling the sky coordinates and proper motions from uncorrelated Gaussian distributions. For each star, we take the mean of the distributions from the value of the coordinates obtained in the \nbody\ simulation, and the standard deviations from the \textit{Gaia} performance models. On the other hand, the heliocentric distances and the radial velocities are estimated from the orbit of the cluster. The phase-space position of the M68 cluster is fixed to the values used in the simulation (Table~\ref{M68_orbital_parameters_table}), and the orbit is computed for each trial potential during the optimisation process described in the following section.

\section{Inverse time integration method with realistic star samples}\label{opt_obs}

In this section we investigate how accurately we can recover the parameters of the reference potential using the \texttt{invi} method with realistic star samples. These samples include the \textit{Gaia} selection function, the observational uncertainties, and the estimates of the heliocentric distances and radial velocities from the cluster orbit. We begin by describing the appearance of outliers in the distribution of stripping points caused by the errors in the estimates of the angles and frequencies of the stream stars. We then evaluate the magnitude of the statistical errors and the bias introduced by optimising using only samples of stars in the observable section of the stream. We then incorporate the observational uncertainties and compare the results obtained for GDR3 and GDR5. We also test the effect of restricting the sample to the brightest stars. Finally, we evaluate the total uncertainty in the free parameter determination and present a method for estimating these uncertainties for a real sample of observed stars.

\subsection{Effect of the observational uncertainties on the stripping point distribution}\label{obs_strpp}

\begin{figure*}
\includegraphics[width=1.0\textwidth]{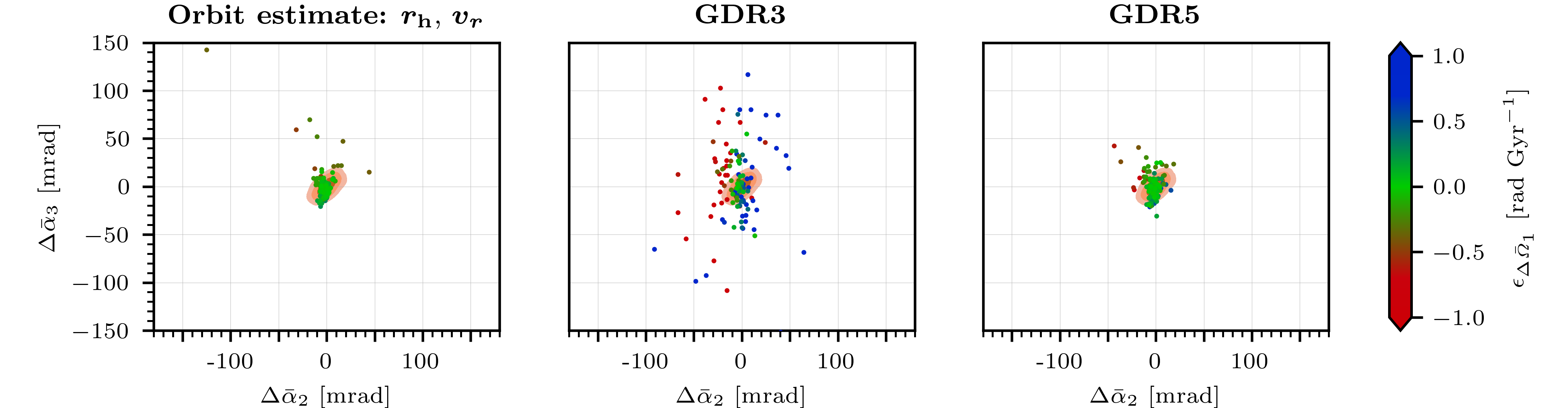}
\caption{Position of the stripping points on the plane perpendicular to the principal axis of the stream with respect to the cluster centre ($\Delta\bar{\alpha}_2, \Delta\bar{\alpha}_3$) for the 116 stream stars of magnitude $G<20$ mag that pass the selection function in the observable section of the stream (coloured dots). The colour of the dots is proportional to the error in frequency along the principal axis of the stream $\Delta\bar{\varOmega}_1$. Green colour corresponds to stars with a small error such that their estimated frequency is roughly compatible with the stream. The orange areas mark the region containing the 99, 95, and 68 per cent of the stripping points of the simulated stream stars. \textit{Left:} Effect of estimating the heliocentric distance $r_{\rm h}$ and radial velocity $v_r$ from the cluster orbit. \textit{Middle:} Example of a realistic star sample as observed by GDR3 with $r_{\rm h}$ and $v_r$ estimated from the cluster orbit. \textit{Right:} Same as the middle panel, but for GDR5.}
\label{stripp_err}
\end{figure*}

Estimating the heliocentric distance and radial velocity from the cluster orbit and introducing the simulated observational errors increases the entropy of the stripping point distribution. In Figure~\ref{stripp_err} we show the effect of different sources of error on the distribution of the stripping points in the plane perpendicular to the principal axis of the stream ($\Delta\bar{\alpha}_2, \Delta\bar{\alpha}_3$). The coloured dots  correspond to the 116 stream stars that pass the selection function in the observable section of the stream. The colour is proportional to the error in frequency along the principal axis $\Delta\bar{\varOmega}_1$. The stars shown in green have errors of $|\epsilon_{\Delta\bar{\varOmega}_1}| \LessSim 0.5$ rad~Gyr$^{-1}$, which implies that they are roughly compatible with the distribution of frequencies of the stars along the stream. The orange areas in the centre of each panel mark the region containing the stripping points of the simulated stream stars shown in Figure~\ref{invi_alpha}.

In the left panel of Figure~\ref{stripp_err} we show the effect of estimating $r_{\rm h}$ and $v_r$ from the cluster orbit. In Appendix~\ref{App7} we find that this method introduces negligible errors in angle space, and frequency errors with a standard deviation of $\sigma_{\Delta\bar{\varOmega}_1}\Approx0.2$ rad~Gyr$^{-1}$. The stars with the smallest errors (green dots) overlap with the reference distribution (orange area), indicating that they return to the cluster after the inverse integration. The outliers are stars with underestimated $\Delta\bar{\varOmega}_1$. In these cases $\Delta\bar{\varOmega}_1 \approx \Delta\bar{\varOmega}_2 \approx \Delta\bar{\varOmega}_3 \Approx 0$ rad~Gyr$^{-1}$, which implies that the estimated integration time from Eq.~\ref{delta_t} is large. According to Eq.~\ref{alpha}, a long integration time amplifies the deviations caused by the frequencies perpendicular to the stream \{$\Delta\bar{\varOmega}_2, \Delta\bar{\varOmega}_3$\}. These deviations make the slow stars appear as outliers in the plane of the stripping points.

In the middle panel of Figure~\ref{stripp_err} we show an example of a distribution of stripping points obtained for a realistic GDR3 sample of stream stars. The sample is generated using the method described in Section~\ref{real_sample}. This simulation includes the observational uncertainties in right ascension and declination, although their effect is negligible, and the estimates of $r_{\rm h}$ and $v_r$ from the cluster orbit. The errors in $\Delta\bar{\varOmega}_1$ are dominated by the proper motion uncertainties. In Appendix~\ref{App7} we estimate that for GDR3 the standard deviation of the errors is $\sigma_{\Delta\bar{\varOmega}_1}\Approx1.0$ rad~Gyr$^{-1}$. This error is significantly larger than the internal dispersion of frequencies along the principal axis of the stream. As a consequence, the resulting number of stars returning to the cluster (green dots) is about $30$ per cent of the total.

\begin{figure}
\includegraphics[width=1.0\columnwidth]{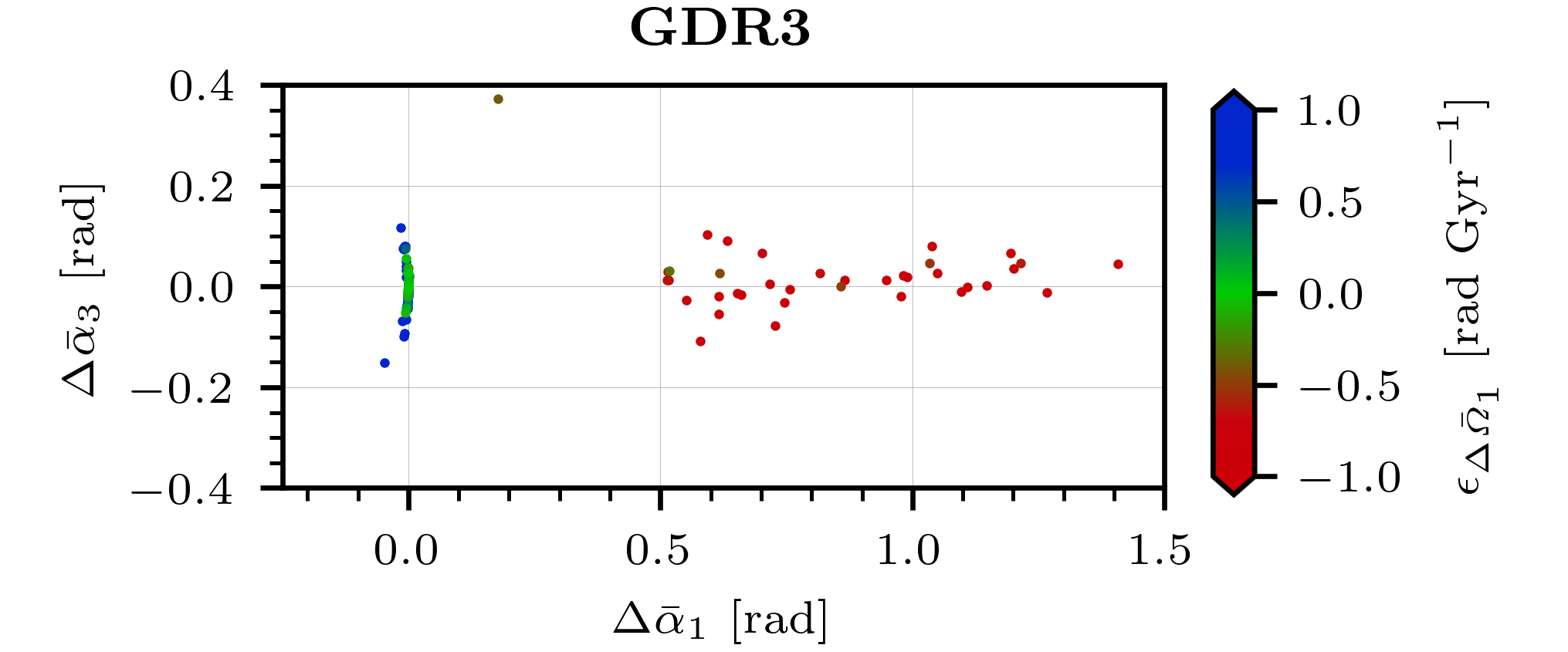}
\caption{Same as the middle panel of Figure~\ref{stripp_err} but for the principal axis of the stream $\Delta\bar{\alpha}_1$ and the perpendicular direction $\Delta\bar{\alpha}_3$.}
\label{stripp_err_A1}
\end{figure}

The large errors in frequency imply that the estimated $\Delta\bar{\varOmega}_1$ can be negative for some stars. Taking into account that for the leading arm $\Delta\bar{\varOmega}_1>0$, according to Eq.~\ref{alpha}, the stars with estimated $\Delta\bar{\varOmega}_1<0$ are integrated in the opposite direction to the cluster. In configuration space, this corresponds to a situation where the relative velocity of the stream stars with respect to the cluster is underestimated. In this case, the distance between the stars and the cluster increases in the inverse integration because the cluster is moving faster than the stars. In Figure~\ref{stripp_err_A1} we show the position of the stripping points along the principal axis of the stream for the GDR3 sample. The stars with larger underestimated values of $\Delta\bar{\varOmega}_1$ (red dots) do not return to the cluster. On the other hand, according to Eq.\ref{alpha_approx}, the stars with overestimated $\Delta\bar{\varOmega}_1$ (blue dots) always return approximately to the plane of stripping points.

\subsection{Optimisation of the free parameters using realistic star samples}

In Section~\ref{opt_diff_samples} we optimise the parameters of the Milky Way potential by minimising the mean of the stripping point distances from the cluster centre (Eq.~\ref{max_dkl_mean}). The presence of outliers in the stripping point distribution caused by observational uncertainties implies that the mean is not a good estimator of the central tendency of the distribution. These outliers are inevitable because an overestimation of the heliocentric distance of $\Approx\Range{50}{100}$ pc is sufficient to obtain errors of $\epsilon_{\Delta\bar{\varOmega}_1}\LessSim-0.5$ rad Gyr$^{-1}$ (Appendix~\ref{App7}). This implies that the results are systematically biased by the stars with underestimated $\Delta\bar{\varOmega}_1$, especially by those with negative estimates, since they do not return to the cluster.

To avoid this problem, we can estimate the central tendency of the distance distribution using a robust statistic in presence of outliers. We use the median and optimise the free parameters of the model using the method in Eq.~\ref{max_dkl_median}. The disadvantage of the median is that we do not use the whole sample to estimate the mean distance, but only the central values. This loss of information affects the sensitivity, especially when determining correlations between parameters. In addition, the median introduces discontinuous variations in the estimated mean distance as the free parameters are varied, especially for small numbers of stars. Depending on the configuration of the stars, this can result in an irregular loss function with multiple relative minima, which is difficult to optimise.

To test the median-based method (Eq.~\ref{max_dkl_median}), we optimise the free parameters $\kappa$ using the entire simulated stream. We list the results obtained in the first and second columns of Table~\ref{opt_median_table_a}. The results are normalised by dividing by the values used in the simulation. In this way, the reference potential is $\hat{\kappa}=1$. In this case, we get a systematic bias of $\LessSim4$ per cent for all parameters. This bias is slightly smaller than that obtained with the mean-based method (Eq.~\ref{max_dkl_mean}). If the orientation of the stripping point plane can be determined, this bias can be corrected using the method presented in Appendix~\ref{App10}. In the following sections we test the median-based method when including the effect of the selection function and the observational uncertainties.

\begin{table}
\caption{Results of the free parameter optimisation $\hat{\kappa}$ for different samples of stars, obtained by minimising the median of the stripping point distance distribution (Eq.~\ref{max_dkl_median}). Estimates of the halo mass $\hat{M}_{\rm h}$ and the mean stripping point distance $\hat{\mu}$ are included. All results are normalised by the values used in the simulation so that the reference potential is $\hat{\kappa}=1$.}
\begin{subtable}[b]{1.0\columnwidth}

\begin{center}
\vspace{1.25em}
\begin{tabular}{ccccc}
\toprule
 \boldmath$\hat{\kappa}$ & \multicolumn{2}{c}{\boldmath$S=5471$} & \boldmath$S=121\pm11$ & \boldmath$S=22\pm5$ \\[0.25em]
 & Eq.~\ref{max_dkl_median} & App.~\ref{App10} & Eq.~\ref{max_dkl_median} & Eq.~\ref{max_dkl_median} \\
\midrule
$\hat{M}_{\rm d}$ & $1.013$ & $1.005$ & $1.009\pm0.010$ & $1.014\pm0.039$\\[0.05em]
$\hat{a}_{\rm d}$ & $1.037$ & $1.005$ & $1.013\pm0.015$ & $1.017\pm0.054$\\[0.05em]
$\hat{q}_{\rm h}$ & $1.031$ & $0.997$ & $1.017\pm0.010$ & $1.023\pm0.035$\\[0.05em]
$\hat{a}_{\rm h}$ & $1.005$ & $1.000$ & $0.996\pm0.006$ & $0.996\pm0.023$\\
\midrule
$\hat{M}_{\rm h}$ & $1.009$ & $1.001$ & $0.993\pm0.011$ & $0.994\pm0.038$\\[0.05em]
$\hat{\mu}$       & $0.919$ & $0.971$ & $0.852\pm0.057$ & $0.725\pm0.120$\\[0.05em]
\bottomrule
\end{tabular}
\end{center}

\caption{The columns below $S=5471$ show the results obtained for the entire sample of simulated stream stars. The other columns show the mean and standard deviation of the results for stars of magnitude $G<20$ mag ($S=121 \pm 11$) and $G<18$ mag ($S=22 \pm 5$) in the observed section of the stream that pass the selection function. The results shown below App.~\ref{App10} include the correction for the systematic bias.}
\label{opt_median_table_a}

\end{subtable}
\begin{subtable}[b]{1.0\columnwidth}

\begin{center}
\vspace{1.25em}
\begin{tabular}{ccccc}
\toprule
\boldmath$\hat{\kappa}$ & \textbf{Simu.} & \boldmath$r_{\rm h}$, \boldmath$v_r$ & \textbf{GDR3} & \textbf{GDR5} \\
\midrule
$\hat{M}_{\rm d}$ & $1.035$ & $1.028$ & $0.961\pm0.092$ & $1.001\pm0.011$\\[0.05em]
$\hat{a}_{\rm d}$ & $1.034$ & $0.964$ & $1.071\pm0.085$ & $1.020\pm0.024$\\[0.05em]
$\hat{q}_{\rm h}$ & $1.041$ & $0.994$ & $0.956\pm0.078$ & $1.011\pm0.027$\\[0.05em]
$\hat{a}_{\rm h}$ & $0.990$ & $0.964$ & $0.981\pm0.063$ & $0.993\pm0.012$\\
\midrule
$\hat{M}_{\rm h}$ & $0.983$ & $0.940$ & $0.969\pm0.106$ & $0.988\pm 0.020$\\[0.05em]
$\hat{\mu}$       & $0.754$ & $0.913$ & $6.434\pm1.172$ & $1.310\pm 0.092$\\[0.05em]
\bottomrule
\end{tabular}
\end{center}

\caption{Results for the 116 stars of magnitude $G<20$ mag. The columns show the results obtained with simulated stars (Simu.), including the estimates of $r_{\rm h}$ and $v_r$ from the cluster orbit, and including the observational uncertainties for GDR3 and GDR5.}
\label{opt_median_table_b}

\end{subtable}
\begin{subtable}[b]{1.0\columnwidth}

\begin{center}
\vspace{1.25em}
\begin{tabular}{ccccc}
\toprule
\boldmath$\hat{\kappa}$ & \textbf{Simu.} & \boldmath$r_{\rm h}$, \boldmath$v_r$ & \textbf{GDR3} & \textbf{GDR5} \\
\midrule
$\hat{M}_{\rm d}$ & $0.996$ & $1.090$ & $0.978\pm0.060$ & $1.001\pm0.022$\\[0.05em]
$\hat{a}_{\rm d}$ & $1.003$ & $0.980$ & $1.024\pm0.072$ & $0.983\pm0.028$\\[0.05em]
$\hat{q}_{\rm h}$ & $1.023$ & $1.108$ & $1.039\pm0.057$ & $1.062\pm0.022$\\[0.05em]
$\hat{a}_{\rm h}$ & $0.999$ & $0.956$ & $1.001\pm0.047$ & $0.989\pm0.019$\\
\midrule
$\hat{M}_{\rm h}$ & $0.998$ & $0.927$ & $1.003\pm0.079$ & $0.981\pm 0.031$\\[0.05em]
$\hat{\mu}$       & $0.457$ & $0.526$ & $1.747\pm0.476$ & $0.684\pm 0.081$\\[0.05em]
\bottomrule
\end{tabular}
\end{center}

\caption{Same as (b) but for the 19 stars of magnitude $G<18$ mag.}
\label{opt_median_table_c}

\end{subtable}
\label{opt_median_table}
\end{table}

\subsubsection{Optimisation using the observable section of the stream}\label{opt_obs_sec}

The observable section of the M68 stream is disconnected from the globular cluster and comprises about three quarters of the leading arm. From the results of this simulation, we expect to be able to identify about $5$ per cent of the stars of magnitude $G<20$ mag in this section using data from GDR3. In order to determine whether this selection cut introduces systematic biases, and the size of the expected statistical errors, we optimise several star samples. These samples can be generated by multiple \nbody\ simulations with different initial conditions for the stars. In practice, this is not possible with our computational resources. Instead, we generate random samples by shuffling the absolute magnitudes obtained from the synthetic population (Section~\ref{sim_stell_pop}) of cluster stars, and apply the selection process described in Section~\ref{gaia_sf} to each of them.

Using this method, we generate $350$ random samples with a number of stars of $S=121\pm11$ on the observable section of the stream. We optimise with Eq.~\ref{max_dkl_median} and list the mean and standard deviation of the results in the third column of Table~\ref{opt_median_table_a}. The systematic bias is slightly reduced and the statistical errors are of about $\Range{0.6}{1.5}$ per cent of the mean values. The results are approximately compatible with the reference potential for all the free parameters, and do not show strong correlations. We therefore conclude that the selection function does not significantly bias the results of the optimisation. Note that these results are not directly comparable to those obtained in Section~\ref{opt_diff_samples} with the mean-based method for samples of $S=100$ stars along the entire stream. In the present case, the stream is much shorter and more sensitive to the presence of outliers. The use of the mean-based method results in larger variances than those obtained in this section using the median-based method.

\subsubsection{Optimisation including observational uncertainties}\label{opt_obs_unc}

In this section we study the sample of 116 stars of magnitude $G<20$ mag located in the observable section of the stream that pass the selection function (Section~\ref{obs_section}). First, to obtain a reference for the following analysis, we optimise with this sample without simulated uncertainties using the median based method. The resulting optimal configuration is given in the first column of Table~\ref{opt_median_table_b}. This result is roughly compatible with the distribution obtained for the simulated samples studied in the previous section. For each pair of free parameters, this result lies within the range containing 95 per cent of all marginal distributions. This implies that the sample studied in this section is not significantly biased.

When the $r_{\rm h}$ and $v_r$ are estimated from the cluster orbit, the result of the optimisation varies, but remains within the 95th percentile of the distribution of simulated samples. We list the result obtained in the second column of Table~\ref{opt_median_table_b}. Note that the $r_{\rm h}$ and $v_r$ are estimated for each trial potential during the optimisation process. This means that larger deviations for the trial potential cause larger errors in the estimates of $r_{\rm h}$ and $v_r$. In this way, for incorrect configurations, not only is the potential incorrect, but also the distances and radial velocities of the stars. Adding the constraint that the stars have $r_{\rm h}$ and $v_r$ compatible with the cluster orbit, in addition to returning to the cluster for the correct potential, can increase the sensitivity of the method and reduce the systematic bias, as happens in this case. However, if the phase-space location of the cluster is not properly measured or the potential model cannot reproduce the real potential for any parameter configuration, this method can introduce another source of systematic bias.

The realistic sample of 116 stream stars as observed by GDR3 includes the observational uncertainties. This means that the phase-space positions of the stars are distributions. We cannot translate these distributions directly into angle-action coordinates because the AvGF method (Section~\ref{stream_aaf}) numerically estimates the angles, actions and frequencies for points individually. Alternatively, we can generate random samples according to the simulated uncertainty distributions as described in Section~\ref{real_sample}, and optimise each one of them using the median-based method of Eq.~\ref{max_dkl_median}.

\begin{figure}
\includegraphics[width=1.0\columnwidth]{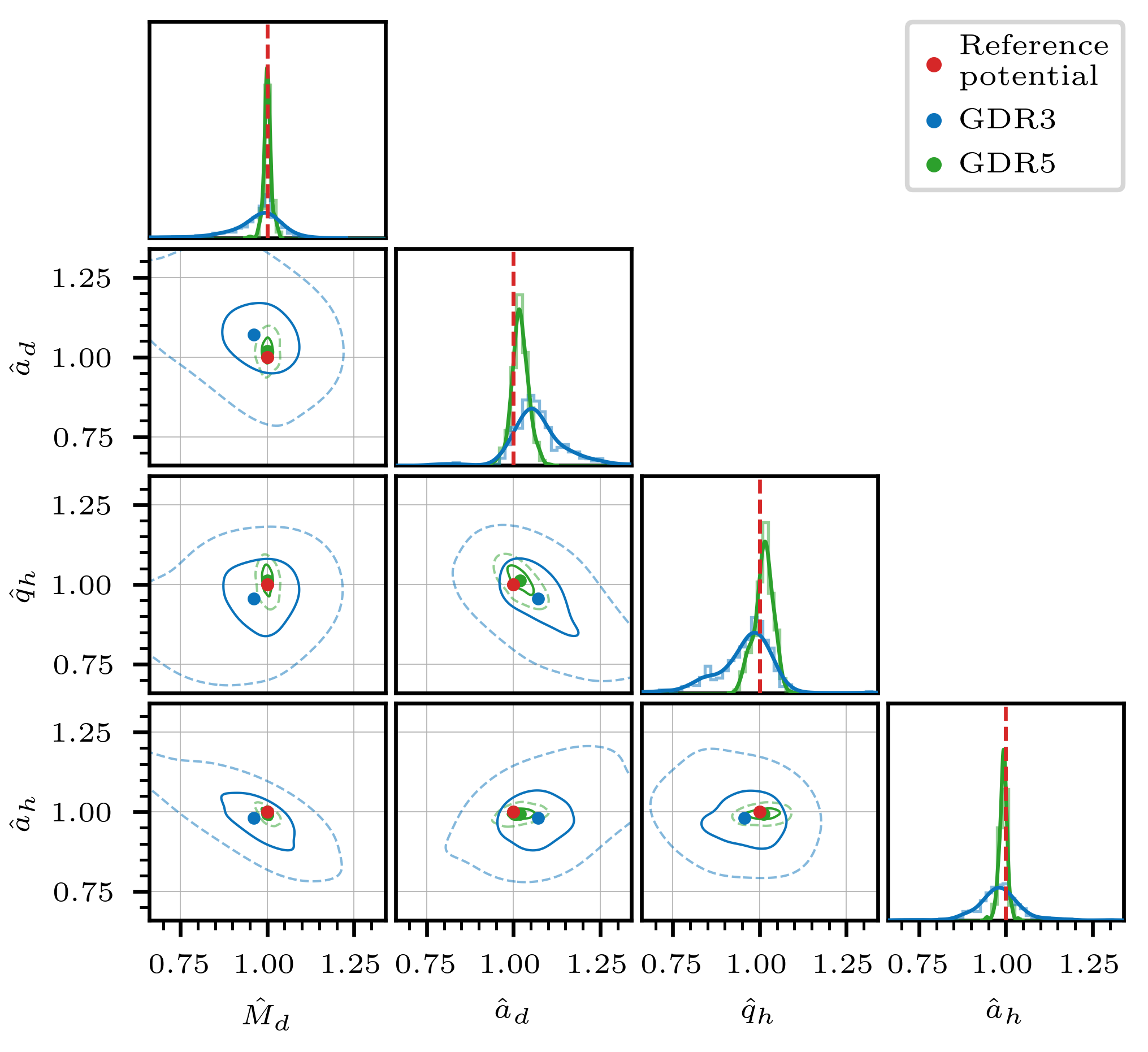}
\caption{Marginal distributions of the results of the optimisation of Eq.~\ref{max_dkl_median} using the 116 stream stars of magnitude $G<20$ mag that pass the selection function in the observable section of the stream with. The blue distributions are obtained with samples simulating GDR3 observations, and the green distributions with GDR5. The distributions are estimated using a KDE method. The values of the parameters are normalised so that the reference potential is $\hat{\kappa}=1$. \textit{Diagonal panels:} The solid lines show the estimated distributions and the light lines show the histogram of the results. The vertical dashed red line indicates the reference potential. \textit{Inferior panels:} The solid lines mark the boundary containing 68 per cent of the distributions and the dashed lines the 95 percent. The large coloured dots mark the mean of the distributions and the large red dot marks the reference potential.}
\label{corner_obs_err}
\end{figure}

Figure~\ref{corner_obs_err} shows in blue the marginal distributions of the results of the optimisations for different combinations of free parameters. The distributions are estimated using a KDE method from a set of $700$ optimisations. The diagonal panels include a histogram of the results and a vertical dashed red line marking the reference potential. In the lower panels, the solid blue line marks the region containing 68 per cent of the distribution, and the dashed blue line the 95 per cent. The mean is marked by a large blue dot, and the reference potential by a large red dot. The numerical values of the mean and standard deviation of each parameter are given in the third column of Table~\ref{opt_median_table_b}. The results are compatible with the reference potential, and the standard deviations are in the range of about $\Range{6}{9}$ per cent for all free parameters.

Given that the dominant source of error is the proper motions, we expect significant improvements in these results with future Gaia catalogues. Here we test the uncertainties expected from the GDR5, which are estimated using the performance models used in Section~\ref{gaia_obs_un}. In the right panel of Figure~\ref{stripp_err} we show the distribution of the stripping points obtained for the GDR5 samples. In this case, about $70$ per cent of the stars return to the cluster after the inverse integration. In addition, we observe that the number of outliers is significantly reduced. Most of the outliers are not caused by the errors in the proper motions, but by the estimation of $r_{\rm h}$ and $v_r$ from the cluster orbit. In Figure~\ref{corner_obs_err} we show in green the marginal distributions of the results of the optimisations for GDR5 samples. The numerical values of the medians and standard deviations are given in the fourth column of Table~\ref{opt_median_table_b}. In this case, the results are compatible with the reference potential for all parameters, and the deviations are reduced to less than about $3$ per cent of the reference values.

\subsubsection{Optimisation using the brightest stars}

We test whether these results can be improved using only the brightest stars in the sample. We choose the stars with magnitude $G<18$ mag in the observable section of the stream. We repeat the process described in Section~\ref{opt_obs_sec} and obtain samples of $S=22\pm5$ stars. We optimise with Eq.~\ref{max_dkl_median} and give the results in the fourth column of Table~\ref{opt_median_table_a}. In this case, the statistical errors increase to $\Approx 4$ per cent of the reference values as a result of the reduction in the sample size. We repeat the study of the effects of the observational uncertainties of Section~\ref{opt_obs_unc} for the subsample of 19 stars in the observable section of the stream. We test the expected observational uncertainties for the GDR3 and GDR5 catalogues and list the results in Table~\ref{opt_median_table_c}. In this case, the standard deviations of the results of the optimisation are reduced for GDR3, but do not improve for GDR5. In the following section we examine the variances of the free parameters when the statistical and observational sources of error are combined.

\subsection{Determination of the variances of the optimised parameters}

In this paper we consider two sources of randomness that determine the variability of the results of the optimisation. The first one is the selection function that introduces a variance $\sigma_{\rm stat}^2$ caused by optimising using a subsample of the stream stars. The second one is the observational uncertainties that introduce a variance $\sigma_{\rm obs}^2$ because the phase-space position of the stars is not a single point but is given by a probability distribution. In Section~\ref{opt_obs_sec} and~\ref{opt_obs_unc} the variances of the results are estimated by generating several random samples and optimising each one of them. In Table~\ref{opt_median_table} we list the normalised deviations of the results of the optimisation for each free parameter of the model, and in Table~\ref{sigma_table} we give their average for the statistical and observational cases.

\begin{table}
\caption{Average of the standard deviations of the results of the free parameter optimisations for the statistical $\sigma_{\rm stat}$ and observational $\sigma_{\rm obs}$ sources of variability. Limit $\mu_{1\sigma}$ containing 68 per cent of the distribution $E$\scalebox{0.9}{$(\epsilon_{\mu})$} for each source of variability.}
\begin{subtable}{0.47\columnwidth}

\begin{center}
\vspace{0.0em}
\addtolength{\tabcolsep}{-0.35em}
\begin{tabular}{ccc}
\toprule
\boldmath$S$ & \boldmath$\sigma_{\rm stat}$ & \boldmath$\mu_{1\sigma}$\\
&\units{per cent}&\units{mrad}\\
\midrule
$121\pm11$ & $1.1$ & $1.22$\\[0.06em]
$22\pm5$ & $3.8$ & $2.39$\\
\bottomrule
\end{tabular}
\end{center}

\caption{Statistical source.}
\label{sigma_table_a}

\end{subtable}
\begin{subtable}{0.49\columnwidth}

\begin{center}
\vspace{-0.0em}
\addtolength{\tabcolsep}{-0.35em}
\begin{tabular}{rccc}
\toprule
& \boldmath$S$ & \boldmath$\sigma_{\rm obs}$ & \boldmath$\mu_{1\sigma}$\\
&&\units{per cent}&\units{mrad}\\
\midrule
\multirow{2}{*}{ \textbf{GDR3} } & 116 & $7.9$ & $13.38$\\[0.05em]
& 19 & $5.9$ & $6.30$\\[0.35em]
\multirow{2}{*}{ \textbf{GDR5} } & 116 & $1.8$ & $1.70$\\[0.05em]
& 19 & $2.3$ & $2.87$\\
\bottomrule
\end{tabular}
\end{center}

\caption{Observational source.}
\label{sigma_table_b}

\end{subtable}
\label{sigma_table}
\end{table}

We can obtain a simple estimate of the total variance by assuming that the results of the optimisation follow a compound distribution of total variance:
\begin{equation}
 \sigma^2_{\rm tot} = \sigma_{\rm stat}^2 + \sigma_{\rm obs}^2.
\end{equation}
For a simulated sample of GDR3 stars in the observable section of the stream with magnitude $G<20$ mag we obtain $\sigma_{\rm tot} \simeq 8.0$ per cent. This value is dominated by the $\sigma_{\rm obs}$. It is therefore preferable to restrict the sample to the stars of magnitude $G<18$ mag, since the observational uncertainties of the brightest stars are significantly smaller. In this case, the $\sigma_{\rm stat}$ increases because there are fewer stars, but the combined effect reduces the total variance to $\sigma_{\rm tot} \simeq 7.0$ per cent. On the other hand, for a sample of GDR5 stars we get $\sigma_{\rm tot} \simeq 2.2$ per cent. In this case, restricting to the brightest stars does not improve $\sigma_{\rm obs}$. Instead, it worsens the total variance to $\sigma_{\rm tot} \simeq 4.4$ per cent because the statistical error increases. Thus, for high quality measurements such as those expected for GDR5, it is better to use the largest possible sample to minimise the total variance of the estimated parameters.

In practice, the $\sigma_{\rm stat}$ cannot be determined from the observed real sample of stream stars, and estimating the $\sigma_{\rm obs}$ by sampling the phase-space distributions of the stars is computationally expensive. We therefore introduce an alternative method to approximate the variances of the results of the optimisation given a single sample of stream stars. This method is based on the variability of the value of the loss function for the optimal configuration. In Section~\ref{const_pot} we introduce the loss function, which is given by the mean of the distribution of the stripping point distances $\mu$ (Eq.~\ref{mu}). For a single sample of stars, we define the error:
\begin{equation}
 \epsilon_\mu \equiv \mu\var{\kappa_0} - \mu\var{\hat{\kappa}} \hfill \geqslant0,
\end{equation}
where $\kappa_0$ are the reference free parameters and $\hat{\kappa}$ is the configuration that minimises the loss function. When this error is evaluated for several samples, we obtain a distribution of errors $E\var{\epsilon_\mu}$. The value $\mu_{1\sigma}$ is defined as the limit that contains a fraction of $1\sigma\simeq0.683$ of the distribution:
\begin{equation}
 \int_0^{\mu_{1\sigma}} E\var{\epsilon_\mu} \, d\epsilon_\mu = 1\sigma.
\end{equation}
In Table~\ref{sigma_table} we give the numerical values of $\mu_{1\sigma}$ for the statistical and observational cases obtained in Section~\ref{opt_obs_sec} and~\ref{opt_obs_unc} respectively. For a given sample of stars, the $1\sigma$-confidence interval of the results of the optimisation, assuming uncorrelated variances, is computed by solving the following equation for each free parameter $\kappa_i$:
\begin{equation}
 \mu\var{\kappa_i} = \mu\var{\hat{\kappa}_{\rm s}} + \mu_{1\sigma},
\end{equation}
where $\hat{\kappa}_{\rm s}$ are the parameters that minimise the loss function for the given sample of stars. In this way, assuming that the real results follow the error distributions of the loss function given by this simulated model, we can obtain an approximation of the statistical and observational variances from the values of $\mu_{1\sigma}$.

\section{Conclusion}\label{conclusion}

In angle-action coordinates, the dynamics of the system formed by a stellar stream and its progenitor cluster within a host potential is greatly simplified. In this framework, a stellar stream in a simple potential appears as a highly symmetrical structure and its internal components are easily identifiable. In addition, the stream stars follow a uniform rectilinear motion along the stream. These orbits can be integrated backwards in time to recover the position where the stream stars were stripped from the progenitor cluster. Determining the distribution of the stripping points relative to the cluster provides a simple method for constraining the potential of the host galaxy: in the potential that generated the stellar stream, all the stars return approximately to the centre of the cluster. In a different potential, the stars are redirected to different locations. We call this method Inverse Time Integration or \texttt{invi}.


In this paper, we use the mean of the distribution of stripping point distances relative to the cluster centre to evaluate the degree to which the stream stars return to the cluster. In this way, we reduce the system from a six-dimensional phase-space to a one-dimensional space of distances. By minimising the mean distance, we obtain the value of the free parameters that characterise the potential of the host galaxy. Thus, the \texttt{invi} method requires the complete phase-space position of the progenitor cluster and the stream stars. It also requires a time-independent potential to compute the angle-action coordinates, but makes no additional assumptions about the formation and evolution of the progenitor cluster.

Streams with a known progenitor that are long, thin, dynamically cold, and close to the Sun are best suited for use with the \texttt{invi} method. Good candidates are the streams of M2 (NGC~7089) \citep{2022ApJ...929...89G}, M5 (NGC~5904) \citep{2019ApJ...884..174G} and especially NGC~3201 \citep{2020ApJ...901...23H, 2021MNRAS.504.2727P}. In this paper we focus on the stream generated by the globular cluster M68. We determine its capability to constrain the inner potential of the Galaxy. We choose the following free parameters to characterise the potential of the Milky Way: the mass and scale length of the disc, and the axis ratio $q_{\rm h}$ and scale length of an NFW dark halo. The number of free parameters can be increased by breaking the degeneracies by including additional observational data, such as the rotation curve of the Milky Way or multiple streams together.

We simulate the M68 stream using an \nbody\ code, and using the \texttt{invi} method, we recover the parameters of the simulation with an error $\LessSim2$ per cent for all cases. These results require a correction for the systematic bias introduced by the assumption that the stars return to the centre of the cluster rather than to the centre of their respective arms. In addition, using subsamples of 100 stars, we find that there is a significant correlation between the mass of the disc and $q_{\rm h}$. This is explained because the stream flows almost parallel to the disc, and it is located close to the symmetry plane, at about $5$ kpc. This implies that a model that correctly describes the transition between the disc and the halo is necessary to accurately constrain the Galactic potential with this stream.

In practice, the limitations of the \textit{Gaia} selection function reduce the observable length of the streams and the number of stars that can be used to estimate the stripping point distribution. In the case of M68, we only observe a section of the leading arm where about $100$ candidate stream members have been identified. Furthermore, the observational uncertainties, especially on the parallax, and the absence of radial velocities do not allow us to precisely determine the angles, actions and frequencies of the stream stars. With these limitations, the surface density of the stream cannot be completely resolved and the stripping times of the stream stars cannot be precisely estimated. However, we test the application of the \texttt{invi} method when the distances and radial velocities of the stream stars are estimated from the cluster orbit during the optimisation process.

The inclusion of observational uncertainties leads to the appearance of outliers in the distribution of stripping points. The larger outliers are those stars whose relative velocities to the cluster are severely underestimated. Some of these stars do not return to the cluster after the inverse time integration because they move slower than their progenitor. In this situation, the mean is not a good estimator of the central tendency of the stripping point distribution. This problem can be solved by using a robust estimator such as the median. In this case, we recover the value of the free parameters used in the simulation without significant bias. These results include the additional information that the stars must be compatible with the cluster orbit, which is incorporated into the method when the distances and radial velocities are estimated from the orbit. If the phase-space position of the cluster is not measured with sufficient accuracy, the estimation of distances and radial velocities from the cluster orbit can cause an additional systematic bias.

In this study we consider two sources of variance on the recovered values of the free parameters. The statistical source is caused by using a subsample of about $100$ observed stream stars, and it is of about $1$ per cent of the reference values. On the other hand, the variances caused by the simulated GDR3 observational errors and the estimates of the distances and radial velocities from the cluster orbit are much larger, of about $8$ per cent of the reference values. The total variance can be reduced slightly by limiting the sample to the brightest stars, but a more significant improvement is expected with future \textit{Gaia} releases. We estimate that, using GDR5 data, the \texttt{invi} method can provide estimates of the four free parameters of the Milky Way potential studied in this paper to an accuracy of about $2$ per cent.


\section*{Acknowledgements}

We would like to thank Jordi Miralda-Escudé for his guidance and advice during the early stages of this project. We also thank the \textit{Gaia} Project Scientist Support Team and the \textit{Gaia} Data Processing and Analysis Consortium (DPAC) for the development of the \texttt{pygaia} Python toolkit.

This work is supported by National Key R\&D Program of China (2023YFA1607800, 2023YFA1607801), 111 project (No.\ B20019), and the science research grants from the China Manned Space Project (No.CMS-CSST-2021-A03). We also acknowledge the support of NSFC(12273021), the National Key R\&D Programme of China (2023YFA1605600, 2023YFA1605601) and the Yangyang Development Fund. The computation of this work is done on the \textsc{Gravity} supercomputer at the Department of Astronomy, Shanghai Jiao Tong University.


\section*{Data Availability}

The Python package \texttt{invi} containing the codes used in this paper is available on GitHub: \url{https://github.com/cgpalau-astro/invi}. The following files related to the M68 stellar stream are available on Zenodo \url{https://zenodo.org/records/17020518}:
\begin{enumerate}
 \setlength\itemsep{0.35em}
 \item \textit{N}\!\!\:-body simulation $T=1500$~Myr
 \item Mock \textit{Gaia}-DR3 star catalogue
\end{enumerate}


\vfill
\section*{Software}

The following Tools and Python \texttt{packages} are used in this research:
\begin{center}
\begin{tabular}{lp{3.6cm}r}
\textbf{Package} & \textbf{Reference} & \textbf{Version}\\[0.15cm]

\texttt{agama} & \citet{2019MNRAS.482.1525V} & \href{https://github.com/GalacticDynamics-Oxford/Agama}{1.0} \\[0.125cm]

\texttt{astro-limepy} & \citet{2015MNRAS.454..576G} & \href{https://github.com/mgieles/limepy}{1.2} \\[0.125cm]

\textsc{CMD} & Section~\ref{sim_stell_pop} & \href{http://stev.oapd.inaf.it/cgi-bin/cmd}{3.7} \\[0.125cm]

\texttt{dustmaps} & \citet{2018JOSS....3..695M} & \href{https://github.com/gregreen/dustmaps}{1.0.13} \\[0.125cm]

\texttt{gaia-unlimited} & Section~\ref{gaia_sf} & \href{https://github.com/gaia-unlimited/gaiaunlimited}{0.3.0} \\[0.125cm]

\texttt{galpy} & \citet{2015ApJS..216...29B} & \href{https://github.com/jobovy/galpy}{1.9.2} \\[0.125cm]

\texttt{naif} & \mbox{\citet{2023ApJ...955...38B}} & \href{https://github.com/digitalex/naif}{0.1.0} \\[0.125cm]

\textsc{PeTar} & \citet{2020MNRAS.497..536W} & \href{https://github.com/lwang-astro/PeTar}{1006\_285} \\[0.125cm]

\texttt{pygaia} & Section~\ref{gaia_obs_un} & \href{https://github.com/agabrown/PyGaia}{3.0.3} \\[0.125cm]

\texttt{scikit-learn} & \citet{2011JMLR...12.2825P} & \href{https://github.com/scikit-learn/scikit-learn}{1.5.0} \\[0.125cm]

\texttt{scipy} & \citet{2020SciPy-NMeth} & \href{https://github.com/scipy/scipy}{1.12.0} \\[0.125cm]
\end{tabular}
\end{center}



\bibliographystyle{mnras}
\bibliography{bib/ref.bib}



\clearpage
\appendix


\section{Selection of M68 stars}\label{App1}

We select $3\,677$ stars at the location of the globular cluster M68 passing through the cuts specified in the Code~\ref{adql_m68}. We apply several quality cuts, the most important being the Renormalised Unit Weight Error (\texttt{ruwe}), which excludes stars with poor astrometric measurements, mainly in the bright centre of the cluster. We remove 28 outliers by clustering in \texttt{phot\_g\_mean\_mag} and \texttt{bp\_rp} using the DBSCAN algorithm included in the Python package \texttt{scikit-learn} \citep{2011JMLR...12.2825P} with a threshold $\texttt{eps}=0.368$.

\begin{lstlisting}[caption={\small{ADQL query for the M68 stars.}}, label={adql_m68}]
SELECT *
FROM gaiadr3.gaia_source
WHERE 1 = CONTAINS(POINT('ICRS',ra,dec),CIRCLE('ICRS',189.867,-26.744,0.15))
AND SQRT((pmra+2.739)*(pmra+2.739) +        (pmdec-1.779)*(pmdec-1.779)) <= 1.0
AND parallax BETWEEN -2.0 AND 2.0
AND bp_rp BETWEEN -0.1 AND 1.6
AND ruwe < 1.2
AND visibility_periods_used >= 10
AND duplicated_source = FALSE;
\end{lstlisting}

\begin{tabular}{l}
\textbf{Note:}\\
Host server: \url{https://gaia.aip.de/}\\
\end{tabular}


\section{Dust reddening correction}\label{App2}

Given an observed colour index $M'$, we compute the corrected colour index $M$ by subtracting the colour excess $E\var{M}$:
\begin{equation}
M = M'-E\var{M}.
\end{equation}
We use the colour excess $E\var{\BV}$ for $\BV$ colour from the SFD Galactic dust map \citep{1998ApJ...500..525S} reduced by a factor 0.86 following the recalibration of \citet{2011ApJ...737..103S}. We use the Python interface \texttt{dustmaps} \citep{2018JOSS....3..695M} to obtain the colour excess for each location on the sky. To improve the agreement between the synthetic population generated by CMD 3.7 and the M68 stars observed by \textit{Gaia} (Appendix~\ref{App1}), we include a correction of 0.0175 mag to the colour excess from the recalibrated SFD dust map. The values obtained, expressed in $\BV$ colour, can be related to the \textit{Gaia} colour $\BPRP$, for most common stellar metallicities and surface gravities, using the approximate expression of \citet{2010AandA...523A..48J}, from their Table 3:
\begin{multline}
\BPRP = 0.0981 + 1.429\,(\BV) \\- 0.0269\,(\BV)^2 + 0.0061\,(\BV)^3.
\end{multline}
We also follow the approximation of \citet{2010AandA...523A..48J} that the dust extinction colour excess runs nearly parallel to this
colour-colour relation. Neglecting the coefficients of the second and third order terms in $\BV$, we obtain the approximation:
\begin{equation}
E\var{\BPRP} = 1.429\:\! \left(0.86\,E\var{\BV} + 0.0175\right).
\end{equation}

The extinction correction in the $\Gband$ magnitude $A_{\SM{G}}$ can be approximately expressed in terms of the colour excess $E\var{\BPRP}$. We use the expression calibrated at a typical dust extinction
$A_{\lambda = 550\, {\rm nm}} = 1$ mag, given in Table 13\footnotemark of
\citet{2010AandA...523A..48J}:
\begin{equation}
A_{\SM{G}} = 1.98\, E\var{\BPRP}.
\end{equation}

\footnotetext{\url{https://vizier.cds.unistra.fr/viz-bin/VizieR-3?-source=J/A\%2bA/523/A48/table13}}


\section{Sun phase-space position}\label{App3}

We use the mean value of the observational estimates of the position and velocity of the Sun listed in Table~\ref{Sun_table} to transform between the Heliocetric (ICRS) to the Galactocentric coordinate system. The position is given in galactocentric cylindrical coordinates, and the components of the velocity vector are given with respect to the Local Standard of Rest, where $U$ points to the Galactic centre, $V$ is positive along the direction of the Sun's rotation (clockwise when viewed from the North Galactic Pole), and $W$ is positive towards the North Galactic Pole.

\begin{table}
\caption[]{Sun phase-space position.}
\begin{center}
\begin{tabular}{lllc}
\toprule
\multicolumn{3}{l}{\textbf{Sun}}&\textbf{Ref.}\\
\midrule
$R$&\units{kpc}&$8.275\pm0.042$&[1]\\
$z$&\units{pc}&$20.8\pm0.3$&[2]\\
$U$&\units{km s$^{-1}$}&$11.10\pm1.25$&[3]\\
$V$&\units{km s$^{-1}$}&$12.24\pm2.05$&[3]\\
$W$&\units{km s$^{-1}$}&$7.25\pm0.62$&[3]\\
\bottomrule
\end{tabular}
\end{center}

\begin{tabular}{l}
\textbf{References:}\\
\text{[1]}: \citet{2021AandA...647A..59G}\\
\text{[2]}: \citet{2019MNRAS.482.1417B}\\
\text{[3]}: \citet{2010MNRAS.403.1829S}\\
\end{tabular}
\label{Sun_table}
\end{table}


\section{Classification of simulated stars}\label{App4}

The simulated stars are classified according to the following criteria:

\begin{itemize}
\setlength\itemsep{0.5em}

\item[(i)] \textit{Globular cluster:} stars with a relative Euclidean distance from the cluster centre of less than $r_{\rm trun}$ (Table~\ref{M68_properties_table}).

\item[(ii)] \textit{Escapees:} stars identified as outliers by clustering with the DBSCAN algorithm (Appendix~\ref{App1}) in the normalised subspace $\Delta\bar{\theta}^{\prime\prime}$ with parameter $\texttt{eps}=0.45$ or in $\Delta\bar{J}^{\prime\prime}$ with parameter $\texttt{eps}=0.22$. The normalisation of each coordinate $x$ is defined as a two-step process:
\begin{equation*}
\text{1)} \quad x^{\prime} \equiv \frac{x - {\rm mean}\var{x}}{\std\var{x}} \hfill \text{2)} \quad
x^{\prime\prime} \equiv \begin{cases}
                            \phantom{-}x^{\prime} & \bar{x}_1^{\prime} > 0 \\[0.125cm]
                            -x^{\prime} & \bar{x}_1^{\prime} < 0 \,.
                        \end{cases}
\end{equation*}

\item[(iii)] \textit{Stream:} stars that do not belong to the globular cluster and are not escapees. The leading arm is defined as the stream stars with $\Delta\bar{\theta}_1>0$ and the trailing arm with $\Delta\bar{\theta}_1<0$.

\item[(iv)] \textit{Internal streams (1,\,2,\,3):} For each arm, the stars are grouped by the stripping time $t_{\rm s}$ (Eq.~\ref{ts}) in three groups defined by the boundaries: $[-1500,\,-1055,\,-610,\,-200]$ Myr. The stars outside these limits are not labelled in Table~\ref{number_stars}.
\end{itemize}


\section{Resonant and trapped orbits}\label{App5}

Consider a particle following a regular orbit in phase-space in angle-action coordinates. The actions are integrals of motion, and the angles grow linearly according to Eq.~\ref{int_orb}. Using a system of modular arithmetic, we can describe the angular position with a number in the interval $[0,2\pi)$. In this way, the particle is confined to a three-dimensional topological torus, orbiting at a frequency $\varOmega_i$, where the index $i$ denotes each dimension.

An orbit is defined as resonant if the frequencies are commensurable: $\varOmega_i / \varOmega_j = n_i / n_j$, for integers $n\neq0$. In such cases, the orbit is restricted to a closed spiral that does not completely fill the surface of the torus. Assuming that ${n_i<n_j}$, we can enumerate all possible resonances by taking the new elements that appear in each order of a Farey sequence \citep[e.g.][]{2014PhRvS..17a4001T}. For example, the new rationals appearing in the sequence from order 1 to 6 are:
\begin{gather*}
 F_1 = \left\{ \frac{1}{1} \right\} \quad
 F_2 = \left\{ \frac{1}{2} \right\} \quad
 F_3 = \left\{ \frac{1}{3}, \frac{2}{3} \right\} \quad
 F_4 = \left\{ \frac{1}{4}, \frac{3}{4} \right\} \\[0.5em]
 F_5 = \left\{ \frac{1}{5}, \frac{2}{5}, \frac{3}{5}, \frac{4}{5} \right\} \quad
 F_6 = \left\{ \frac{1}{6}, \frac{5}{6} \right\}.
\end{gather*}
This structure allows us to define the strength of a resonance: the lower the order of appearance in the Farey sequence and the lower the numerator of the rational number, the stronger the resonance.

\begin{figure}
\includegraphics[width=1.0\columnwidth]{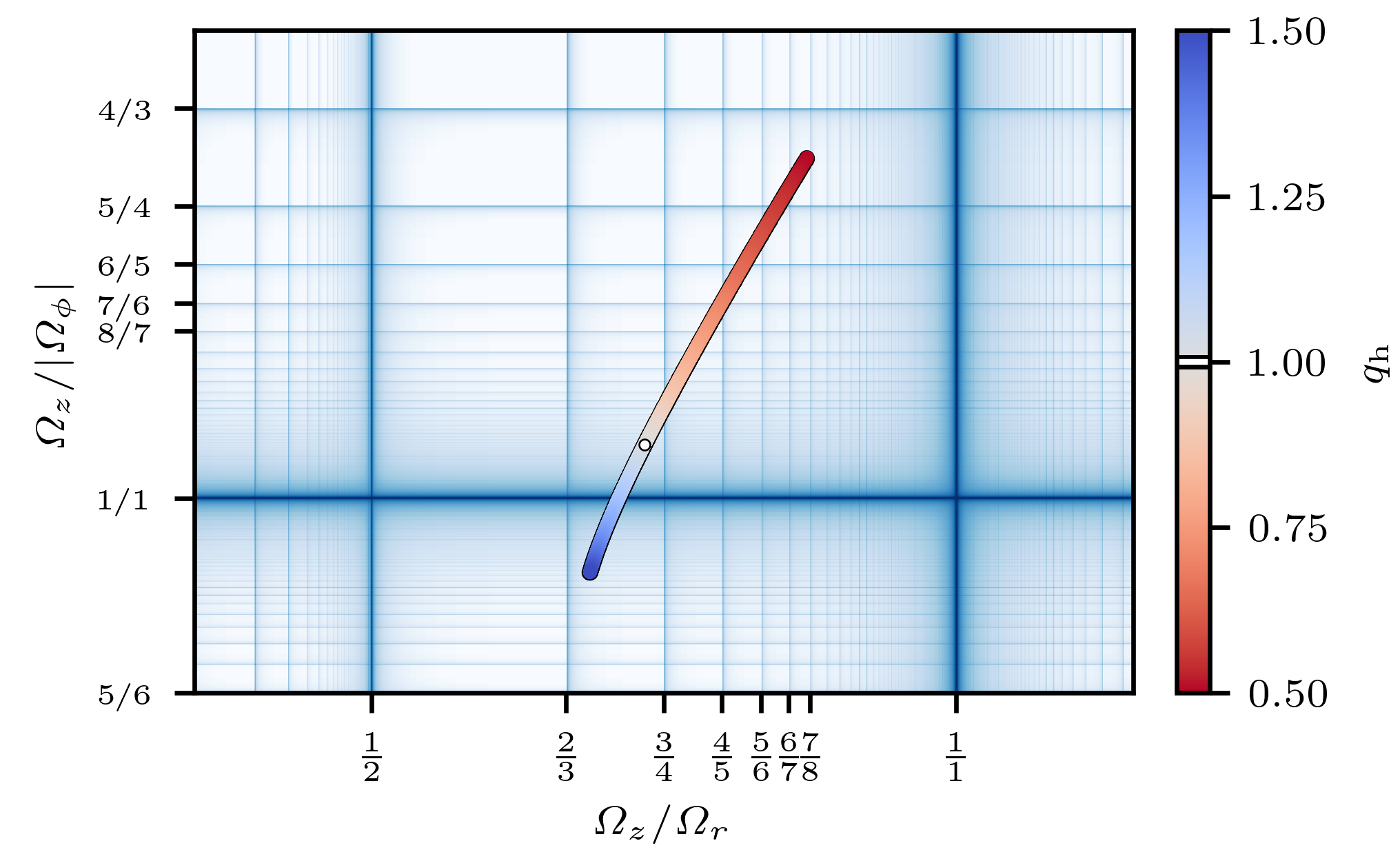}
\caption{Frequency ratios of the M68 globular cluster orbit as a function of the dark halo density axis ratio $q_{\rm h}$. The red colours in the coloured line indicate oblate halos and the blue colours prolate halos. The reference configuration ($q_{\rm h}=1$) is marked with a large white dot. The background colour is proportional to the strength of the resonance. Dark blues indicate strong resonances and light blues indicate weak resonances.}
\label{resonances_map}
\end{figure}

We characterise the orbit of the globular cluster M68 by the frequency ratios $\varOmega_z/\varOmega_r$ and $\varOmega_z/|\varOmega_\phi|$. Depending on the parameters of the potential, the cluster can enter resonances of different strengths. As an example, in Figure~\ref{resonances_map}, we plot as a coloured line the frequency ratios for different potentials characterised by the dark halo axis ratio $q_{\rm h}$ (Eq.~\ref{axis_ratio}) within the interval $\Range{0.5}{1.5}$. The oblate configurations are shown in red, and the prolate configurations in blue. The spherical dark halo used in the simulation of the stream (Section~\ref{mw_pot}) is marked with a white dot. To visualise the strength of the resonances, we plot each frequency ratio with a colour brightness proportional to its strength. Dark blues indicate strong resonances and light blues indicate weak resonances.

\begin{figure}
\includegraphics[width=1.0\columnwidth]{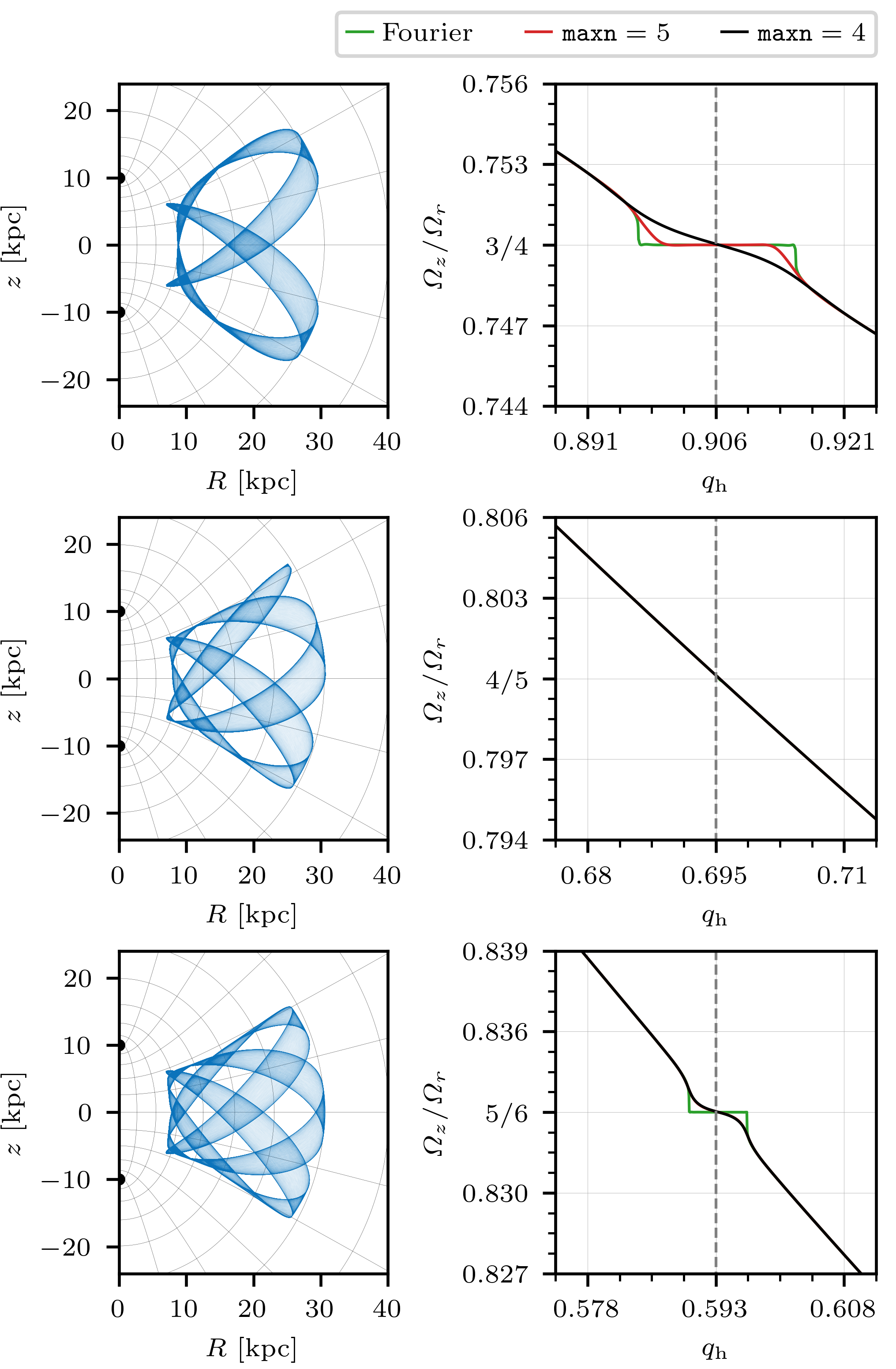}
\caption{\textit{Left:} Orbit of the globular cluster M68 in cylindrical coordinates for the resonant states $\varOmega_z/\varOmega_r$ equal to $3/4$ (\textit{Top}), $4/5$ (\textit{Middle}), and $5/6$ (\textit{Bottom}). The grid shows a prolate spheroidal coordinate system with foci located at $z=\pm10$ kpc. \textit{Right:} Frequency ratio $\varOmega_z/\varOmega_r$ as a function of the dark halo axis ratio $q_{\rm h}$ for the same resonant states. The green solid line shows the frequency ratio calculated from the dominant peaks of the Fourier spectrum of the cluster orbit. The black and red solid lines show the frequency ratio calculated by the AvGF method for $\texttt{maxn}=4$ and $5$ respectively. The vertical dashed line indicates the value of $q_{\rm h}$ for the exact resonant state.}
\label{resonances_orbit}
\end{figure}

For prolate halos, the resonance $\varOmega_z/|\varOmega_{\phi}| = 1/1$ for $q_{\rm h}\simeq1.164$ dominates. This corresponds to an orbit that is approximately contained on a two-dimensional plane. For oblate configurations, the cluster can enter several resonant states, the three strongest being $\varOmega_z/\varOmega_r = (3/4,\,4/5,\,5/6)$ for $q_{\rm h}\simeq(0.906,\,0.695,\,0.593)$ respectively. We plot these orbits in cylindrical coordinates in the left panels of Figure~\ref{resonances_orbit}. Note that, unlike $4/5$, the $3/4$ and $5/6$ resonances correspond to a symmetric orbit with respect to the Galactic plane ($z=0$). In addition, these two resonances generate a family of orbits, which are called resonant trapped orbits (\S3.7.2, \citetalias{2008gady.book.....B}). In general, the trapping resonances satisfy the following condition for $n>0$:
\begin{equation}
 \frac{\varOmega_z}{\varOmega_r} = \frac{2\:\!n-1}{2\:\!n}.
\end{equation}

We can determine the size of the trapped family as a function of $q_{\rm h}$ using the dominant peaks of the Fourier spectrum of the orbit of the cluster. We compute the ratio of the dominant peak of the coordinates $\{r, z\}$ using the method introduced in Section~\ref{sim_orb}. We plot the ratio as a green line in the right panels of Figure~\ref{resonances_orbit}. The ratio is constant for the orbits of the trapped family. We observe that the family generated by the stronger resonance ($3/4$: top panel) is larger in $q_{\rm h}$ space than the family generated by the weaker resonance ($5/6$: bottom panel). On the other hand, the $4/5$ resonance (middle panel) do not generate a trapped family and the frequency ratio varies monotonically around the resonance.

\begin{figure}
\includegraphics[width=1.0\columnwidth]{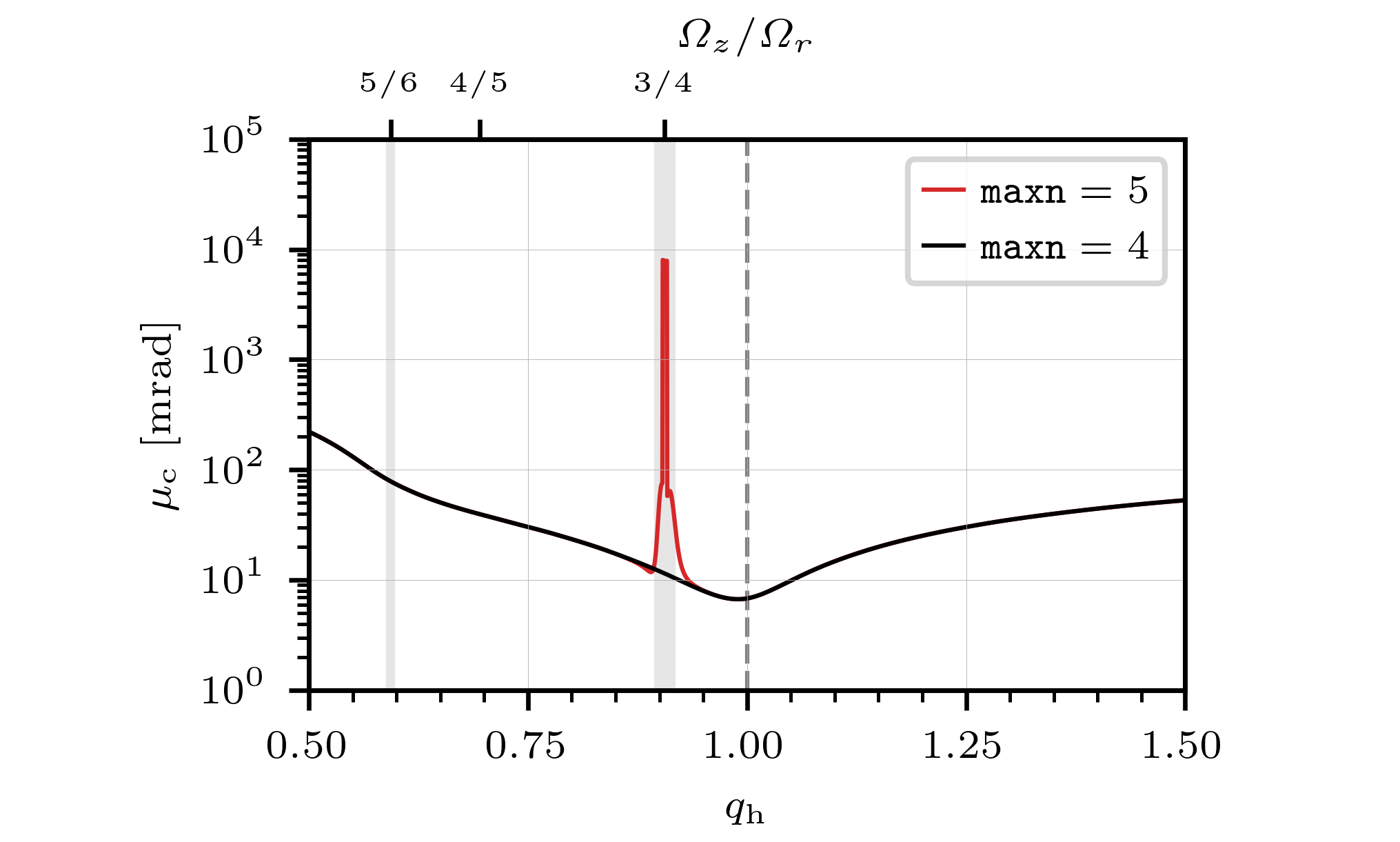}
\caption{Mean value of the corrected stripping point distance distribution of the stream stars $\mu_{\rm c}$ as a function of the dark halo axis ratio $q_{\rm h}$ calculated by the AvGF method for $\texttt{maxn}=4$ (black) and $5$ (red). The shaded areas show the size in $q_{\rm h}$ space of the resonant trapped family $\varOmega_z/\varOmega_r = 5/6$ and $3/4$. The vertical dashed line marks the reference configuration.}
\label{resonances_loss_function}
\end{figure}

Each family of resonant trapped orbits defines a disjoint phase-space where the angles, actions, and frequencies cannot be computed accurately by convergent methods \citep{2016MNRAS.457.2107S}. These errors are significant for the application of the \texttt{invi} method (Section~\ref{invi}) because they tend to spread the estimated distribution of stripping points. This results in an artificially high entropy or low degree of clustering for the configurations where the stream stars enter in a resonant trapped orbit. Figure~\ref{resonances_loss_function} shows the magnitude of this effect on the mean of the corrected stripping point distance distribution $\mu_{\rm c}$ (Eq.~\ref{mu_corr}) as a function of $q_{\rm h}$. The stripping points are calculated from Eq.~\ref{alpha} with the angles and frequencies estimated using the AvGF method configured as described in Section~\ref{stream_aaf}.

When the parameter $\texttt{maxn}$ of the AvGF method is set such that $\texttt{maxn}\geqslant5$, the errors significantly magnify the estimated mean as the stream stars enter the stronger resonance. In Figure~\ref{resonances_loss_function} we show as a solid red line the values of $\mu_{\rm c}$ obtained with $\texttt{maxn}=5$. In this case, the width of the peak is of approximately the size of the $3/4$ trapped family, which is marked with a grey area. A similar peak is not present for the $5/6$ resonance because the size of the trapped family is smaller and the resonance is not as strong. If we set the parameter $\texttt{maxn}=4$, we obtain a smooth variation of $\mu_{\rm c}$ as a function of $q_{\rm h}$. We plot this case as a solid black line. This setup therefore eliminates the resonant family.

Similarly, with the frequency ratio $\varOmega_z/\varOmega_r$ of the globular cluster orbit, we can determine whether the AvGF method resolves the presence of a resonant family for different values of $\texttt{maxn}$. In the top right panel of Figure~\ref{resonances_orbit} we plot the frequency ratio as a function of $q_{\rm h}$ around the $3/4$ resonance for $\texttt{maxn}=5$ (red line). The ratio is constant within the trapped family, similar to that estimated from the dominant peaks of the Fourier spectrum (green line). On the contrary, for $\texttt{maxn}=4$ (black line) the ratio varies smoothly as a function of $q_{\rm h}$, with only a small variation of the slope within the trapped family. This behaviour is similar to that of the $4/5$ resonance, which does not give rise to a trapped family. We plot this case in the middle right panel. The bottom right panel shows the $5/6$ resonance. Here the slope of the frequency ratio varies considerably within the trapped family, and is not constant for any value of $\texttt{maxn}$.

Although it is unlikely that the errors can increase the degree of clustering of the stripping points for a given parameter configuration and thus bias the results, in this paper we set $\texttt{maxn}=4$ to eliminate the families of trapped orbits. If an analysis of the real M68 stream data using the \texttt{invi} method concludes that the best-fitting configuration is compatible with the cluster or some stream stars being in a resonance trapped orbit, the family of trapped orbits must be analysed separately from the rest of the phase-space. In general, if it can be shown that a progenitor or its stream is trapped by a resonance, then the parameter space of the host potential would be extremely constrained.


\section{Diagonalisation of the Hessian matrix}\label{App6}

The Hessian matrix of the Hamiltonian function expressed in terms of the actions for the potential defined in Section~\ref{mw_pot} is not known. However, we can estimate its eigenvalues and eigenvectors from an \nbody\ simulation of the stellar stream. The Hessian matrix is symmetric, so it can be diagonalised by an orthogonal transformation. Restricting to rotations, we define the linear transformation:
\begin{equation}
R \equiv R_{\!\:r} \, R_{\!\:\phi} \, R_{\!\:z},
\end{equation}
where $R_{\!\:i}$ is a three-dimensional rotation matrix of angle $\varphi_i$ around the axis $i=(r,\phi,z)$, defined with respect to the original angle-action coordinates $(\theta,J)$. A rotation of the phase-space coordinates from $(\theta,J) \rightarrow (\theta',J')$, and from $(\varOmega,0) \rightarrow (\varOmega',0)$ is defined by the Jacobian matrix:
\begin{equation}
K \equiv \begin{pmatrix}
R & 0 \\
0 & R
\end{pmatrix}.
\end{equation}
This transformation is canonical because $\omega=K\,\omega\,K^{T}$, since $R^{-1}=R^{T}$, where $\omega$ is the symplectic matrix (\S D.4.3, \citetalias{2008gady.book.....B}).

Given a sample of $N$ simulated stream stars, we can rotate the actions and frequencies with respect to the centre of the cluster with the transformation $K$, and calculate the following ratios for each star:
\begin{equation}
\Lambda_j\var{\varphi} \equiv \left\{ \frac{\Delta\varOmega_{j,1}'}{\Delta J_{j,1}'}, \dots, \frac{\Delta\varOmega_{j,N}'}{\Delta J_{j,N}'} \right\}.
\end{equation}
Approximately, in the rotated coordinate frame where the Hessian is diagonal $\{\bar{\theta}, \bar{J}\}$, the ratios $\Lambda_j$ are the same for all the stars, and by Eq.~\ref{Hessian} they are equal to the eigenvalues:
\begin{equation}\label{ratio}
\lambda_j \approx \frac{\Delta\bar{\varOmega}_j}{\Delta \bar{J}_j},
\end{equation}
where the coordinates are labelled with a $j=(1,2,3)$ such that $|\lambda_1| > |\lambda_2| > |\lambda_3|$. In this way, we can estimate the eigenvalues $\lambda_j$ by calculating the rotation angles that minimise the dispersion of $\Lambda_j$. The optimal rotation angles $\hat{\varphi}$ are computed by maximising the Kullback-Leibler divergence between the distribution of $\Lambda_j$ and a uniform distribution covering the entire $\varphi$ space. This is equivalent to Eq.~\ref{dkl_uni}, and under the assumption that the ratios $\Lambda_j$ are independent, we obtain from Eq.~\ref{max_dkl}:
\begin{equation}\label{Lk}
\hat{\varphi} = \underset{\varphi\in[0,\pi/2)}{\mathrm{argmin}} \!\left( \:\! \sum_{j=1}^{3} h\var{\varphi|\Lambda_j} \right),
\end{equation}
where the differential entropy $h$ is numerically estimated with the method \texttt{ebrahimi} \citep{Ebrahimi1994} implemented in the \texttt{differential\_entropy} function included in the \texttt{scipy} Python package.

Finally, the eigenvalues are estimated using the median because it is a robust statistic in presence of outliers:
\begin{equation}
\hat{\lambda}_j = \median(\Lambda_j\var{\hat{\varphi}}).
\end{equation}
To quantify the deviations from the median, we use the Median Absolute Deviation, defined for a random variable $x$ as:
\begin{equation}\label{MAD_def}
{\rm MAD}\var{x} \equiv \frac{1}{\sqrt{2}\,\erf^{-1}\!\var{1/2}}\,\median(|x-\median\!\var{x}|),
\end{equation}
where $\erf^{-1}$ is the inverse of the error function. This statistic is a robust estimate of the standard deviation, assuming that $x$ follows a Gaussian distribution.

For the \nbody\ simulation of Section~\ref{sim_str}, using Eq.~\ref{Lk}, we obtain:
\begin{equation*}
\begin{tabular}{lll}
$\hat{\varphi}$ &\hspace{-0.7em}$\simeq$\hspace{0.3em} $(-1.3958,\,\, 0.5708,\,\, 0.6112)$ & \hspace{-0.5em}rad.\\
\end{tabular}
\end{equation*}
For this angle of rotation, the distribution of $\Lambda_2$ has a single peak, but $\Lambda_1$ and $\Lambda_3$ follow bimodal distributions. There is also a significant presence of outliers for $\Lambda_2$ and $\Lambda_3$. This may be due to small inaccuracies in the angle-action determination by the AvGF method, resulting in a slightly asymmetric stream, or to significant deviations from the assumption of a linear Hamiltonian (Eq.~\ref{Hessian}) for some stars. Finally, the numerical estimates of the eigenvalues of the Hessian matrix and their uncertainties are:
\begin{equation*}
\hat{\lambda} \simeq (-10.08 \pm 0.24,\, -0.30 \pm 0.02,\,\, 0.24 \pm 0.01) \,\,  {\rm mrad \,kpc}^{-2}.
\end{equation*}


\section{Stripping time observational estimation}\label{App7}

\begin{figure*}
\includegraphics[width=1.0\textwidth]{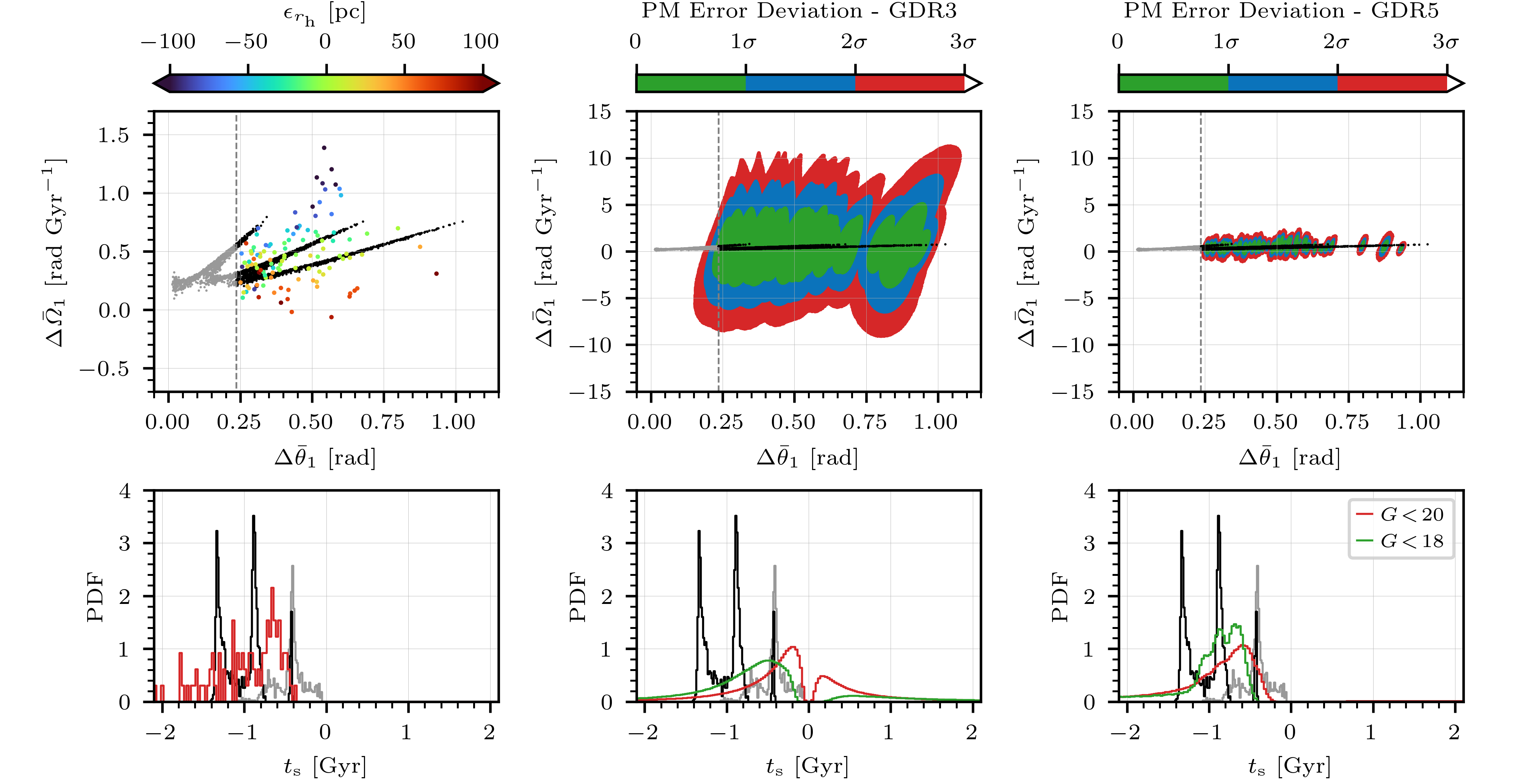}
\caption{Effect of different sources of error on the estimated relative angle and frequency along the principal axis of the M68 stream \{$\Delta\bar{\theta}_1, \Delta\bar{\varOmega}_1$\} (\textit{Top}) and on the distribution of stripping times $t_{\rm s}$ (\textit{Bottom}). The observable section of the stream is shown in black (dots and histograms), and the rest of the leading arm in grey. The two parts of the stream are separated by a vertical dashed line. The red histograms show the results for the stars of magnitude $G<20$ mag and the green histograms for the stars of $G<18$ mag that pass the \textit{Gaia} selection function. The histograms are divided into bins of 14 Myr. \textit{Left:} Effect of estimating the heliocentric distances and radial velocities from the cluster orbit. \textit{Centre:} Distributions obtained by sampling following the observational uncertainties in the sky coordinates and proper motions as expected in the GDR3 catalogue. \textit{Right:} Same as the centre panels, but for GDR5.}
\label{obs_err}
\end{figure*}

The stripping time $t_{\rm s}$ of the stream stars is estimated from the angles and frequencies of the particles according to Eq.~\ref{delta_t} and is therefore sensitive to the precision with which we can determine the angle-action coordinates. In Figure~\ref{obs_err} we show the effect of different sources of error on the estimated relative angle and frequency along the principal axis of the stream \{$\Delta\bar{\theta}_1, \Delta\bar{\varOmega}_1$\} and on the stripping time. In the top panels we plot the angles and the frequencies of the observable section of the stream as black dots. We also show the rest of the stars in the leading arm as grey dots. The small internal streams generated by tidal shocks at each pericentre passage are clearly visible. In the bottom panels we plot the distribution of the stripping times using the same colour scheme. In the correct Milky Way potential, $t_{\rm s}$ is clustered into peaks corresponding to the increase in mass loss at each pericentre passage.

In the top left panel we show the effect of estimating the heliocentric distance and radial velocity from the cluster orbit (Section~\ref{rh_rv_est}) on the sample of $116$ stars of magnitude $G<20$ mag that pass the selection function (Section~\ref{gaia_sf}). In this case, we take the exact values of the sky coordinates and proper motions from the simulation. We obtain a standard deviation of the error for the angles of $\sigma_{\Delta\bar{\theta}_1}\Approx6.3$ mrad, which we consider negligible. On the other hand, for the frequency we get $\sigma_{\Delta\bar{\varOmega}_1}\Approx0.2$ rad Gyr$^{-1}$. This value is dominated by the error in the distance estimate $\epsilon_{r_{\rm h}}$. We thus plot the stars as coloured dots, where the colour is proportional to $\epsilon_{r_{\rm h}}$. The error in the frequencies is large enough to blur and distort the peaks. The resulting stripping time distribution is shown as a red histogram in the lower left panel of Figure~\ref{obs_err}.

In the middle panels we show the effect of including the observational uncertainties of the sky coordinates and proper motions as expected from the GDR3 catalogue (Section~\ref{gaia_obs_un}). As in the previous case, the distances and radial velocities are taken from the orbit of the cluster. We plot the distribution of angles and frequencies for each star, obtained by sampling the sky coordinates and proper motions following Gaussian distributions as described in Section~\ref{real_sample}. The shape of the distributions is dominated by the uncertainties in the proper motions, while the effect of the uncertainties in the sky coordinates is negligible. To visualise the magnitude of the error in angles and frequencies caused by the proper motion error, we divide the distribution into three regions, marked in green, blue, and red. Each indicates the area of the distribution generated by samples with deviations up to 1, 2, and 3-$\sigma$ in proper motion space.

The obtained standard deviation of the error of the angles is $\sigma_{\Delta\bar{\theta}_1}\Approx13.5$ mrad. For the frequency we get $\sigma_{\Delta\bar{\varOmega}_1}\Approx1.0$ rad Gyr$^{-1}$, which is about $2$ times larger than the size of the stream in $\Delta\bar{\varOmega}_1$. When the frequencies are overestimated, the distribution of stripping times is systematically biased towards zero, as it is shown by the red histogram in the middle lower panel. Negative values of $t_{\rm s}$ can be obtained when the frequency is underestimated. This corresponds to a non-physical situation where the stars do not return to the cluster, and the estimated stripping times are positive, or equivalently, the stars are stripped in the future. The number of stars with a negative frequency can be reduced by restricting the sample to the brightest stars, since they have smaller observational uncertainties. We plot as a green histogram the distribution of stripping times obtained by sampling within the observational uncertainties the stars with G<18 mag, which corresponds to a sample of $19$ stars. In this case, the distribution of $t_{\rm s}$ is also biased towards zero and the presence of stars with negative frequency is not completely eliminated.

We estimate that to recover the peaks in $t_s$ it is necessary to measure the distance of the stream stars with a precision of $\LessSim 10$ pc, the radial velocity with $\LessSim 1.0$ km s$^{-1}$, and the proper motion with $\LessSim 10$ \textmu as yr$^{-1}$. Such precision is achievable for bright stars, except for the distance. The method presented in Section~\ref{rh_rv_est} for estimating distances from the orbit of the cluster cannot give estimates with the required precision. It gives estimates with a deviation of $\Approx 60$ pc (Table~\ref{str_orb_offset}), assuming the exact position of the globular cluster and the correct potential of the Galaxy. We therefore conclude that the stripping time cannot be accurately measured with the expected observational uncertainties on the distances and velocities of the stream stars.

\subsection{Average stream age estimation}

The observational uncertainties of the proper motions are expected to improve significantly in future \textit{Gaia} catalogues. As an example, in the right panels of Figure~\ref{obs_err} we plot the distribution of angles, frequencies, and stripping times for simulated GDR5 observations. In this case, $\sigma_{\Delta\bar{\theta}_1}\Approx4.8$ mrad and $\sigma_{\Delta\bar{\varOmega}_1}\Approx0.2$ rad Gyr$^{-1}$. The size of these errors allows us to estimate the average age of the stream from the distribution of stripping times in a similar way as \citet{2018ApJ...859L..13B}. If we restrict to the stars of magnitude $G<18$ mag, the stars with estimated negative frequencies are eliminated. The resulting distribution of stripping times has a peak comparable to the dispersion of the observable section of the stream, and a tail corresponding to the stars with underestimated frequency. This distribution is shown as a green histogram in the right bottom panel of Figure~\ref{obs_err}. We estimate the age of the stream $T_{\St}$ as the median and MAD (Eq.~\ref{MAD_def}) of the distribution of $t_{\rm s}$, and obtain $T_{\St}\simeq0.91 \pm 0.23$ Gyr. This value is compatible with the age of $T_{\St}\simeq1.02 \pm 0.18$ Gyr computed using the exact coordinates of the $1\,752$ simulated stars in the observable section of the stream.


\section{Distance distributions}\label{App9}

Consider a 2-dimensional random variable $x=(x_1, x_2)$ that follows a bivariate Gaussian distribution with mean $\mu=0$, variances $\sigma^2_1$ and $\sigma^2_2$, and correlation coefficient $\rho\in[0,1)$. Then, the random variable $R\equiv\sqrt{x^2_1+x^2_2}$ follows the Hoyt distribution with probability density function:
\begin{multline}
 f_{\Ht}\var{R\:\!|\:\!\sigma_1, \sigma_2, \rho} \equiv 2\:\!\varGamma \exp\Bigl( -\varGamma^2 \! \left(\sigma^2_1 + \sigma^2_2\right) \Bigr) \\ \IO\! \left( \varGamma^2 \sqrt{(2\:\!\rho\:\!\sigma_1\sigma_2)^2 + \left(\sigma^2_1 - \sigma^2_2\right)^2} \right),
\end{multline}
where $\varGamma \equiv R/(2\:\!\sigma_1\sigma_2\sqrt{1-\rho^2})$, and $\IO$ is a modified Bessel function of the first kind given by the series:
\begin{equation}
 \IO\:\!\!\var{z} \equiv \sum_{k=0}^{\infty} \frac{ (z^2/4)^k }{ (k\:\!!)^2 }.
\end{equation}

When $\sigma_1=\sigma_2\equiv\sigma$ and $\rho=0$, the Hoyt distribution is equivalent to the Rayleigh distribution with probability density function:
\begin{equation}\label{ray}
 f_{\RL}\var{R\:\!|\:\!\sigma} \equiv \frac{R}{\sigma^2} \exp\:\!\!\left( -\frac{R^2}{2\sigma^2} \right).
\end{equation}
The Rayleigh distribution is characterised by the mean:
\begin{equation}\label{mean_ray}
 \mu_{\RL} = \sigma\sqrt{\frac{\pi}{2}},
\end{equation}
median:
\begin{equation}\label{median_ray}
 \nu_{\:\!\RL} = \sigma\:\!\sqrt{2\ln{2}},
\end{equation}
and differential entropy:
\begin{equation}\label{entropy_ray}
 S_{\RL} = 1 + \frac{\gamma}{2} + \log\:\!\!\left(\frac{\sigma}{\sqrt{2}}\right),
\end{equation}
where the Euler's constant is $\gamma\simeq0.5772$.


\section{Bias correction}\label{App10}

The systematic bias obtained when optimising the parameters of the potential can be significantly reduced if the orientation of the plane of the stripping points can be determined for each parameter configuration. In this case, we can redefine the distance function $\Delta d$ (Eq.~\ref{delta_d}) as the distance from the stripping points of the stream stars $\Delta \alpha$ (Eq.~\ref{alpha}) to the centre of their corresponding arm $\Delta c$:
\begin{equation}
\Delta d_{\rm c} \equiv \norm{\Delta\alpha-\Delta c\:\!},
\end{equation}
and the mean as:
\begin{equation}\label{mu_corr}
\mu_{\rm c}\equiv\mean\!\var{\Delta d_{\rm c}}.
\end{equation}

In the reference frame defined by the principal axis of the stream, the position of the centre of the arms can be approximated by:
\begin{equation}
 \Delta \bar{c}\, \Approx \frac{\mu_{\rm h}}{\sqrt{\pi}} \, \Delta\bar{\delta},
\end{equation}
where the position on the plane of the stripping points is $\Delta\bar{\delta}=(0,-1,-1)$ for the leading arm and $\Delta\bar{\delta}=(0,1,1)$ for the trailing arm, and the mean of the distribution of distances on the plane $\mu_{\rm h}$ is estimated according to Eq.~\ref{alpha_mass}. The proportionality factor $1/\sqrt{\pi}$ is determined assuming that the stripping points on the plane follow a Rayleigh distribution (Eq.~\ref{ray}). The peak of this distribution is located at a distance from the centre of the cluster $R=\sigma$, or equivalently by Eq.~\ref{mean_ray}, at $R=\mu_{\rm h}\sqrt{2/\pi}$. From this radius, we obtain the proportionality factor by the Pythagorean theorem.



\bsp	
\label{lastpage}

\end{document}